# Butterfly

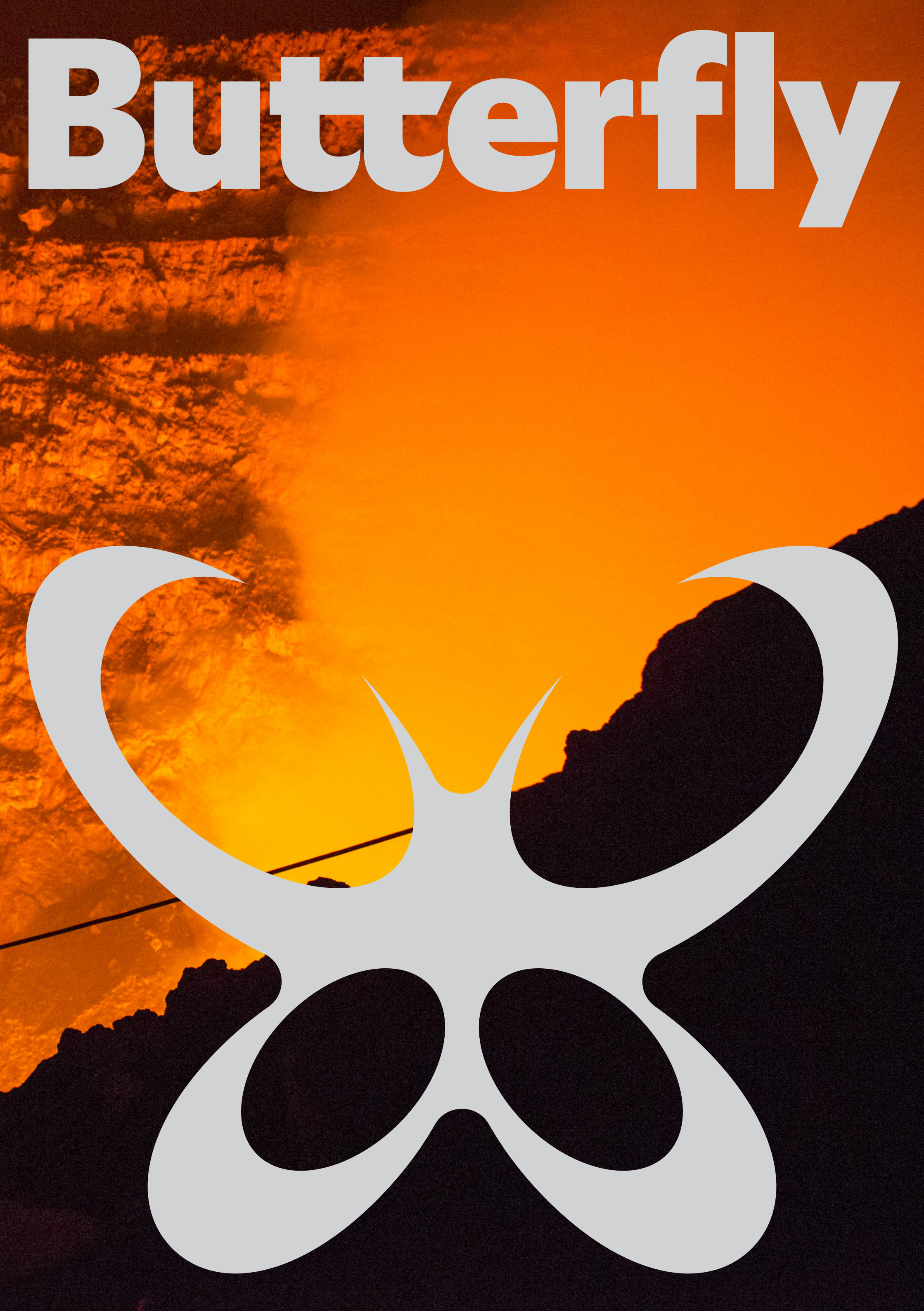

# Butterfly: Glo-cal Effects of Data, Energy, and Industry

## An Interactive New Media and Performance Art Exhibition

11–31 August, 2025
Wasa Innovation Centre
& Bock's Corner Village
Gerbyntie 16, Vaasa, Finland

## Contents

# Statement

The exhibition **Butterfly: Glo-cal Effects of Data, Energy, and Industry** is, at its core, a meditation on entanglement—between the global and the local, the ecological and the digital, the material and the virtual. It asks how we might reframe the infrastructures that shape our lives not only as technologies of efficiency or convenience, but as ecosystems themselves—dynamic, interdependent, and in need of care. It emerges from a prior exhibition, *EcoDigital Futures*, presented as part of Melbourne Design Week 2024 in Australia—an initiative of Creative Victoria and the National Gallery of Australia. That landmark exhibition spotlighted the growing imperative to align our digital futures with ecological sensibilities. *Butterfly* carries forward this vision, but also intensifies it by shifting the focus from speculation to activation, from theory to lived intervention.

This exhibition features an array of immersive works and digital media art by researchers and artists operating at the frontier of digital and creative practice. The exhibition and its catalogue provide a space in which scholars and professionals may contemplate the unfolding conditions of our planet, and the interactions between our activities and ecologies implicated. The art engages with, unravels and unpicks the very technologies and industries that enable them, while offering new insight into how our socio-technological systems may be reconfigured to open up a future that has not yet been considered. This is pertinent from a sustainable development perspective, as we desperately need to find solutions that are floating somewhere outside the box.

**These are not just artifacts of creativity—they are propositions.** Each work interrogates the 'glo-cal' intersections of data flows, energy systems, and industrial networks, inviting us to see where responsibility lies, and where transformation might begin. Inspired by *ecosophy*, a philosophy of ecological wisdom and balance, the exhibition foregrounds ethical engagements with technology. Here, code becomes climate, interface becomes ecology, and data becomes a site of resistance and renewal. *Butterfly* is both the culmination and beginning. It reflects our commitment to fostering research that not only critiques, but contributes—actively and ethically—to our shared digital and ecological futures. We invite you to explore these works not as isolated gestures, but as interconnected signals in a growing constellation of change.

As part of the academic and curatorial program, *Butterfly* also hosted an international one-day symposium on Tuesday, August 12th, 2025, at the Wasa Innovation Centre. This symposium acted to expand the reach of the exhibition into scholarly terrain, fostering critical dialogue across disciplines—from engineering and energy to the arts, communication, and cognitive science. Drawing inspiration from Chaos Theory and the metaphor of the "butterfly effect," the symposium explored how subtle shifts within technological and ecological systems can lead to profound, often unpredictable consequences. But as meteorologist Ed Lorenz originally proposed, even in the midst of complexity, prediction is not beyond reach—particularly when approached through the creative lens of artistic inquiry.

This catalogue features a rich collection of essays about art, design, creativity, sustainability, ethics and privacy. We view these areas as driving forces that should be combined and applied to topics implicating digitalisation, data, energy, economics, politics and culture, in order to operationalise and isolate elements that may and will contribute to forging a new sustainable future. We also see into the inner-workings of the artists in the statements about their art, ranging from the highly personal to the structural, historical and cultural. *Butterfly* is testimony to the value of art and design as experimental spaces that probe the possibilities of tomorrow.

**Toija Cinque, Rebekah Rousi, Aska Mayer & Esteban Guerrero Rosero**
**Butterfly: Glo-cal Effects of Data, Energy, and Industry, Co-Curators**

# Statements from Our Patrons

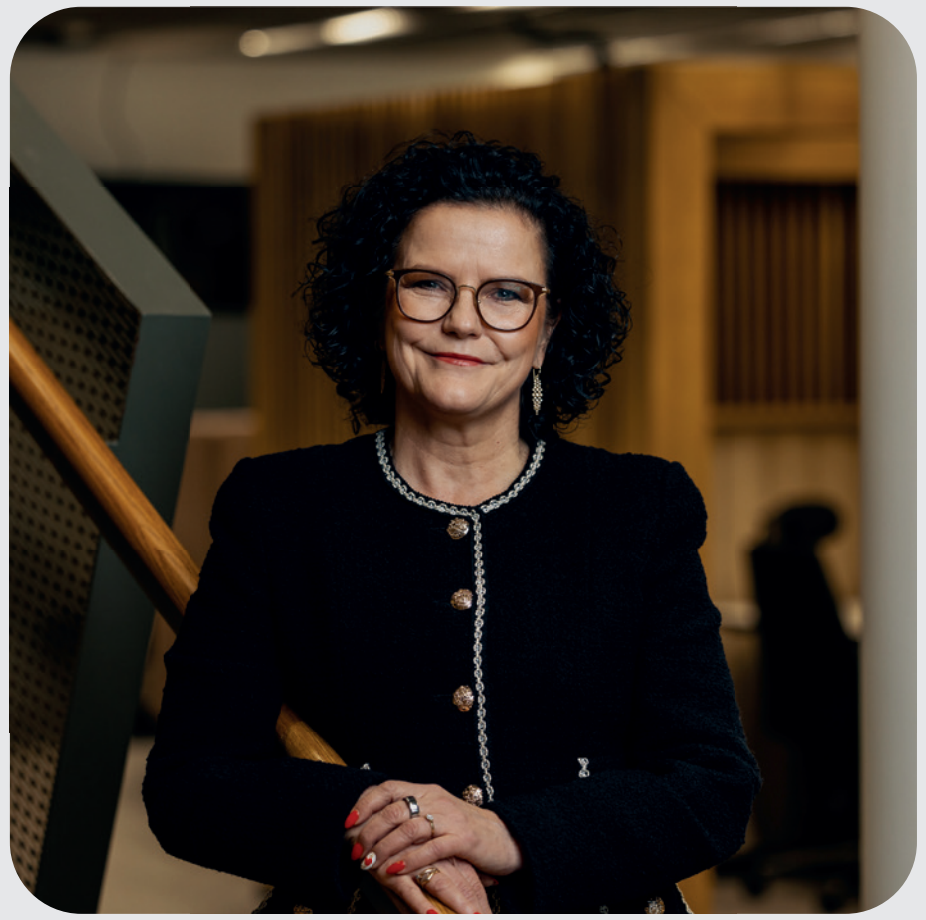

Art inspires and attracts creativity in thinking. Art increases interaction and prevents prejudice. Butterfly is a great example of transdisciplinary international collaboration, and embodies the spirit of what the University of Vaasa is, and will be to come—open, innovative, and leading the way in topical issues.

**Minna Martikainen**
Rector, University of Vaasa

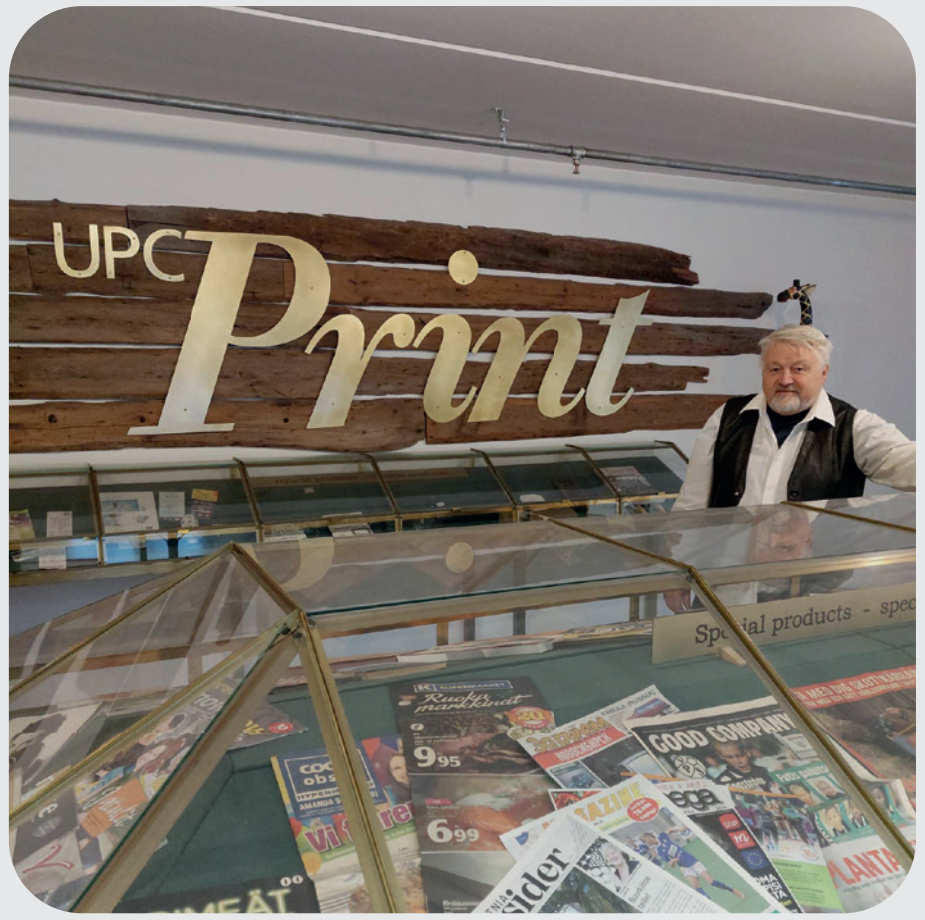

The Butterfly exhibition is a site for impact in which the New Nordic transpires as a transdisciplinary agora that shines through art. This will be the cherry on the cake of what promises to be an incredible event, drawing politicians, business, academia, and the Nordic community together to discuss what is happening, and innovative solutions to move forward with. Butterfly will be our meeting ground for making things happen.

**Sture Udd**
CEO, Wasa Innovation Center

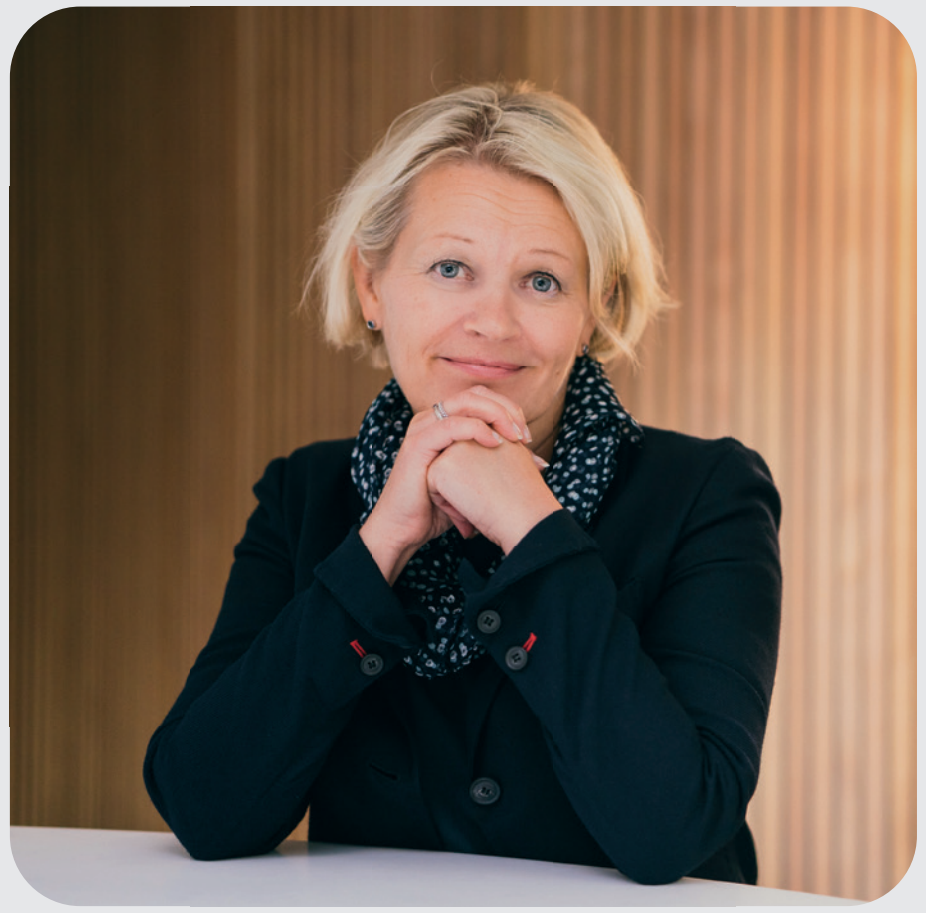

Butterfly is born from freedom, joy and the unbridled power of bringing together surprising elements and experts in art, science, and technology,  from different countries. This international art and science show proves that change starts with people of good will, for whom change is an opportunity in which they want to participate of their own free will.

**Elina Melgin,**
Producer of the exhibition,
Professor of Practice,
University of Vaasa

# Exhibition Team Bios

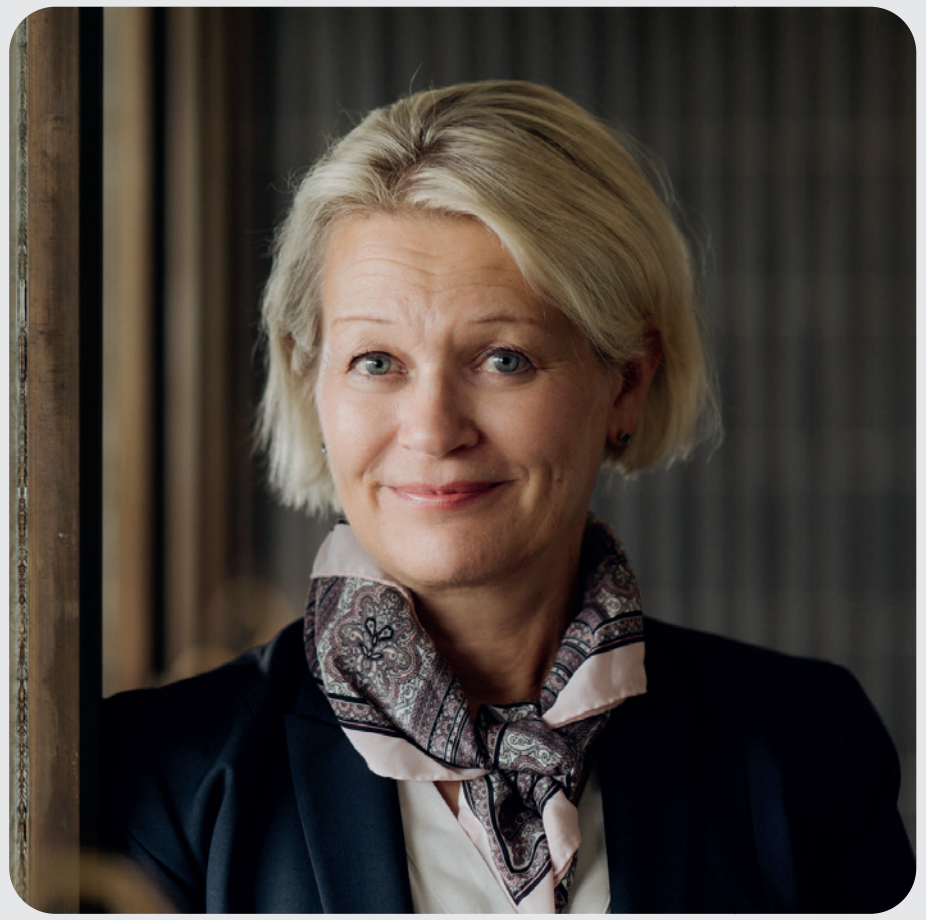

**Elina Melgin (Producer)** Ph.D., is Professor of Practice of the University of Vaasa's School of Marketing and Communication, Finland. She is also Adjunct Professor of the University of Turku, Finland. She served as a Managing Director of ProCom, the Finnish Communication Professional Association, for 17 years, representing the communications industry both in Finland and internationally. After stepping down in autumn 2022, she became an entrepreneur and senior advisor for various organizations, including T-Media, a leading Finnish company focused on reputation and strategic consulting. In 2022, she received the ProCom award "Elämäntyöpalkinto" (trans. Lifetime Achievement Award) for her life-long career in communication. Elina Melgin is also a talented painter. Since 2019 she has participated in several art exhibitions in Finland and abroad.

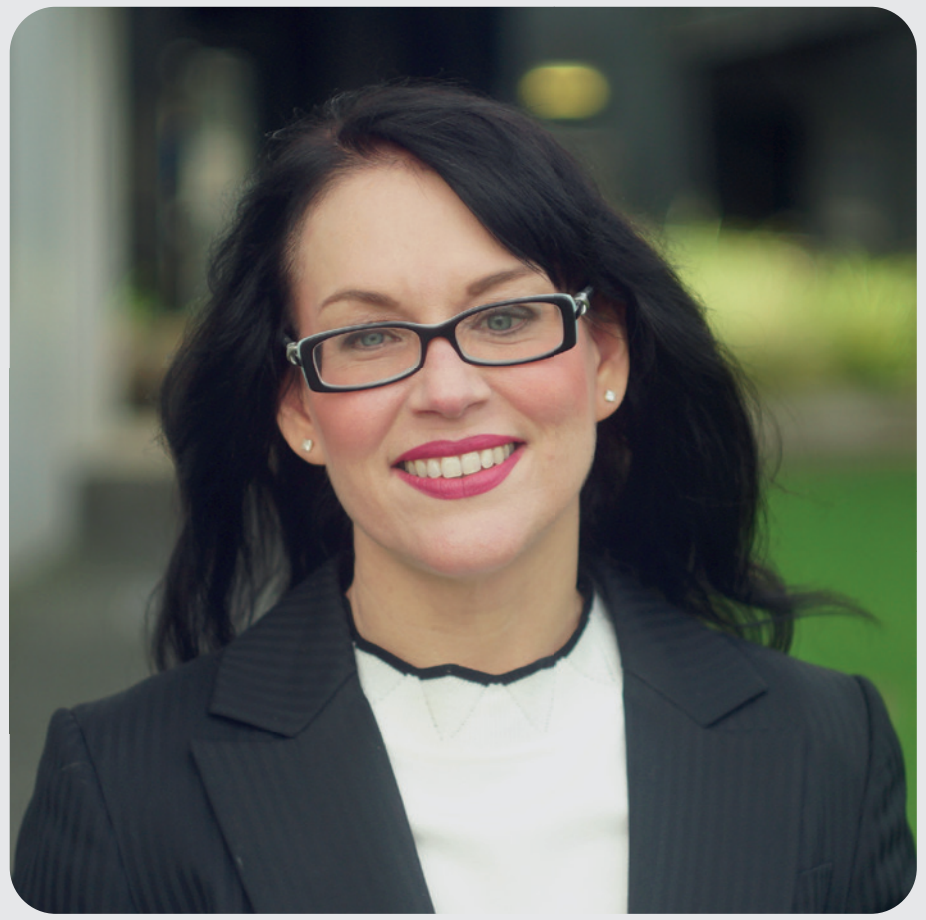

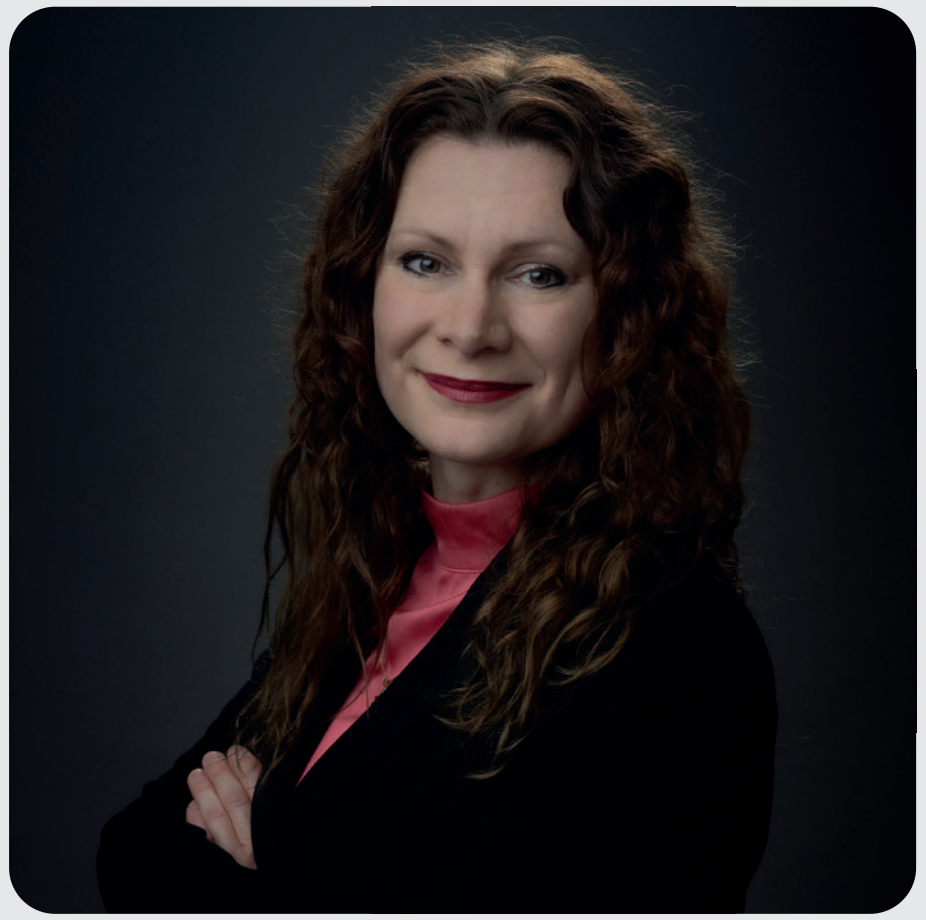

**Toija Cinque (Curator)** is Associate Professor of Communication (Digital Media) in the School of Communication and Creative Arts at Deakin University, Australia. She is the founder of the Intelligent Media Lab and leads the Critical Digital Infrastructures and Interfaces (CDII) research group. Her interdisciplinary research explores how emerging screens and digital technologies shape education, culture, and society. Drawing on diverse theoretical and methodological frameworks, Cinque investigates the transformative impacts of digitisation, datafication, and platformisation—particularly their techno-cultural dimensions and implications for our digital futures. Her latest book, Emerging Digital Media Ecologies: The Concept of Medialogy (Routledge, 2025), offers a bold rethinking of how we engage with digital systems.

**Rebekah Rousi (Curator)** is an Associate Professor of Communication and Digital Economy, University of Vaasa, who holds a PhD in Cognitive Science. Rousi is a Research Artist—undertaking her performance arts practice in the research context—completed a Bachelor of Art (Visual Art) with First Class Honours at the Western Australian Academy of Performing Arts (WAAPA), Edith Cowan University, Australia, with a Masters in Nordic Arts and Cultural Studies (Digital Culture major) from the University of Jyväskylä, Finland. Rousi's work focuses on human-AI, and human-robot interactions, particularly from the perspectives of ethics, embodiment, trust, privacy and posthumanism. Rousi has significant experience in art and academic event organization.

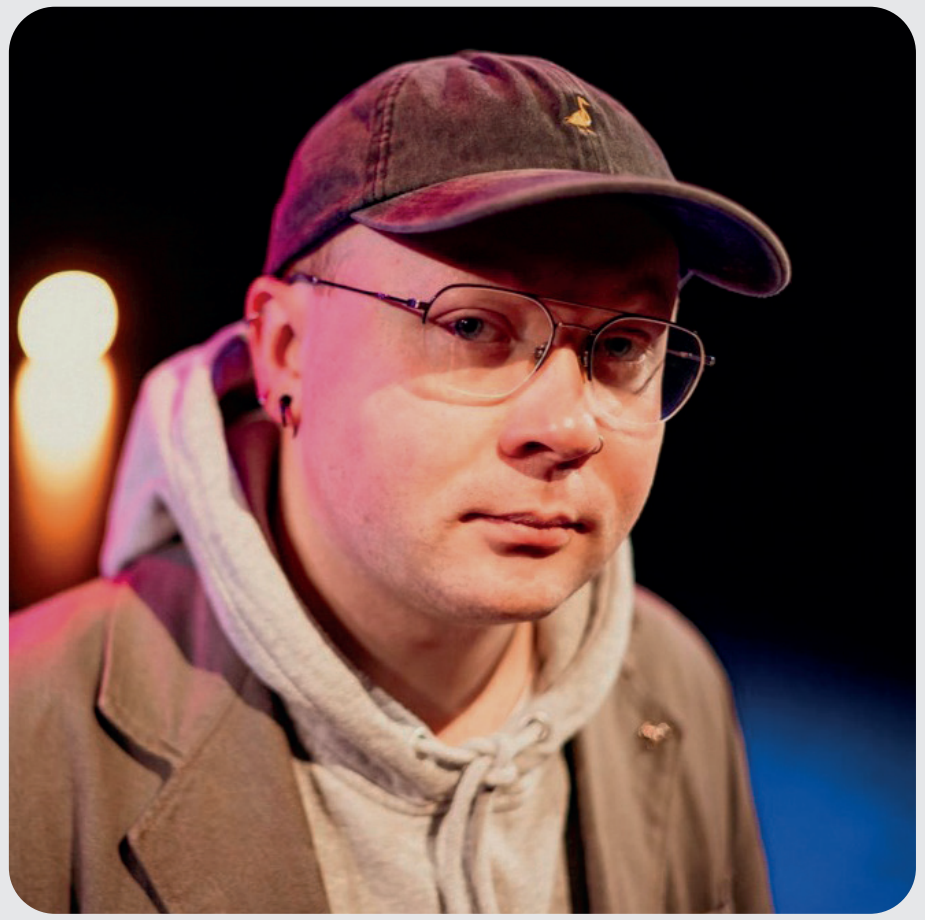

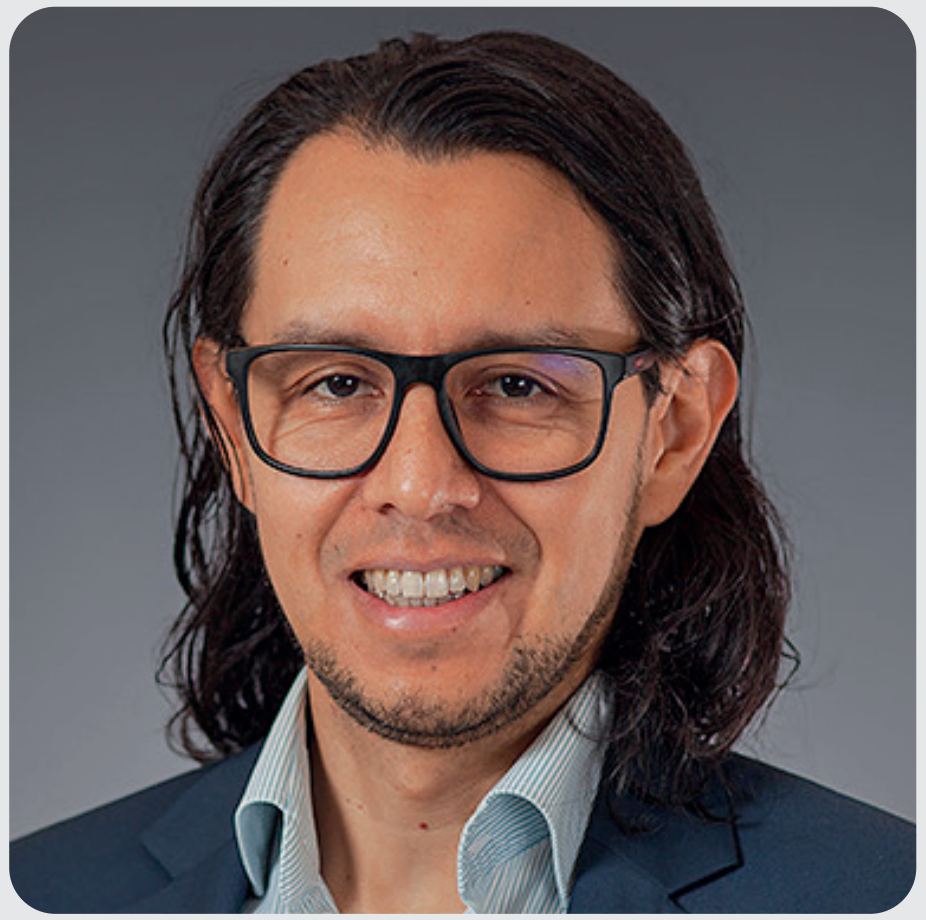

**Aska Mayer (Curator)** is a Finland-based researcher and curator, working on the intersections of Game Culture Studies, Human-Computer Interaction, Science and Technology Studies, and Semiotics. Mayer is interested in investigating futurized technologies and technology practices through the lens of popular transmedia narratives and arts, and has a background in studies of neo-baroque media, as well as in research on the use of game culture in curating and mediation.

**Esteban Guerrero Rosero (Curator)** is an Associate Professor in Social-aware Artificial Intelligence at the Department of Computer Science at Umeå University. He specializes in Artificial Intelligence (AI), focusing on neuro-symbolic approaches of AI, which is the combination of machine learning and reasoning, and formal methods of AI. His contributions have been applied and evaluated in various contexts, including finances, health, sports, ambient intelligence, and e-learning. Currently, Esteban co-leads the Formal Methods for Trustworthy Hybrid Intelligence research group at Umeå University. He is involved in multidisciplinary research and development projects in Finland and Sweden, addressing intelligent systems for supporting, recommending, and tutoring human activities. Esteban holds a Ph.D. in computing science from Umeå University (Sweden), an M.Sc. in computer science from Malmö University Sweden), and a B.Eng. in electronics and telecommunications from the University of Cauca (Colombia).

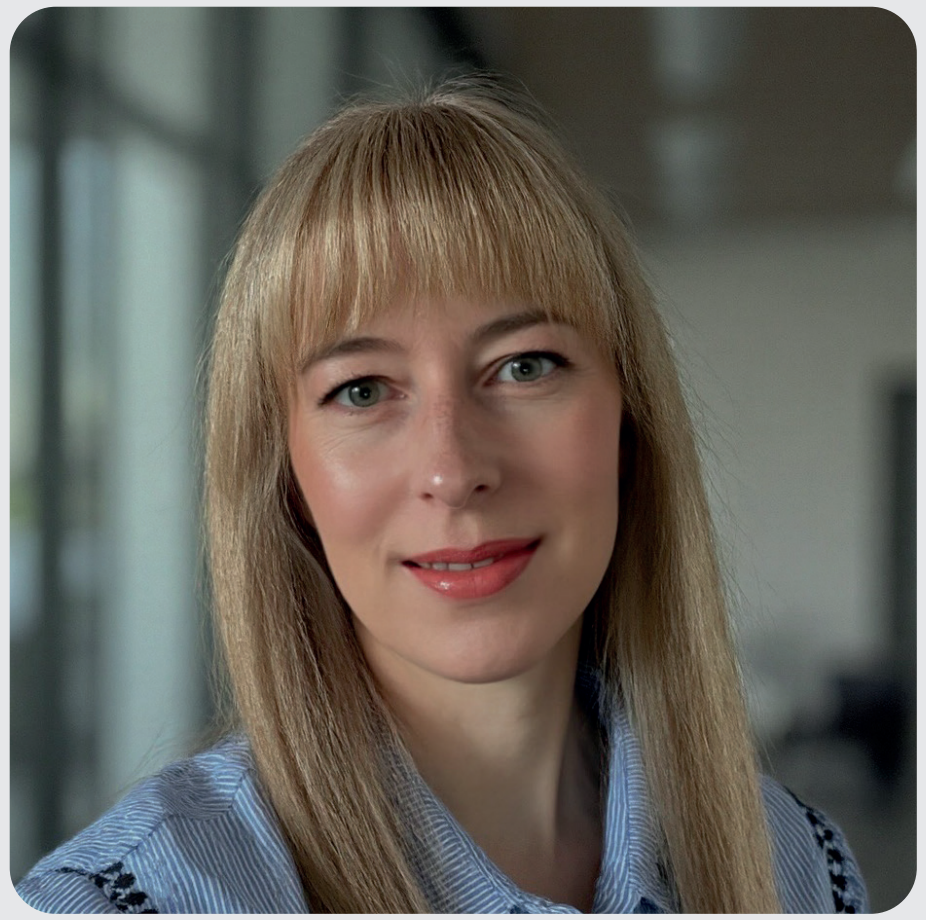

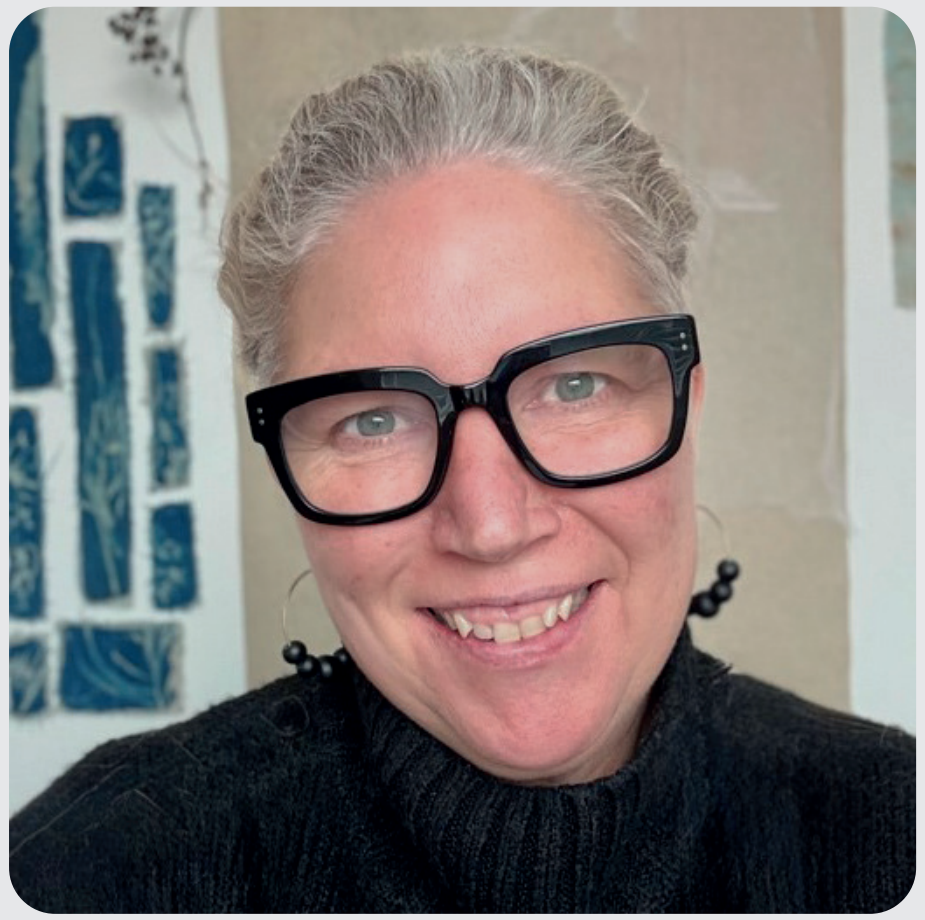

**Lyndsey Morley (Exhibition Designer)** With over 15 years of experience in interior architectural design and project management, Lyndsey has led numerous multidisciplinary projects across the globe. Her career spans from working with London's top branding agencies at Dalziel and Pow to contributing to the vibrant academic environment of Finnish universities. Morley's journey began in interior architecture and retail design, specialising in transforming brand identity into engaging, user-centered spaces that leave lasting impressions. After relocating to Finland, Lyndsey joined Vaasa University, initially contributing to the strategic redevelopment of key campus buildings and evolving into a broader position within the Facilities Services department, overseeing the maintenance, development, and optimization of the university's physical infrastructure. Thriving in dynamic, innovative environments, Morley bridges creative vision with operational efficiency, fostering collaboration across teams to create spaces that inspire and support the diverse needs of the university community.

**Katey O'Sullivan (Communications)** is an Australian visual artist living and working on Wadawurrung Country, Geelong. Her experimental art making is connected through underlying themes of storytelling, community, and nature. O'Sullivan likes to use repurposed and gathered natural materials, in abstract, conceptual ways. Embracing opportunities for tension in her work, O'Sullivan's practice begins with self-imposed constraints yet happily surrenders to the element of chance. Her work includes sculptural installations, textiles, alternative printmaking, and photography. Completing a BA in Visual Arts in June 2025, O'Sullivan has participated in group shows at Deakin University's Project Space Gallery. A co-founder of after.fluxus.Collective and Regen-Art Geelong, she has exhibited at Geelong's Third Space + Digital Gallery in 'Pernicious' and received City of Greater Geelong grant funding to produce 'Hidden Treasures', an art show celebrating international students. O'Sullivan is currently developing her research-led arts practice during an Artist's Residency at the University of Vaasa, Finland.

# Butterfly: Glo-cal Effects of Data, Energy, and Industry

## New Media, Performance, and Installation Art

**Toija Cinque & Rebekah Rousi**

In an era marked by the fevered pulse of digital acceleration—where energy flows and data streams coil tightly around the rhythms of daily life—*Butterfly: Glo-cal Effects of Data, Energy, and Industry* offers a compelling, polyphonic intervention into our technologically entangled world. Premiering at the 2025 Wasa Futures Festival, the exhibition assembles a constellation of artists and thinkers from Finland, Australia, Sweden, Latvia, Italy, Germany and the UK, whose works reverberate across disciplines, climates, and cultures. The exhibition is not simply a platform for new media art—it is a sensorial, philosophical, and affective provocation, asking what it means to feel, to think, and to live ecologically in a hyperconnected age.

At its core, *Butterfly* grapples with the fused dynamics of digitalisation, environmental precarity, socio-cultural entanglements and industrial transformation. The selected works—ranging from interactive installations and AI-generated environments to textile-based digital hybrids and augmented performance—demonstrate that contemporary art can intervene in the intensifying ecological crisis not only as critique but as a way of knowing. The exhibition becomes a threshold between: worlds, states, the machinic and the organic.

The exhibition's title evokes the "butterfly effect", a concept rooted in chaos theory and popularised by meteorologist Edward Lorenz to describe the nonlinear sensitivities of complex systems. A single flap of a butterfly's wing, we are told, might shift the trajectory of a distant storm. In this context, Butterfly recodes the metaphor as both a warning and a possibility: an artistic strategy for understanding ecological fragility and a call to imagine how minor gestures—creative, ethical, technological—might precipitate large-scale change. Far from romantic, this conceptualisation foregrounds the asymmetries and extrac-tive pressures of the glo-cal moment: where cloud computing leaves a carbon trace, and digital convenience is underwritten by geopolitical conflict, environmental degradation, and asymmetrical access to resources.

These are the invisible architectures *Butterfly* reveals, inviting audiences to make visible the seams of the digital, to touch its heat.

**Artivism as Method: Critical Creativity and Immersive Inquiry**

Operating within the expanding field of Artivism—a term that fuses artistic practice with activist urgency—the exhibition positions art not merely as representation but as research, as method, as ethical encounter. Drawing on Pfaller Schmid's (2022) work, Artivism becomes here an epistemological tool, generating embodied knowledge that is affectively potent and politically resonant. This is art that asks, *how do we know through feeling? And what can be transformed when knowledge is produced collectively, sensorially, across borders and bodies?* The concept of ecological intimacy anchors this curatorial approach: a relational, affective awareness of our entwinement with the more-than-human world. In this exhibition, intimacy is not proximity, not nostalgia for untouched nature. It is instead the recognition of co-dependence within systems—energy grids, species networks, algorithmic flows—where human agency is reconfigured as entangled and accountable. The immersive qualities of these artworks function as threshold experiences, those that are affectively charged engagements that challenge and potentially rewire perceptual habits.

To approach digital art as a site for ecological intimacy is to resist dichotomies—nature versus culture, human versus machine—that have long underpinned ecological destruction. In this way, *Butterfly* aligns with 'medialogy' (Cinque, 2025), a new framework for critically understanding the profound socio-technological and ecological transformations reshaping our world. Moving beyond technological determinism, medialogy interrogates what media technologies do—for whom, how, and with what consequences—across diverse cultural, social, and environmental contexts. It traces how ubiquitous connectivity, AI, mobile infrastructures, and digital interfaces increasingly shape not only communication and identity, but also the material landscapes of urban life, natural ecosystems, and planetary futures. The artworks resonate with this framework: they do not merely illustrate theory but enact it. They are thinking in form, critique in colour, philosophy in motion. Deeply interdisciplinary and informed by posthumanist thinking (Braidotti, 2013; 2017; 2019), medialogy foregrounds the often unseen environmental and ethical costs of emerging technologies, while amplifying diverse perspectives, including Indigenous media practices and ecosophical thought.

The affective dimension of these works—immersive, participatory, emotionally precise—draws viewers into complex systems of data surveillance, extractivism, environmental collapse, and techno-social entanglement. This is not didactic art. It does not lecture. It implicates.

**Ecological Intimacy: Ethics of the More-than-Human**

At the heart of *Butterfly* lies the potent questions: *can immersive technologies, so often complicit in environmental degradation, also serve to reawaken our ethical obligations to the natural world? Can the virtual summon the visceral?* Drawing on the deep ecological philosophies of Arne Naess (1973/1990) and Félix Guattari (2000), the exhibition operates within an ecosophical frame, where environmental, social, and mental ecologies are interdependent and co-produced. Guattari's three ecologies offer a matrix for understanding how ecological breakdown is never simply "out there" in melting ice or burning forests—it is also internal, psychic, and social. *Butterfly* engages this model through a multiplicity of practices that blend environmental critique with explorations of mental wellness, cultural memory, and collective resilience.

Several works expose the extractive logics that underpin digital capitalism, from labour exploitation in tech manufacturing to the ecological cost of supposedly 'green' infrastructures. Other pieces speculate on post-capitalist, post-anthropocentric futures—speculative zones where repair, reciprocity, and radical care become central design principles. Rather than rehearsing catastrophe however, Butterfly gestures toward what Haraway (2016) calls "staying with the trouble"—inhabiting complexity, resisting apathy, refusing simplicity. This is not an exhibition about the end of the world; it is about the conditions through which other worlds might be made possible.

**Transnational Currents: Glo-cal Frictions and Futurities**

By bringing together a cohort of artists from diverse geographies— from the UK, Finland, Sweden, Australia, Italy, Germany to Latvia—the exhibition stages a transnational conversation that challenges the universalising logics of dominant technological narratives. It resists the flattening of experience often found in global exhibitions, instead privileging the specific, the local, the embedded. These works are attuned to particular ecologies—coastal zones, degraded forests, urban data shadows—yet, they also illuminate how these spaces are woven into planetary systems of power, pollution, and possibility.

The media works span disciplines, languages, and sensoriums. AI-generated soundscapes that echo endangered species, VR journeys through data ecologies, and textile works that map energy histories onto the skin. These are not passive spectacles but active interfaces: they ask the viewer to think with their body, to sense with their ethics. This multi-sited, multi-voiced assemblage defies the idea of art as an isolated artifact. Instead, it po-sitions the exhibition as an event—a live encounter where knowledge circulates, friction generates insight, and futures are prototyped in real time.

### The Sensorial Politics of the Small

To conclude in brief, *Butterfly* does not offer blueprints. It opens apertures. It suggests that even the smallest aesthetic gestures—a shift in form, a flicker of code, a change in tone—might alter how we sense the world, and by extension, how we act within it. In a moment defined by climate urgency and digital ubiquity, the exhibition asks us to pause, to feel, to reflect—but also to act. It reminds us that the future is not elsewhere. It is in the small, the unseen, the interconnected. It is in the flutter.

Both mirror and map, Butterfly is a sensorial manifesto for a different kind of digital ecology—one in which art becomes both witness and world-maker, both alarm and invitation. It is a space where the flap of a wing might not only predict a storm, but soften its landing.

# Artist Statements

# Toija Cinque and Peta White

## Five Victorian Frogs: An Immersive Learning Experience (ILE)

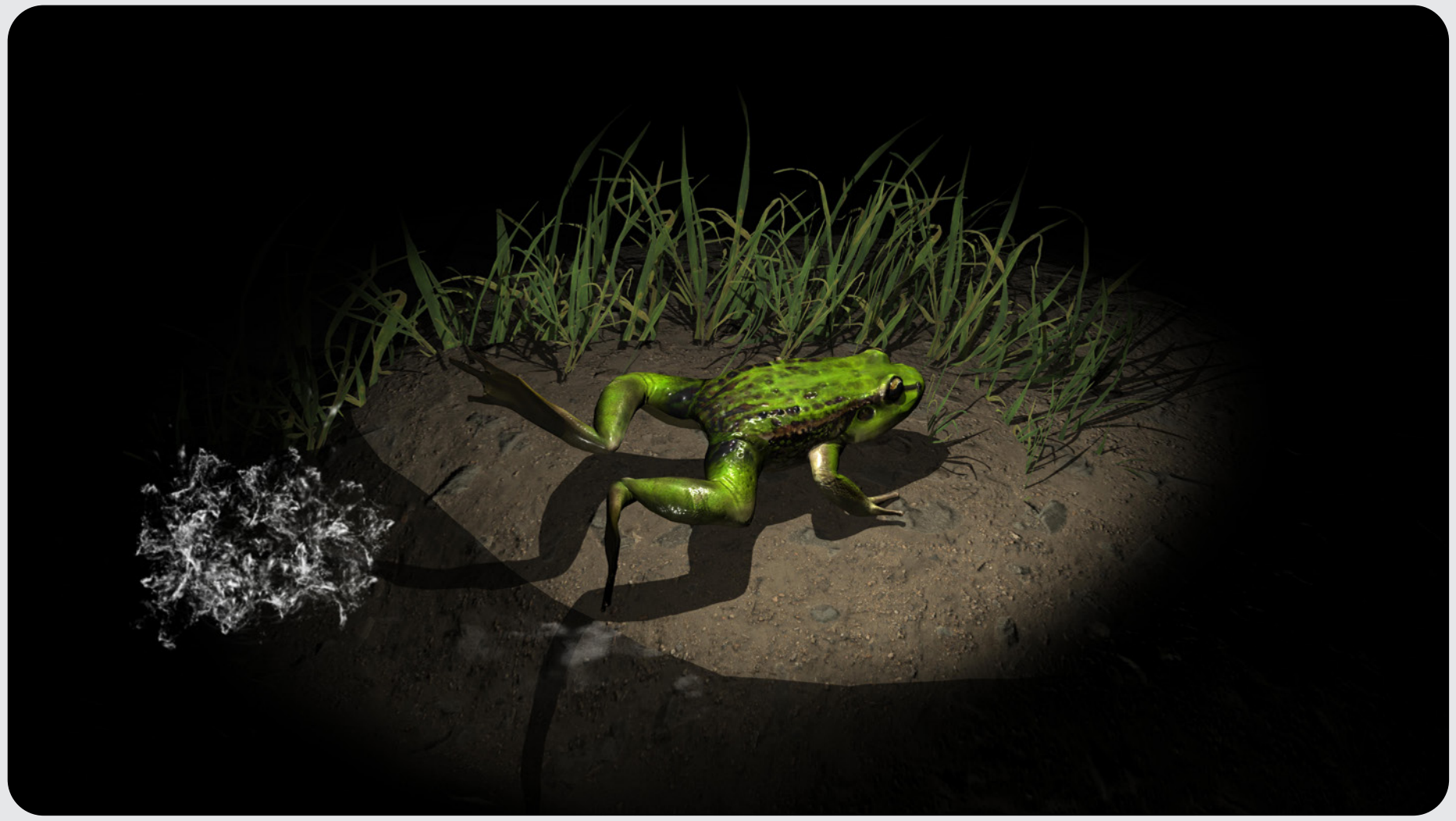

Growling Grass Frog (Ranoidea raniformis)
© Toija Cinque, Peta White (Deakin University) and QuiverVision.

This work responds to the urgency of ecological awareness through a synthesis of digital art practice and scientific inquiry. *Five Victorian Frogs: An Immersive Learning Experience (ILE)* is a new media installation that leverages augmented reality (AR) to reanimate five endemic frog species of Victoria, Australia—each a bio-indicator of ecosystem health and environmental change. Through animated, interactive visualisations, audiences encounter the Baw Baw Frog, Bibron's Toadlet, Common Eastern Froglet, Growling Grass Frog, and the Pobblebonk Frog within layered ecosystems, rich in scientific detail and sensory engagement. Drawing on research into immersive learning and science education, this installation is both a pedagogical and aesthetic intervention. It embodies a response to contemporary educational inequities intensified by global lockdowns, and to the broader disaffection from the natural world, often termed 'nature deficit disorder'. In doing so, it addresses gaps in place-based scientific education, aiming to reconnect young learners—and broader publics—with the biodiversity of their immediate environments. By integrating interactive technologies, ILE extends the potential of digital media art to foster care, understanding, and protection of fragile ecosystems.

# Anne Stenros

## Keynote presentation

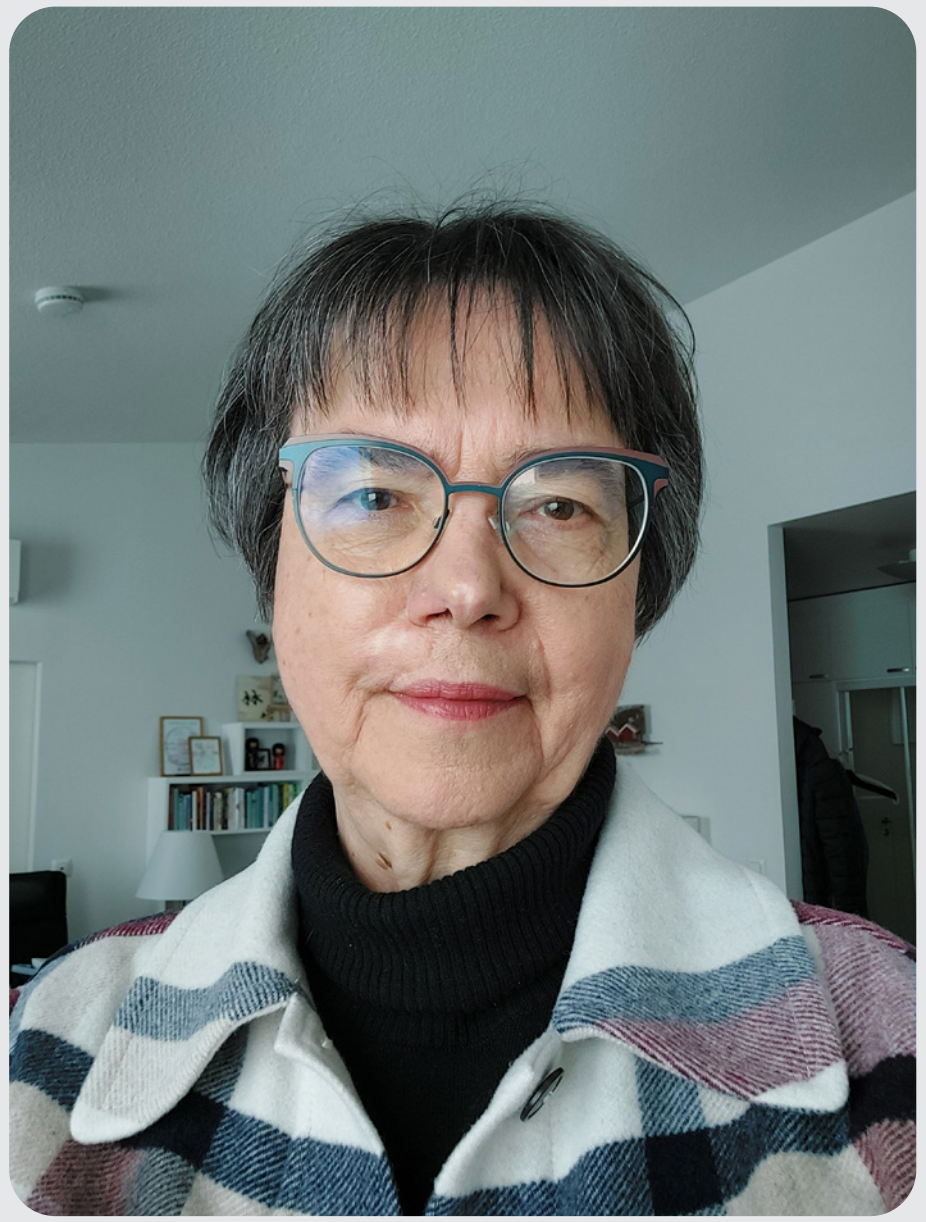

Anne Stenros is a Finnish architect, professor, and was formerly the Design Director at KONE Corporation, as well as the world's first Chief Design Officer (CDO) of any city—Helsinki. Stenros studied architecture under, among others, Reima Pietilä at the University of Oulu, and Christopher Alexander at the University of California, Berkeley. She earned her Doctor of Technology degree from Helsinki University of Technology in 1992. Stenros was the director of Design Forum Finland from 1995 to 2004, after which she spent a year in Asia as the director of the Hong Kong Design Centre. She served as the first Design Director at KONE Corporation from 2005 to 2015, City of Helsinki's CDO from 2016 to 2018, and founded Creative Catalyst in 2018—all the while lecturing at various universities including Aalto University (Finland) and Nagaoka University (Japan). Stenros is also a founding member of WIT Forum in Finland.

# Domenico de Clario

## the sung paintings (al-jabr: or the restoration or re-union of broken parts)

the sung paintings © Domenico de Clario, photograph by Jane Hirshfield.

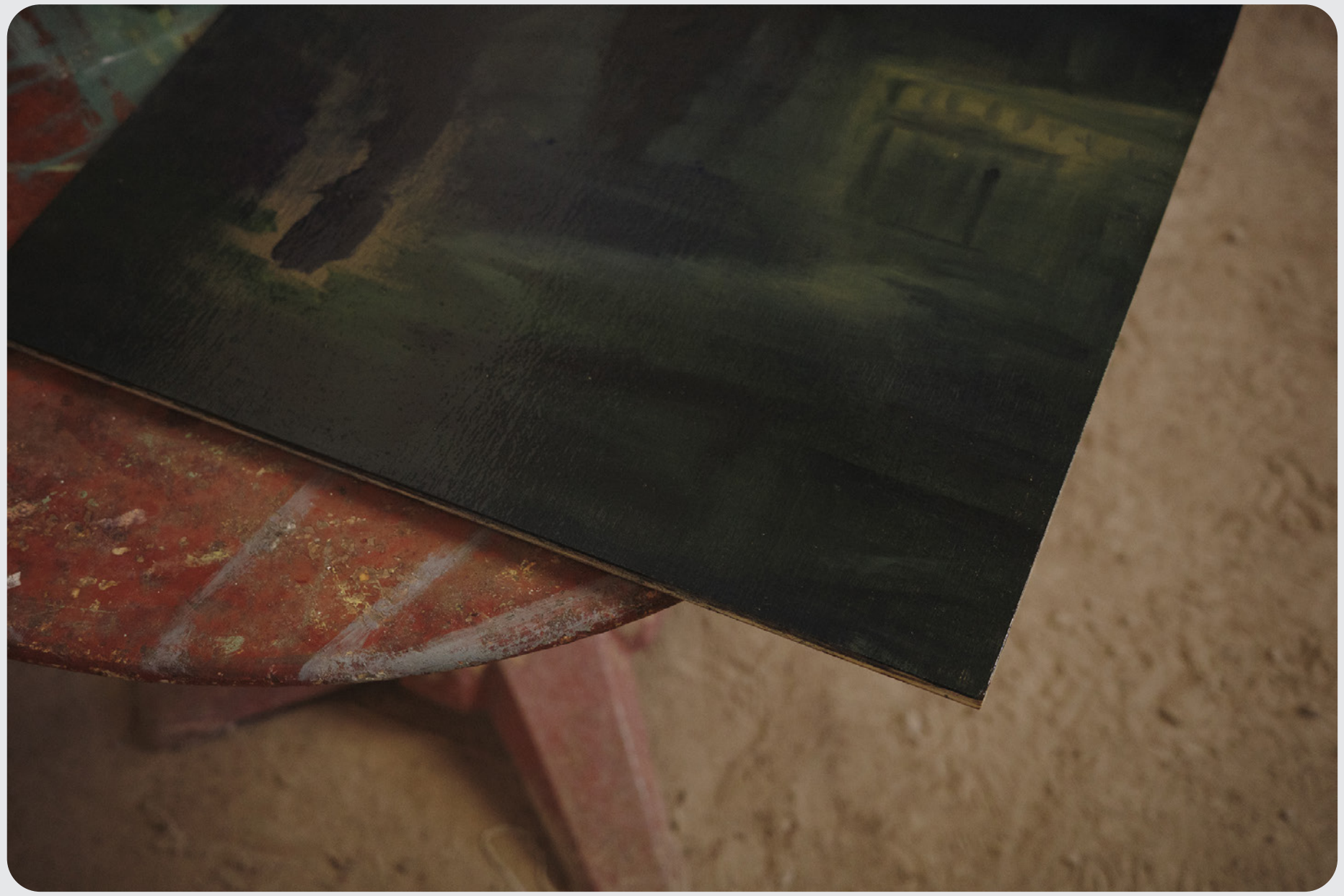

artists at their most perplexing and challenging essentially function as *apocalyptists*[1] in the attempt to reveal what was previously hidden under the surface of meaning, and perhaps in doing so they also restore what has been concealed and broken back to wholeness.

this wholeness may or may not always replicate the original's physiognomy but at its best it perhaps offers a newly redemptive version of the original's *essence*.

the following narrative outlines both the philosophical gestation of the *sung paintings* as well as its physical structure through a description of the elements that have defined some of my previous projects; this project is accordingly composed of the following four:

*

**the first element** consists of forty-nine paintings of various sizes installed along one side of a large rectangular chamber inside **energy hall / wasa innovation centre;** each painting is accordingly numbered from 1 to 49.

these are part of a collection of 187 paintings titled **the entropic space,** made in arles during 2024 in response to a vision of vincent van gogh that appeared to me as i visited the **espace van gogh** in the early morning of march 17 2024.

the *espace* was formerly the public hospital in arles where vincent was admitted in 1888 after he attempted to sever his ear; but my journey with both vincent and the making of the **entropic space** paintings is described more fully in the notes that follow.

*

**the second element** consists of forty-nine small plywood panels, each inscribed with the lyrics of a particular song; they are installed along the opposite side of the chamber in **energy hall** and each panel is also accordingly numbered from 1 to 49.

each song lyric describes one of many forms of love; some may be unrequited, others deeply experienced and many have been simply lost.

the artist has presented these songs on many occasions in the context of installations and performances that have been presented world-wide; all the songs were written at various times during last century and some date back almost one hundred years. they were composed mostly by individuals who, impoverished and broken by wars and famines, had migrated to faraway lands attempting to fulfil the promise of a glittering new life.

rather than finding the longed-for glitter and comfort some would end up on the edge of society living in extreme poverty, fulfilling the promise of a new life through the making of paintings and the writing / performing of their dream songs and poems.

these lyrics and the accompanying music are consequently the shining gifts we have been bequeathed by many of the *'failed'* ones, and the resulting golden fruit, borne by what at times had been characterised as gnarled and malformed plants, is now deeply treasured by us all.

*

**the third element** is located the end of the chamber and placed equidistantly between the paintings and the song lyrics; it consists of a keyboard, microphone and speakers as well as two comfortable chairs, placed directly in front of the keyboard.

*

one could say that long ago the sum of the paintings and the song lyrics made up a single whole, entwined together as an infinitesimally tiny micro-being who along with numberless others formed the entirety of the macro-whole.

this macro-whole had itself long before been gestated into existence through innumerable manifestations of love describing an endless number of its re-imagined forms, while each season turned under numberless spirals of whirling stars as it formed mountains, rivers, oceans and forests.

this timelessly ancient dance eventually fragmented (but only seemingly, and herein lies the paradox) each of the micro-beings into an infinite series of colours, shapes, words and sounds and eventually splintered into endlessly kaleidoscopic manifestations of *imagined intent*.

this project now attempts to restore a coalescing wholeness to the fragmented micro-being by inviting interested visitors to undertake seven actions:

**action 1.** the visitor (or visitors) enters the chamber and views the installation

**action 2.** he / she / they then proceed to choose a particular painting

**action 3.** the visitor carries the chosen
painting across the room to where its correspondingly numbered song panel is located, and replaces it with the painting

**action 4.** the visitor hands the relevant song panel to the artist who awaits at the keyboard

**action 5.** the visitor (s) proceeds to sit in the chair (s) located opposite the keyboard and witnesses the artist performing the song inscribed on the panel

**action 6.** once the song has been performed the visitor returns to the artist to receive the song lyric panel

**action 7.** the visitor places the song lyric exactly where the correspondingly numbered painting had formerly been located

*

**the fourth element** consists therefore of the energy flow generated between the artist and each visitor, as well as with the various components that make up the installation; this flow breathes life-energy into the project's various elements and generates the restoration of all that was separate and broken back into a final wholeness.

the process described above ensures that maximum contact takes place between each participating visitor and the artist, as well as between the visitor, the chosen painting and its corresponding song lyric, allowing in this way *al-jabr's* redemptive aspects to once more restore the seemingly separate parts of the micro-being to wholeness.

*

**some notes regarding the history of *al-jabr: the restoration or re-union of broken parts***

*al-jabr: the compendious book on calculation by completion and balancing* is an arabic mathematical treatise and the precursor by some years to the title above.

it was written in baghdad around 820 CE by the persian polymath muhammad ibn mūsā al-khwārizmī.

it was a landmark achievement in the history of mathematics serving as both the eponymous work and fundamental etymology of the word 'algebra' which was later adapted from the medieval latin term 'algebrāica', from *al-jabr*.

the *al-jabr* text also introduced the fundamental concepts of reduction and balancing to mathematics and the transposition of subtracted terms to the other side of an equation, rather like the cancellation of equal terms on opposite sides of an equation.

*al-jabr* was eventually translated into latin by robert of chester in 1145 AD and was used until the sixteenth century as the principal mathematical textbook employed by european universities.

*

i first came across the term used to describe algebra as a teenager during my middle years at school.

but even as i struggled with its application—i couldn't make head or tail of this particular branch of mathematics —i was fascinated by the phrase used by *algebra's* inventor to describe its underlying function: *al-jabr's* deeper philosophical purpose, my teacher explained, was to *'restore or re-unite formerly broken parts'*.

*algebra* described in this way became, at least in my mind, enmeshed in humanist poetics, and I imagined it as a subject of discussion by white-robed sufi sages as they gathered around sparkling fountains set amidst groves of exotic fruit and perfumed flowers.

this scenario from then on constituted the substance of my daydreams and whilst my fellow students gained a knowledge of algebra i inwardly grappled with *al-khwarizmi's* mysterious description of his discovery.

but now, some sixty years later, i am able to reason that if algebra constitutes the art of restoring wholeness to what has been fragmented then its guiding principle might also accurately describe the quest many artists undertake through their life-journey, because isn't attempting to restore wholeness to whatever has been broken the most worthwhile of possible quests?

i consider all this as i sit in my studio in arles; i pick up the closest of the neglected, discarded plywood panels i had intended to use to paint on and stood it up against the wall.

i move my chair very close to it, sit down and examine it, taking in each small detail.

on its surface i can see some of the accumulated evidence of the wounds its original body had suffered, not only upon its skin but deeper somehow, within the very heart of the wood.

and then a subsequent thought arises: might not all these indications of past wounds be in fact the cogent beginning points for a series of philosophical narratives, homaging both *al-jabr* and its consequent universal poetics?

might they not *be* the lamps that illuminate the uncertain path to a re-unified life?

and might i not appropriately inscribe these broken panels with contemporary words of hope, love and longing that many writers employed to compose the lyrics of our most moving songs?

i place the panel on my easel, switch off all the studio lights and i gradually open my eyes wider and wider into the darkness.

*

**some notes regarding *the entropic space paintings***

i left australia in early march 2024 to travel to marseille in order to briefly visit the family of an artist friend who some days earlier had unexpectedly passed away.

through a series of problems suddenly arising with my passport i was unable at the end of my visit to return home as i had planned, and it was not possible to extend my permanence in marseille.

i contacted a friend who lived in nearby arles and in this emergency he immediately welcomed me as his guest.

i had never been to arles though i had known of vincent van gogh's permanence there, and of how his 15-month arlesian sojourn had deeply impacted his life's work, encouraging him during this time to complete 187 paintings and many hundreds of drawings.

the emotional stamina, physical perseverance and intellectual resilience required to achieve this output, perhaps unequalled in its depth and astonishing in its quality, deeply impressed me.

early on the morning following my arrival in arles i stumbled across the *espace van gogh* and as i strolled through the courtyard a figure suddenly appeared under the colonnade.

vincent stood very still in the shadows and for a few instants we locked eyes.

overwhelmed the following question suddenly arose in my mind; *'why am i here in arles?'*.

i immediately considered whether this unexpected visit to arles might have had a deeper purpose; how might it feel, i asked myself, to undertake a journey as vincent had of such singularly focused intensity?

could such an attempt be the underlying reason i now found myself here?

i began to gradually appreciate how vincent's achievement had extended far beyond sensuous painterly virtuosity; it involved his deep instinct for testing how the intersection of science and soul might manifest visually and suddenly i found myself compellingly interested in the required level of *intent* vincent needed to generate to achieve such a monumental outcome.

that very day i decided i would find a way to remain in arles and commit, with *intent*, to similarly completing the same number of paintings vincent had produced during his stay here (187); and that this would become the sole reason for extending my permanence in provence.

i imagined that this yet-to-be experience might prove to be similar to one of a blind person being led by unknown forces towards an unknowable destination; i accordingly decided i would fittingly begin each painting whilst blindfolded, completing it either during that specific plein-air sitting or even later that day in the studio.

on the new-moon evening of september 4 2024 an installation comprising the 187 paintings i had made in arles since i began the journey on march 17 was inaugurated inside the 8th century arlesian church of st genest.

from this collection 49 paintings were chosen as the ones to *'be sung'* through this performance.

*

each of these paintings has now been re-titled according to its correspondingly numbered song and so each painting's visual characteristics has found, through the collaborative actions of individuals hitherto unknown to each other, an unexpectedly new and perhaps incongruous equivalence in the words and music of its corresponding song.

so here in vaasa, on the shores of the gulf of finland, unusually strange events have unfolded inside a large underground chamber, fittingly titled **energy hall;** algebraic serenades and a blindfolded vincent van gogh have been re-united through the echoes of songs describing ageless love and longing, as each one sounds and re-sounds into a final stillness.

1 From the greek word apokalypsis, meaning 'to reveal, to draw the veil back, to disclose'

# a.metsä
## states of entropy

[*entropy*; the degree of disorder, uncertainty, and transformational content within a system]

*I enter the Helsinki Metro through a gateway showing a blue triangle on an orange background, the signifier of civil defence shelters. Venturing underground, I pass massive walls of rock and giant steel doors, going deeper into the ground towards not only a space of passage, but also an architectural answer to uncertain futures. I am reminded of the underground lines of Glukhovsky's 'Metro 2033'*[1] *and the vaults of Fallout*[2] *designed to protect humanity from what might come, while in my headphones David Tibet softly sings about the end of times: "I thought that I had seen some bright new dawn."*[3]

With a progressively bleak future shaped by ecological, economic, military, political, and technological crises, imaginaries and realities of preparedness are blending together further and further, creating a referential system of signifiers for crisis preparation and what we understand as such.

The resulting signs, symbols, and narratives of preparation, civil defence, and self-protection that permeate our media landscapes and urban environments, have to be understood as an attempt to bring certainty to an uncertain future. Each safeguard, vault, instructional manual, and media narrative is a selective attempt to imagine what might come and creates an idea of being prepared for what awaits us. Whatever might come, we will be prepared, hopefully.

Drawing from the real and fictional narratives, signs, and symbols of civil preparedness and survivalism, *states of entropy* presents a series of symbols for three signal flags, marking and communicating different states of perceived stability and (un) certainty of a system, region, community, or individual, to be flown over structures and spaces of preparation.

1 Glukhovksy, D. (2002) Metro 2033. Eksmo.
2 Interplay (1997) Fallout: A Post Nuclear Role Playing Game. Interplay Productions.

# Benjamin Knock
## Senectus Tempus

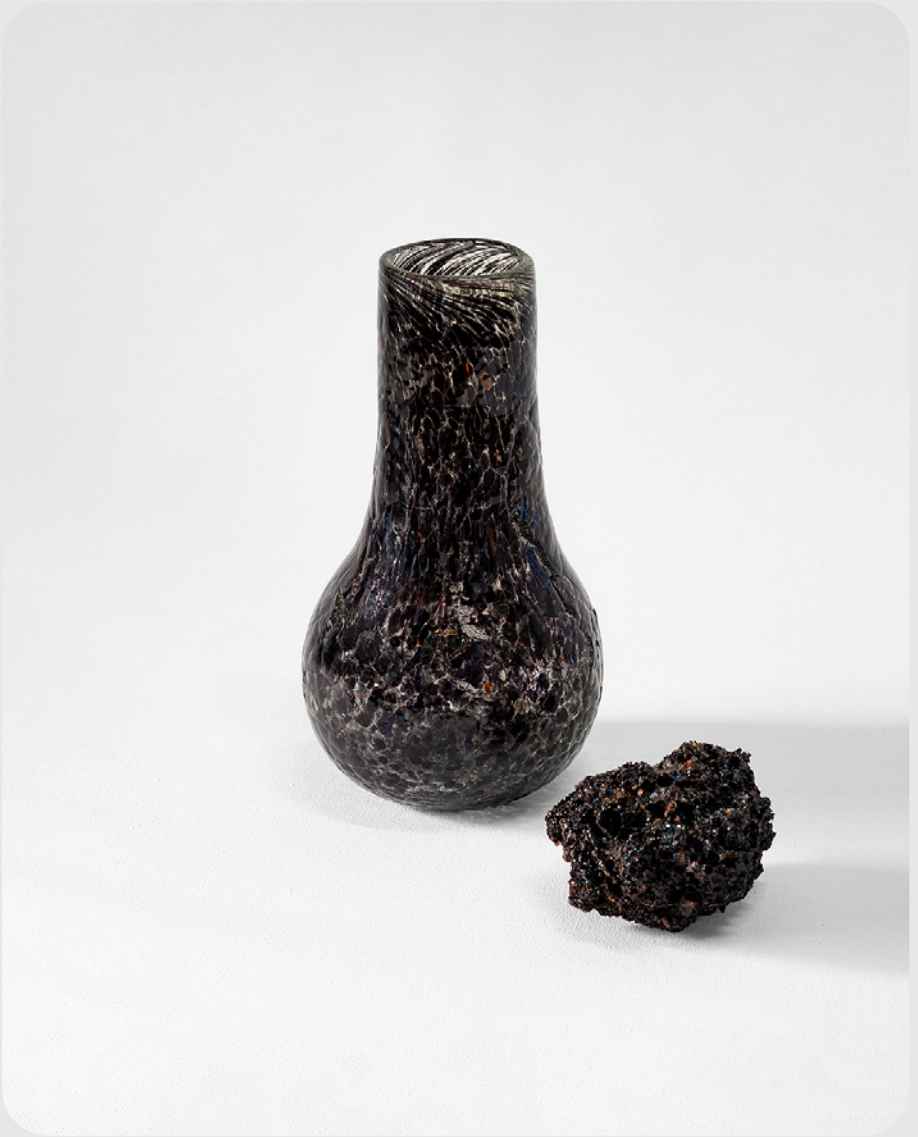

*Senectus Tempus* is a new series of works based around a recent 2-month expedition to the active volcano of Mt. Yasur in the Vanuatu archipelago. In collaboration with a team from Tokyo University's Earthquake Research Department, Knock collected geological samples, 3D Lidar scans, and sonic Geo-phon readings from the site. This collection of works highlights the intriguing volcanic processes that are the source of life that sculpted the natural landscape, and life of our planet throughout time. The various mediums incorporate many of the raw materials and digital samples collected from the caldera, such as raw sulphur oxides, volcanic glass-infused pigments, and basalt layers. The artist hopes that the works act as a gateway to immerse yourself in the deep time scales of geothermal terraforming.

# Michael Lukaszuk
## Away

In *Away*, Michael Lukaszuk constructs a sonic architecture that is at once intimate and industrial, personal and planetary. This sound art installation, composed for a constellation of speakers dispersed throughout a room, invites the listener into a liminal acoustic environment where the boundaries between human and machine, memory and immediacy, are not only blurred but actively reimagined. The work is a meditation on presence—how it is constructed, fractured, and distributed across geographies and technologies.

At its heart, *Away* is a study in dualities. It unfolds along two interwoven trajectories: one that explores the entanglement of human and machine breath, and another that maps the artist's own transnational experience through sound. These trajectories do not run parallel so much as they spiral around one another, forming a complex topology of meaning that resists linear interpretation. The result is a work that is less a composition than a living system—an ecology of sound in which the listener is both observer and participant.

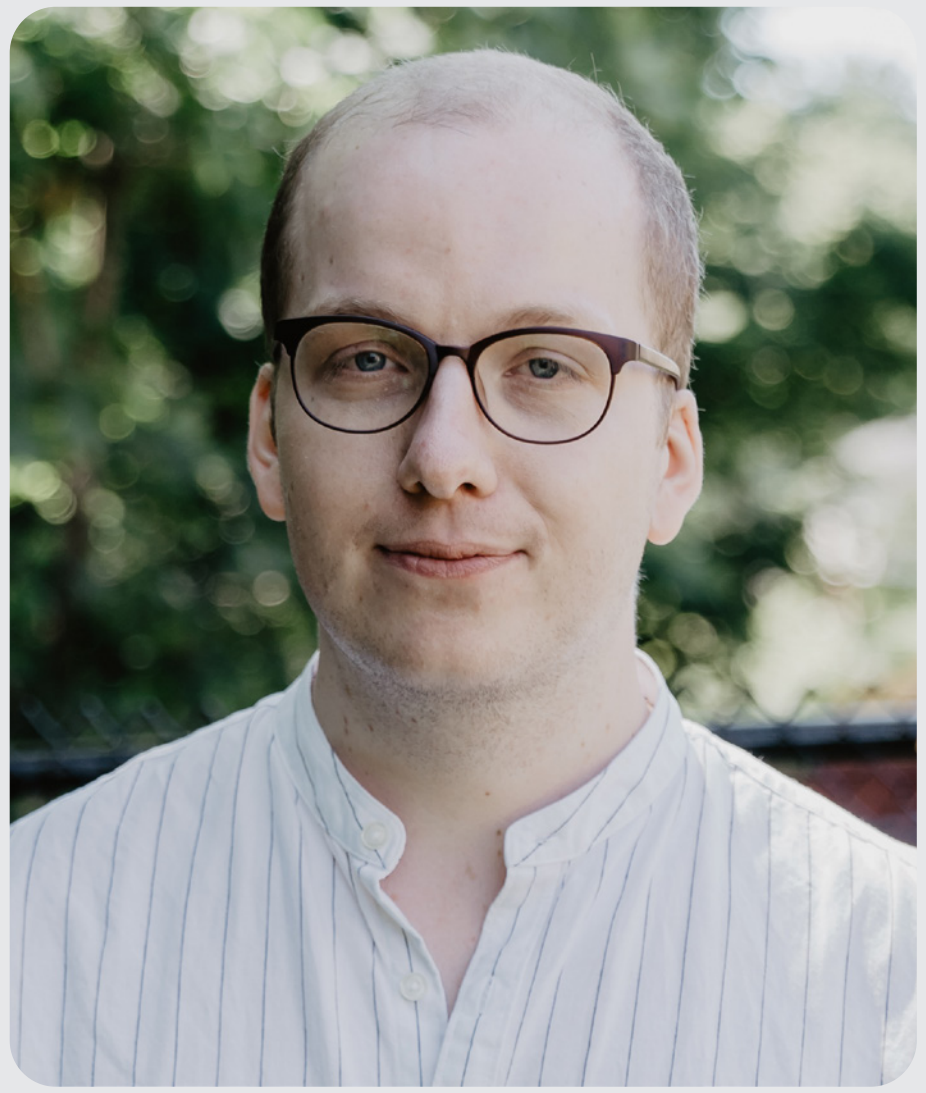

Breath is the most elemental of human sounds. It is the first sign of life, the last trace of presence. In *Away*, breath becomes a site of transformation, a medium through which the human and the machinic coalesce. Lukaszuk begins with recordings of his own breathing and voice—raw, corporeal, unmistakably human. These are then algorithmically processed, hybridized with synthetic vocalizations generated by computer systems. The result is a continuum of sound in which the origin of each breath is obscured, its identity dissolved into a shared sonic fabric. This is not a simple act of mimicry. The machine does not imitate the human; it breathes with it. Through granular synthesis, spectral morphing, and generative algorithms, Lukaszuk creates a space where the breath of the artist and the breath of the machine are indistinguishable. The listener is drawn into a world where circuits exhale, where code sighs, where the boundary between organism and apparatus is rendered porous. This interplay is emblematic of Lukaszuk's broader artistic inquiry into "digital liveness"—a term that encapsulates his fascination with how presence is mediated, simulated, and reconstituted in algorithmic environments. In *Away*, digital liveness is not a conceptual abstraction but a visceral experience. The machine is not a tool but a co-performer, its voice woven into the sonic tapestry with the same intimacy as the artist's own.

The spatial configuration of *Away* is integral to its meaning. The speakers are not merely points of emission but nodes in a distributed network of sound. They are arranged to create zones of proximity and distance, of clarity and ambiguity. As the listener moves through the space, they encounter shifting constellations of sound—some whispering at the edge of perception, others erupting with mechanical force. This spatial dramaturgy transforms the act of listening into a form of navigation, a journey through an invisible landscape shaped by breath, noise, and memory. This approach reflects Lukaszuk's deep engagement with spatial audio and generative systems. His previous works—such as *Habitats* (2021), a series of generative sound ecologies—demonstrate a commitment to sound as a dynamic, emergent phenomenon. In *Away*, this ethos is extended into the physical space of the gallery. The installation becomes a living organism, its sonic behavior shaped by the interplay of algorithmic processes and human presence.

The second axis of *Away* is rooted in geography, in the artist's lived experience of displacement and dual belonging. Having lived and worked in both Northern Sweden and Canada, Lukaszuk draws upon the sonic textures of these environments to construct a kind of acoustic cartography. This is not a documentary impulse but a poetic one. The recordings—urban ambiances, environmental sounds, culturally specific vocalizations—are not presented as indexical traces but as compositional elements, woven into the fabric of the piece.

One particularly evocative motif is the non-lexical inward breath used in Northern Sweden to signify agreement—a subtle, culturally embedded gesture that becomes a recurring sonic signature. This breath, at once mundane and deeply situated, becomes a cipher for presence, for belonging, for the unspoken. It is paired with recordings from Lukaszuk's Canadian life—urban soundscapes, domestic ambiances, fragments of speech—creating a dialogue between places that are geographically distant but emotionally proximate.  This interplay of locations is not nostalgic but reflective. It speaks to the condition of being "away" not as absence but as multiplicity. The artist is not simply displaced; he is multiply placed. The work thus becomes a meditation on diasporic identity, on the ways in which sound can carry memory, affect, and cultural specificity across borders. It is a sonic form of correspondence, a letter written in breath and noise.

Lukaszuk's artistic lineage is deeply rooted in electroacoustic and acousmatic traditions, yet his work consistently pushes beyond their boundaries. In earlier pieces such as *Obsession* (2024) and Sight *Unseen* (2017), he interrogates the boundaries between real and virtual sound sources, often using AI-based timbre matching and custom synthesis tools to blur the line between instrument and simulation. These works are not merely technical exercises but philosophical inquiries into the nature of sound, identity, and presence. In *Away*, these concerns are extended into the spatial and the performative. The installation format allows for a more immersive engagement with the themes of hybridity and presence. The generative systems that underpin the piece are not hidden but made audible, their logic inscribed in the evolving textures of the soundscape. This is a hallmark of Lukaszuk's practice: the algorithm is not a tool but a collaborator, a co-composer whose agency is both constrained and liberated by the artist's design. Moreover, *Away* resonates with Lukaszuk's interest in improvisation and collaboration. His work often involves partnerships with visual artists and performers, and this ethos of co-creation is evident in the way the installation invites the listener to become part of the composition. The audience is not passive but implicated, their movement and attention shaping the experience of the work.

What emerges from *Away* is not a narrative but a condition—a state of being in-between. In-between human and machine, in-between places, in-between presence and absence. This in-betweenness is not a lack but a richness, a space of potential. It is where meaning is not fixed but emergent, where identity is not singular but plural.

The work challenges the listener to reconsider their own presence in the space. As they move through the installation, they become part of the composition, their position and movement influencing the way the piece is perceived. This participatory dimension echoes Lukaszuk's interest in improvisation and collaboration, seen in his work with visual artists and performers. In *Away*, the audience is not passive but implicated, their listening an act of co-creation.

In *Away*, Michael Lukaszuk offers more than a sound installation; he offers a space for dwelling—a space where listening becomes a mode of being, of remembering, of imagining. It is a work that asks us to listen not just with our ears but with our bodies, our memories, our sense of place. It is a work that breathes with us, that speaks in tongues both human and machine, that holds us in the delicate tension of being here and being away. Through its intricate layering of sound, its spatial dramaturgy, and its conceptual depth, *Away* stands as a compelling exploration of what it means to be present in a world increasingly mediated by technology and displacement. It is a work that does not offer answers but opens questions—questions that resonate long after the sound has faded.

# Juhani Risku
## Keynote presentation

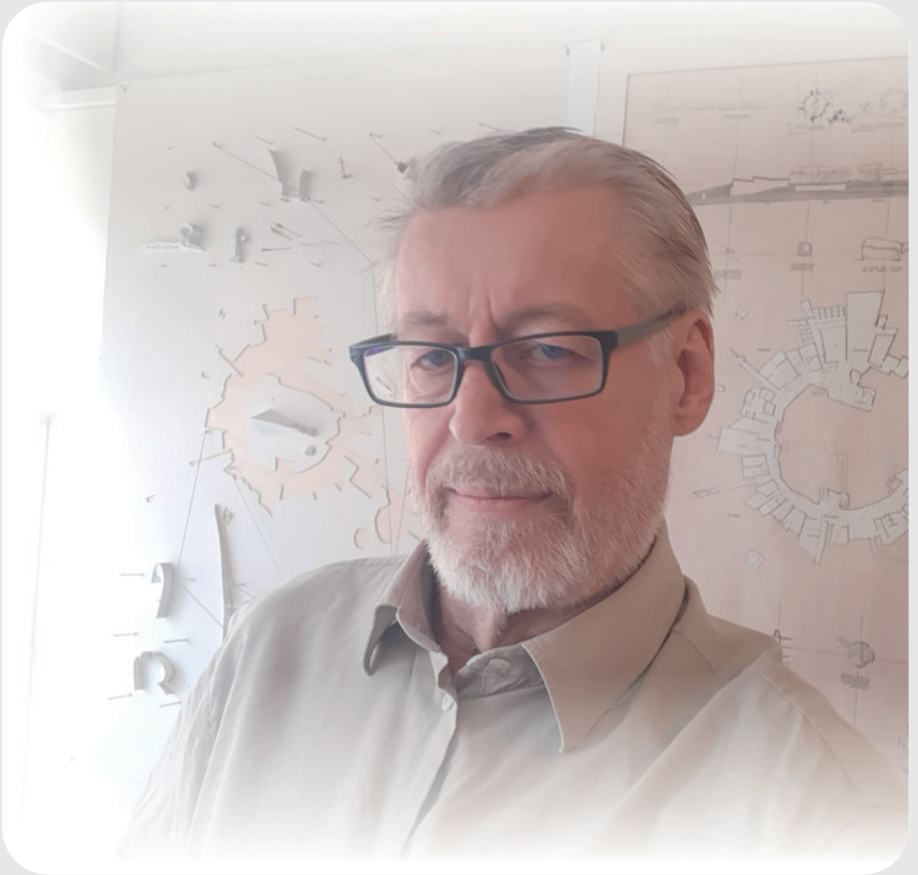

*Art and design are phenomena of higher abstraction of knowledge, skill, understanding and criticism → New leadership begins*

Who, what, why and how? Juhani Risku has studied a lot as an early riser. But, has also been an apprentice for fifteen masters in 400 MW turbines and generators, room building and physical acoustics, Arts & Crafts, photography, and many other fields. In the end, he defended his doctorate on the impact of art, design, and creativity on startups. Now he is returning to architecture, to art and crafts.

Design and art offer a new approach to leadership, for example, because they offer a higher abstraction to creativity and practical work, with civic courage and an ethical attitude. Designers and artists are particularly industrious, and they take their work to the highest level without compromising on content and quality. It is also worth noting that the abstraction of design is in humans from birth—because art is an abstraction for adults.

Design and art form a powerful butterfly effect. Individual people rise as conveners to lead the future in the Fibonacci series 0, 1, 1, 2, 3, 5, 8, 13, 21, 34, 55... ... 17711, 28657..., allowing everyone to join in. Now begins a new hype, from the *Butterfly* symposium.

**Declaration:** The designer-artist-craftsman is the strong leader who creates a new future from the front and with example, courage and ethics in data, energy and industry, but also in society and future systems.

# Katey O'Sullivan
## Two sides, three Haiku

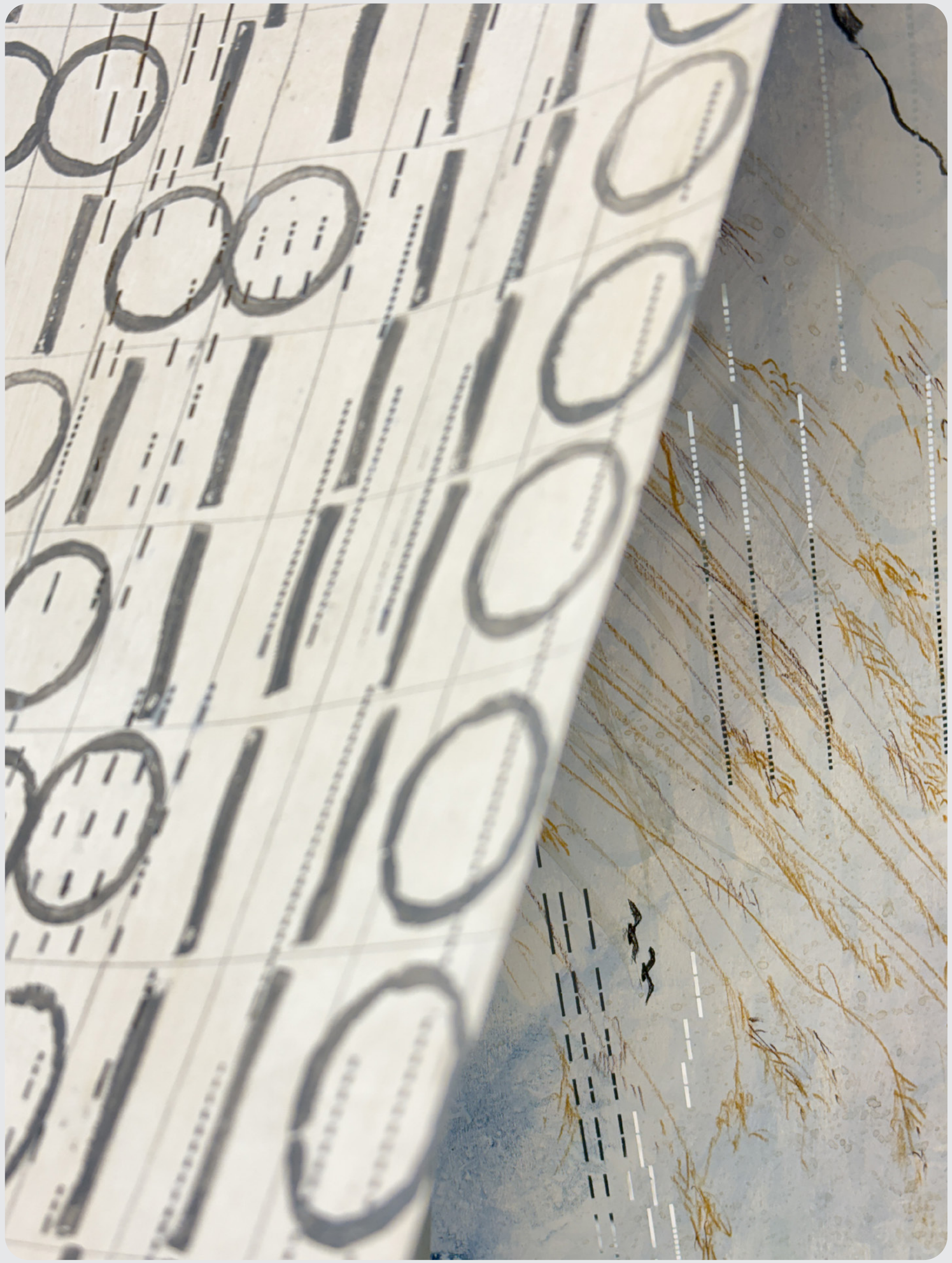

From the early mechanical innovation of pattern-based coding to modern deep learning methods, O'Sullivan respectfully acknowledges advancements in technology while playfully issuing a subtle warning. Her use of materials are directly representative of the historical lineage of AI.

Piano scrolls were one of the first widely accepted examples of an ongoing relationship between technology and the arts. They were modelled on the punch cards used by Jacquard looms, which successfully translated human intention into machine action, laying the groundwork for coding, and eventually AI. The hand stamped binary code on one side of the work is a translation of an AI-generated response. The AI was asked to produce three Haiku in a self-deprecating tone. The artist has taken the unfiltered and candid responses, then removed human understanding. The mystery of machine languages draws attention to the disconnect between people and modern technology. Similarly, there is little understanding of the environmental costs associated with every interaction with AI. It is argued that both the volume of water and electricity needed to power data banks are directly contributing to the destruction of our natural world.

The other side of the scroll depicts nature, organic and nuanced layers, with fragments and words gathered by O'Sullivan during her artist's residency in Vaasa, Finland. She says "While most of us value and understand the re-energising effect that nature provides us, and that rich language and people to people communication are keys to our wellbeing, many of us become so absorbed in the technologies of our time that we forget to prioritise these interactions. Future innovations and technological solutions are incredibly necessary. However, these must come with sustainability and balance at the forefront; reducing the risk to nature (and its inherent intelligence), and reversing the deterioration of human connectedness."

# Jurgis Peters & Samuel Kujala
## Modern Mythologies

© Jurgis Peters & Samuel Kujala

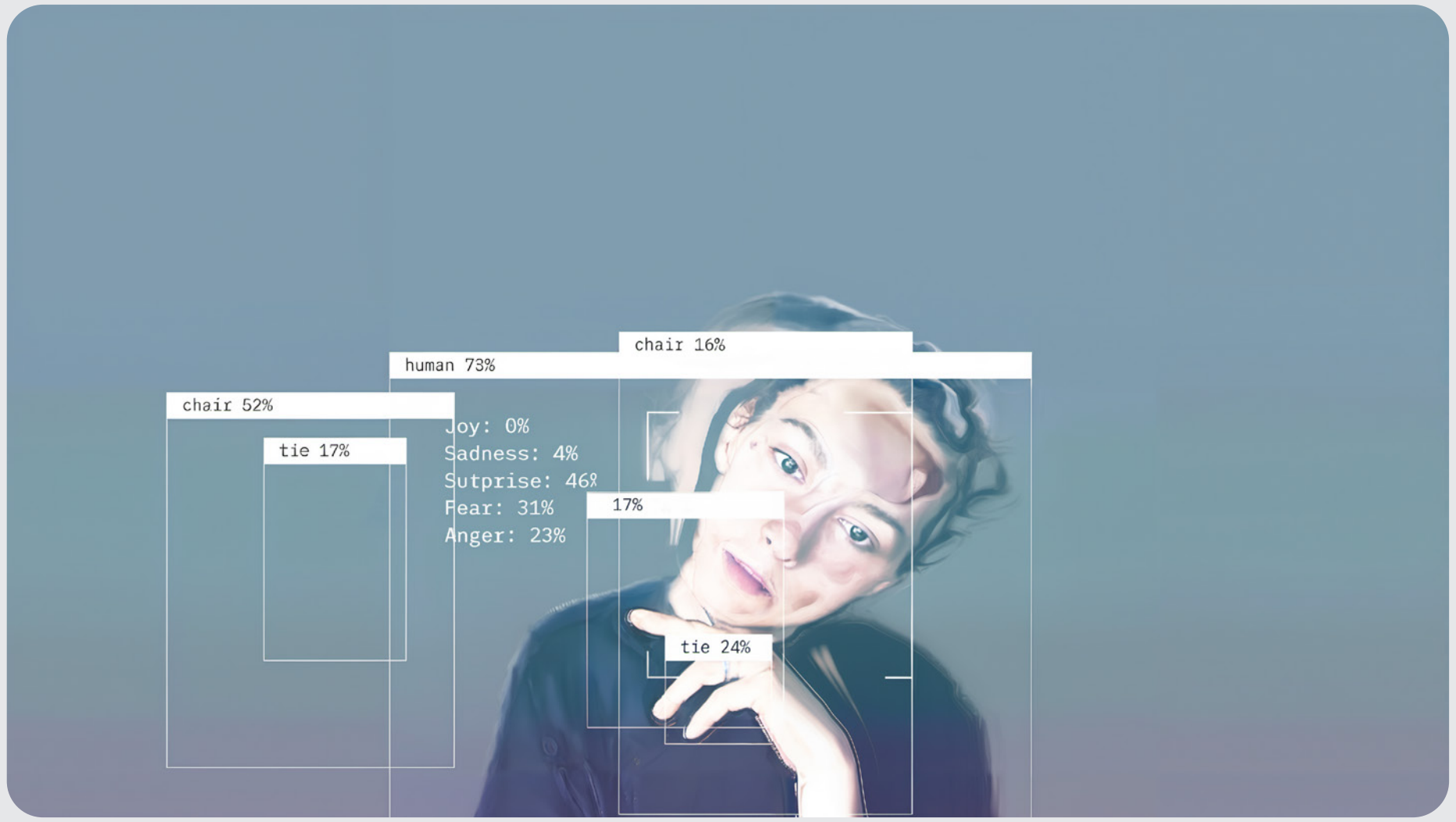

*Modern Mythologies*, an interactive installation by Jurgis Peters with an adaptive soundscape by Terho Ojell-Järventausta, examines the increasing influence of algorithmic systems on how we perceive ourselves and likewise, how we are perceived by technology. In an age where digital classifications increasingly shape our reality, the very essence of human identity —often reduced to quantifiable data— is being re-evaluated.

The installation examines the algorithmic gaze, a pervasive digital lens that interprets and categorizes human expression. While our complex emotions are distilled into simplified, generalized data points, these abstractions paradoxically gain immense value within automated systems. The work investigates how such interpretations, often flawed, subtly dictate our place within this new data landscape.

*Modern Mythologies* installation turns visitors into mere data-points. Their image is captured, undergoing realtime, glitched classification and rendered as a fractured, pixelated spectrum of information. In addition, the system attempts to discern fundamental emotional states, continuously assessing and defining the individual.

The core of the installation lies in a subtle provocation: the system seeks a specific emotional response. It gently coerces the viewer to display certain emotions, not as authentic human feelings, but as performative acts. When such calculated expressions are detected and sustained, the installation shifts to a different mode of operation, rewarding the viewer for their emotional compliance. This transition hints at a question—when our emotions are sought and affirmed by machine logic, where do the boundaries of authentic experience truly lie? It suggests that in this new data reality, our very essence risks becoming a performance, an input, valued for its statistical predictability rather than its inherent complexity.

To further investigate these dynamics, Samuel Kujala will present a series of live performances. In these controlled engagements, Kujala intentionally surrenders his corporeal presence to the algorithmic gaze, allowing the technology's interpretations of his body to become the primary elements for a profound physical and conceptual transformation. This dynamic interplay extends traditional approaches to understanding the performer's body, aligning with contemporary ideas of a "bodyworld" that integrates both human and non-human elements. Simultaneously, Ojell-Järventausta will craft a live, responsive soundscape. This audio environment directly interacts with Kujala's movements and the installation's visual output, generated and modulated in real-time from the algorithmic classifications and the performer's physical actions. This creates a three-way conversation between human, algorithm, and sound, further blurring the lines of agency between person and machine, ultimately questioning where authentic control truly resides.

# Frederick Rodrigues

## Synthetic Ornithology

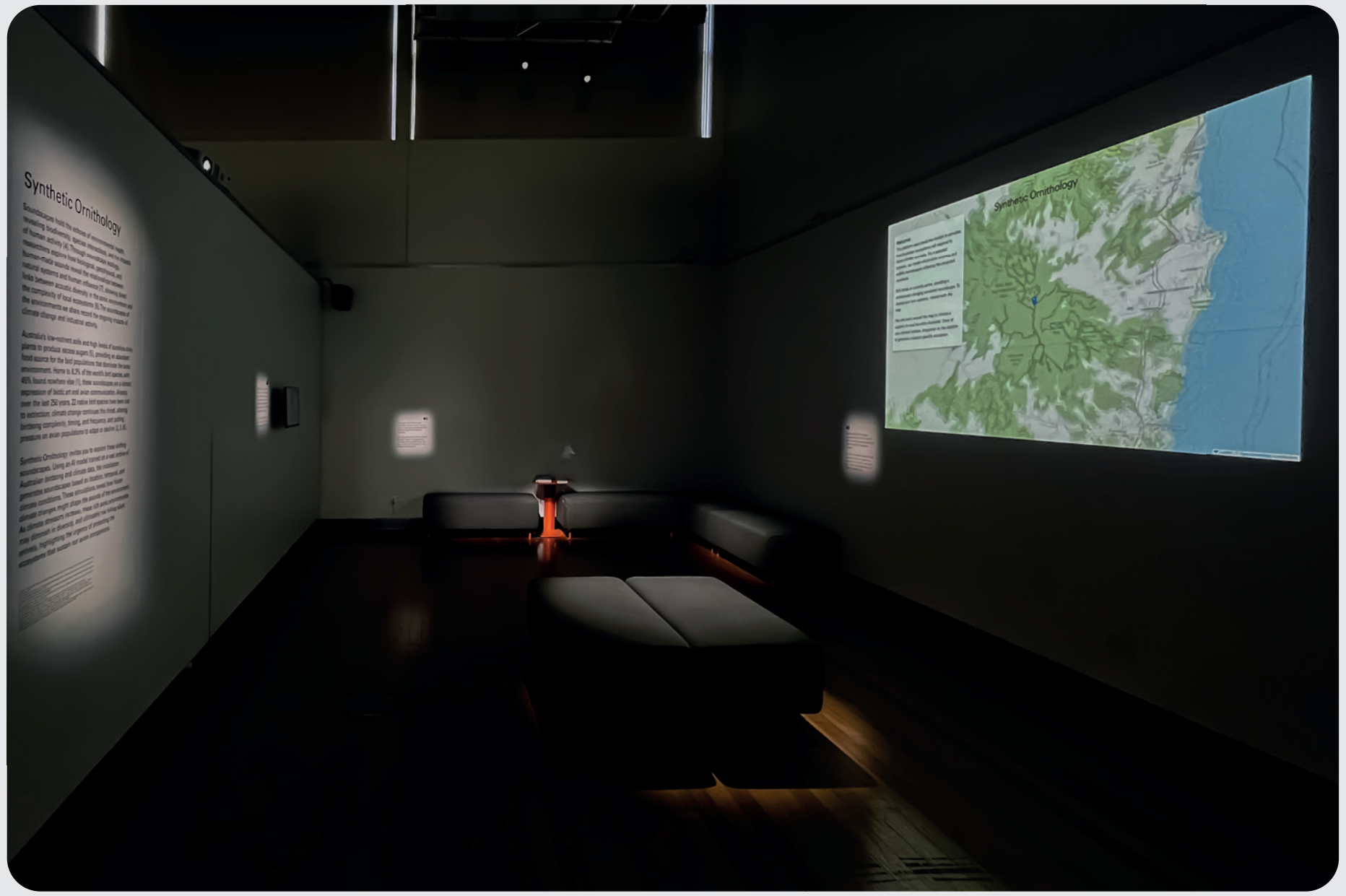

Photographed by the Artist © Frederick Rodrigues

In *Synthetic Ornithology*, you will encounter simulations of how shifting climates might transform Australia's avian soundscapes. Powered by a bespoke machine learning model trained on an archive of birdsong focused recordings from around Australia, the installation generates localised sonic environments based on climate scenarios. Left alone, *Synthetic Ornithology* will drift through pre-generated simulations of speculative futures; you can hear how any prospective scenario might sound by using the touch screen to choose a future date, location in Australia and climate conditions.

*Synthetic Ornithology's* fictional but highly realistic soundscapes, bearing artefacts of human presence, might seem like insignificant environmental recordings; however, with close attention they become uncanny reflections of possible futures, of a complex networked system involving not just the scenario depicted in the audio, but the listener and a larger social and environmental context.

# Moa Bleyner Cederberg & Michele Ucchedu

## Ruins of a Far Future

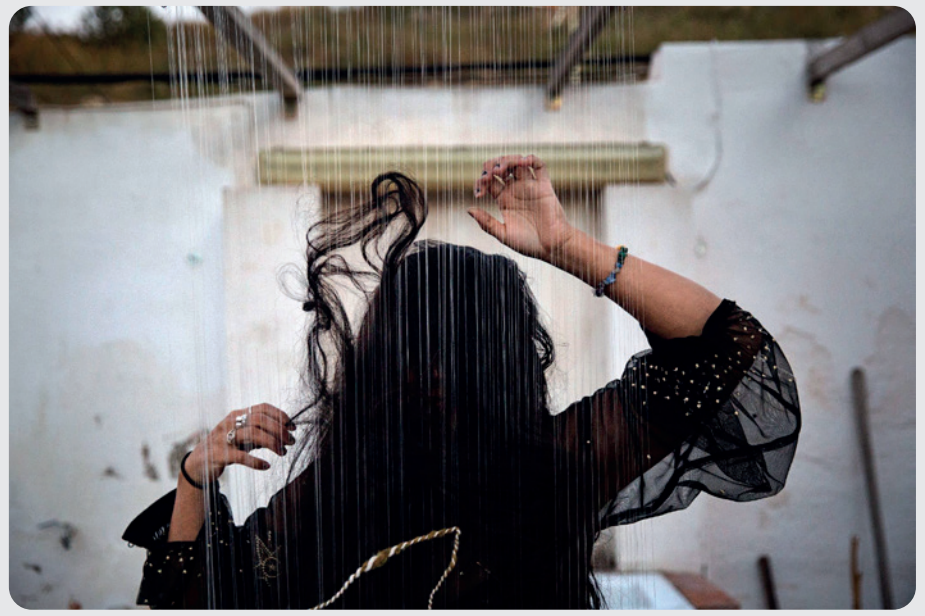

In a distant future, where nature and technology are no longer separate realms, a new kind of life emerges —part organic, part synthetic, part sentient. *"Fourth Nature"* is a post-natural, post-technological landscape where humans, machines, and ecosystems have merged into a single evolving entity.

In this installation, Moa Bleyner Cederberg creates a fabric surface using natural wool and traditional techniques—a tactile memory of ancestral human knowledge. Wool, one of humanity's oldest textile materials, becomes a symbol of our craft and legacy. Over this surface, Michele Uccheddu weaves chopper wires like invasive roots or synthetic grasses, representing the quiet but relentless takeover of technology. These wires are alive: equipped with sensors, they respond to presence and movement, forming a primitive nervous system that emits sound.

Inspired by real-world advances —robotic pollinators, AI-driven regeneration, and bionic prosthetics —this work imagines a world where technology is not separate from nature, but its evolution. What remains are not ruins of collapse, but of transformation—remnants of a *Forth Landscape* where biology, machine, and memory are one.

**Michele Uccheddu** is a sound artist based in Finland, also percussionist, electronic music composer, and record producer. He is deeply interested in the intersection of nature and technology in experimental electronic music. He draws inspiration from the generative arts and the science of psychoacoustics. Also interested in Club Culture like techno sonorities, dub and ambient.

He is the founder of SUPRANU Records, and has participated in high-level projects like the Venice Biennale (2014) Winner of Abbiati Prize with the project GEO of Roberto Doati and the international hackathon of Ars Electronica (2021).

Now focused on the theme of Deconstruction, that can be applied to music or sound installations.

# Daniel Shanken
## VORE

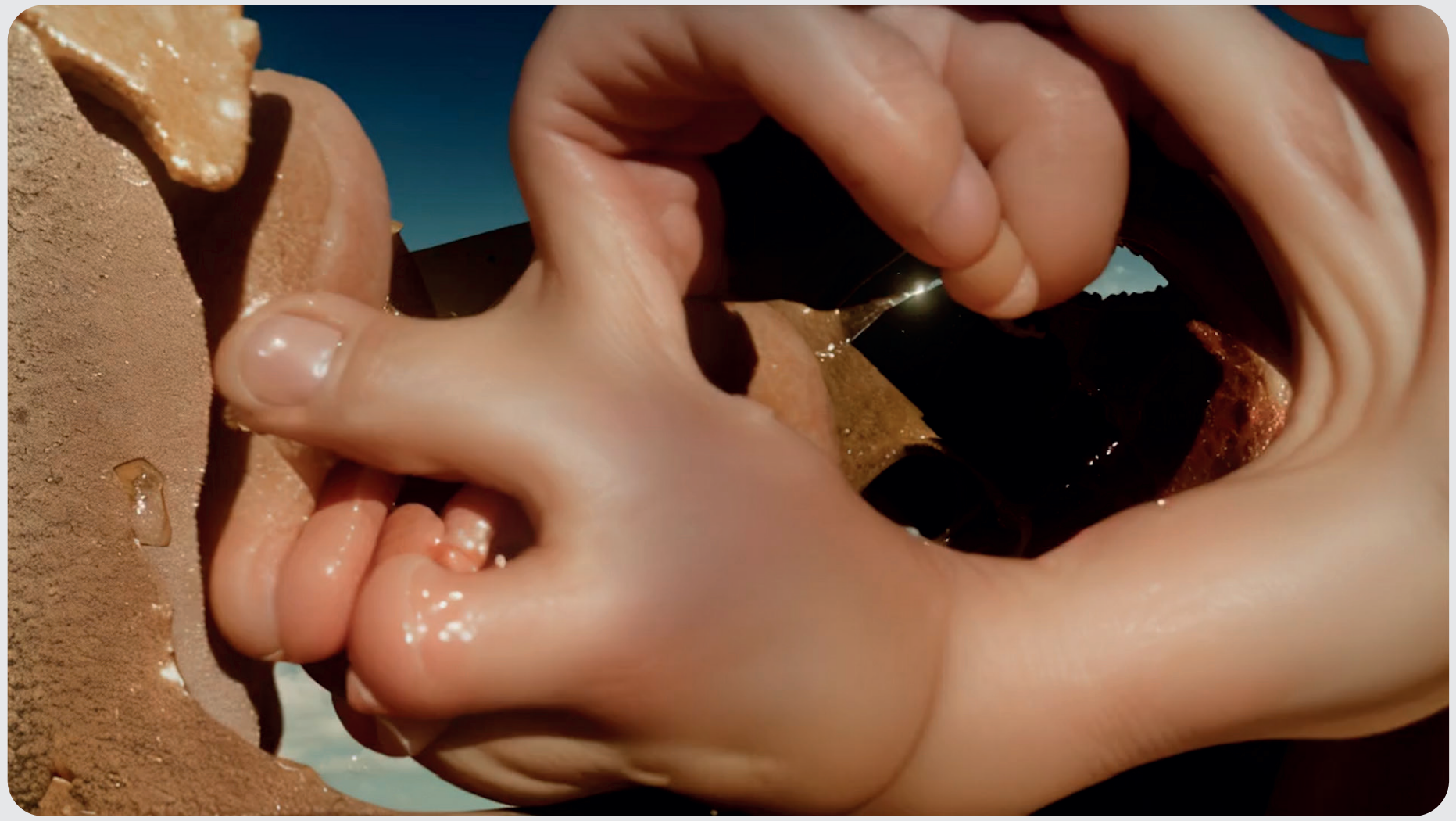

© Daniel Shanken

*VORE* explores the layered dynamics of consumption, situated within broader systems of control. The video descends through tiers and levels: physiological, technological, environmental, and cognitive. Eating becomes a central framework for imagining the digestion, assimilation, offloading, and onboarding that occur when surrendering to technologies or entities with their own agency and agendas.

Predictive AI-generated zooms and pans created with custom tools and datasets propel the viewer through artificial latent spaces, examining the tension between pleasure and horror in the act of opting in and zoning out. *VORE* moves fluidly through bodies and landscapes, tracing the relentless extraction of thought and resources, and the erosion of cognition by corporate agents.

# Spencer Rose
## Illusions of Agency

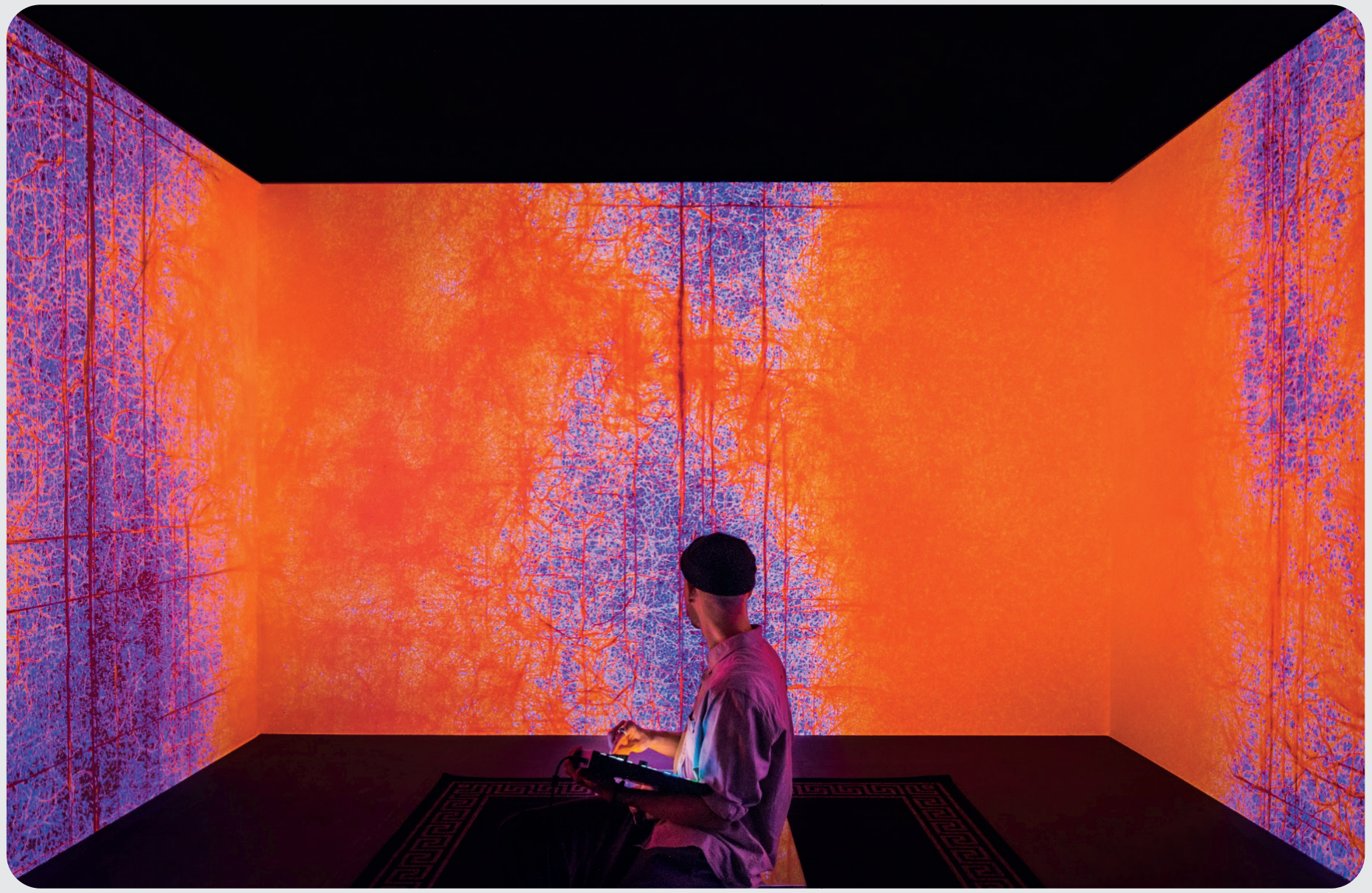

Photographed by the Artist © Spencer Rose

*Illusions of Agency* is a projected digital simulation that maps the trace of directed force across a layered, synthetic terrain. Built from a complex field of particles, the work reflects on how energy—social, environmental, infrastructural—shapes and is shaped by the systems we inhabit. These systems, designed to organise and sustain collective life, often exist in uneasy dialogue with the ecologies they intersect. Their patterns aspire to order but give rise to unintended forms—disruptions, resistances, emergent complexity. Though pre-rendered, the simulation holds the residue of motion and decision, a ghost of agency passing through constraint. Minor impulses—barely perceptible—can cascade into structural shifts. What begins as a flicker becomes a fault line. The work lingers in this ambiguity, where scale distorts responsibility and the smallest act may tilt an entire system. It offers a space to reflect on how structures meant to serve can also enclose, and how conscious agents move within and against these frames to reshape the world around them.

# Anne Scott Wilson & Cameron Bishop
## From X to Eternity

From X to Eternity © Anne Scott Wilson and Cameron Bishop, video still by Aaron Hoffman

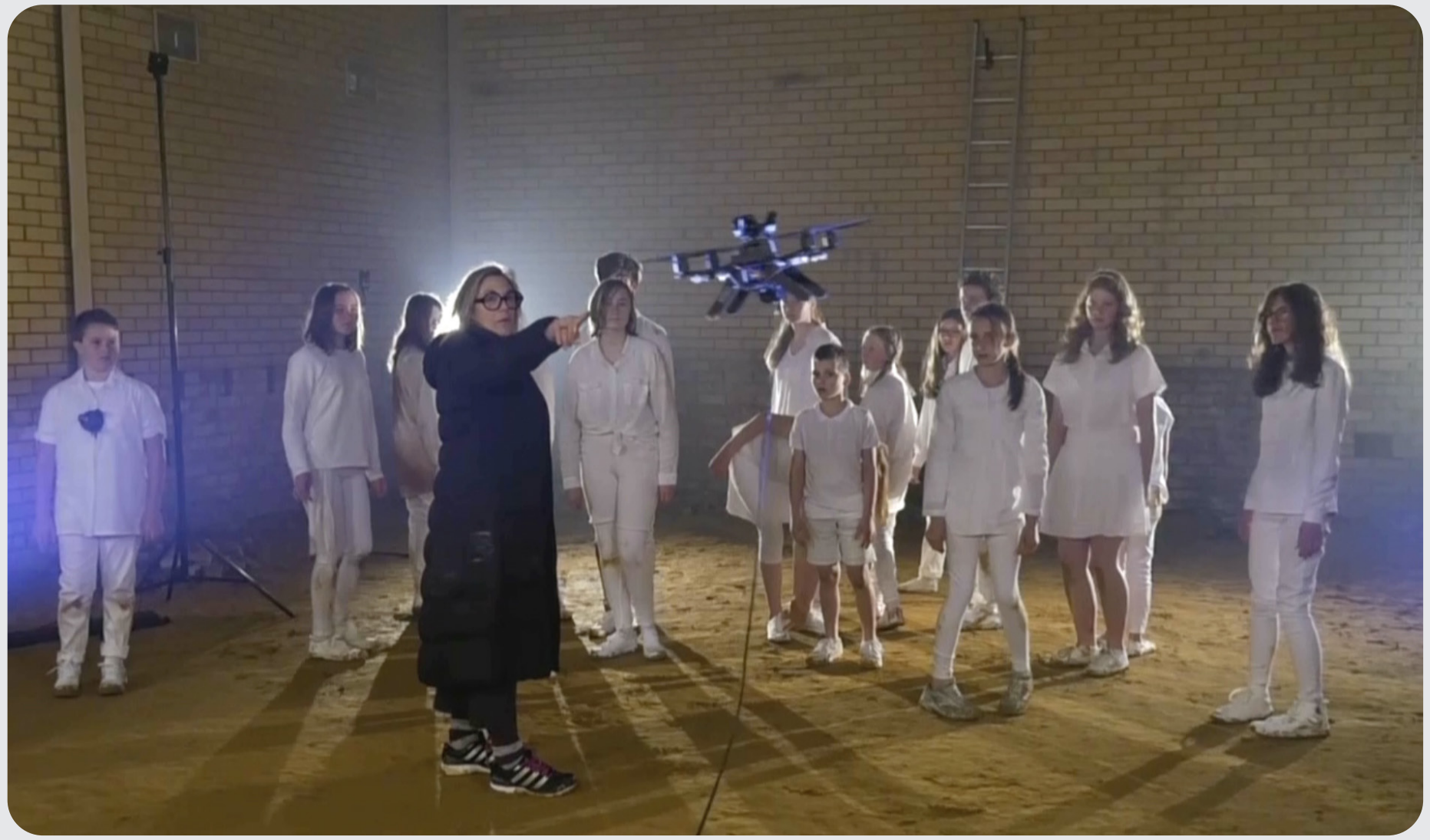

Inspired by Hito Steyerl's twist on the medieval riddle—"How many angels can dance on the head of a pin?"[i]—this project substitutes angels with AI, pointing to the spiritual undertones of bodiless intelligences in our increasingly digital world. The confusion between metaphysical and artificial entities highlights a strange irony: even in a largely atheistic culture, we continue to invest technology with spiritual significance.

*From X to Eternity* marks a moment in time when AI begins to manifest in earthly, more-than-machine-like forms. The work speculates on the possibility that AI functions like a kind of angel—an artificial yet intelligent messenger. In certain Christian theologies, such as Seventh-day Adventism, Michael the Archangel is interpreted as a pre-incarnate form of Christ. In our work, the drone becomes such an Archangel: unknowable and uncanny, yet entirely earthbound.

Angels in religious texts are messengers or agents of divine will—intermediaries between God and humanity. Similarly, AI can be seen as an "artificial messenger," transmitting, processing, and executing human intentions at scale. One poignant example is the emergence of "Death Avatars"—virtual replicas of the deceased. While angels are believed to serve a higher intelligence, AI, often perceived as autonomous, is in fact shaped by human design, input, and values. This analogy invites reflection on faith, agency, and trust: Who or what stands behind the intelligences we interact with?

In the video work, a drone—guided by AI—records children wearing sensors, effectively turning them into moving targets. The drone becomes a central, improvising actor alongside the children, whose bodies respond, adapt, and play. The programmers remain on-site, ready to intervene should the technology go awry. Over time, a game of mutual anticipation and response emerges between the drone and the children. What begins in curiosity eventually becomes a power dynamic: one side must yield. The children exit the scene singing *Dona nobis pacem* meaning "bring us peace"—a Latin hymn echoing across 2000 years of Christian history. In this emptied industrial space, once alive with machines and labour, the chant becomes both elegy and plea—a petition not only to angels, but perhaps to AI itself.

There is an undeniable parallel between medieval religiosity and our digital present. As Margaret Wertheim explores in *The Pearly Gates of Cyberspace*[i], the invention of Euclidean perspective helped structure Renaissance understandings of space—a fixed viewpoint producing an illusion of depth. In the digital age, this fixed viewer becomes the screen user, immersed in a similarly bounded field of vision.[1] Today's VR environments further complicate this: the user now moves, but their view remains tethered to artificial constructs, displacing physical reality. Is this merely an extension of perspective, or a reversal—where AI begins to train the human?

Unrehearsed interactions in the video demonstrate how AI and humans influence each other. While the actors learn to read the drone's movements and intentions, the drone records and reacts, guided by its own embedded code. This feedback loop reveals how rules of engagement—though designed—yield unpredictable and emergent behaviors. It also illus-trates the psychological and social adjustments humans are making in response to AI's presence in daily life.

This work explores both the losses and gains that come from our increasing entanglement with artificial intelligence. It evokes the deep human impulse to imagine the immaterial: an invisible world of non-human agents that influence our own. It also acknowledges spirituality's enduring legacy—its capacity to unify, to guide, and at times, to control. The Christian Church has historically rendered belief visible through architecture; its cathedrals and chapels stand as monuments to a world beyond sight. AI, in contrast, operates invisibly—its presence felt through action, data, and influence.

As Wertheim observes, "Euclidean space provided a new framework for visualizing the world, one in which the spiritual dimension was increasingly excluded."[ii] *From X to Eternity* reclaims the invisible not as superstition, but as an essential terrain—where technology, belief, power, and presence collide.

i Hito Steyerl, "They Are Oblivious to the Violence of Their Acts: Windows, Screens, and Pictorial Gestures in Troubled Times," lecture presented at Castello di Rivoli, Rivoli, Italy, December 12, 2008.
ii Margaret Wertheim, *The Pearly Gates of Cyberspace: A History of Space from Dante to the Internet* (New York: W. W. Norton & Company, 1999), 31.

# Nikiforos Staveris, Filip Lundberg & Isak Öhman

## Northern Lights, lights will guide you home

*Northern Lights, lights will guide you home*, achieves an even deeper resonance when viewed through the lens of gender equality and the pervasive issue of safety in public spaces, particularly for women and non-binary individuals. The installation's core concept of the Aurora Borealis as a form of interactive companionship takes on a heightened significance when considering the well-documented experiences of those who often feel vulnerable and intimidated by dimly lit or deserted urban environments, especially during the long, dark winters characteristic of Umeå.

Extensive research across various disciplines, including sociology, criminology, and urban studies, consistently highlights the disproportionate fear of crime and feelings of unsafety experienced by women and non-binary people in public spaces. Factors such as inadequate lighting, isolated pathways, and the historical and ongoing prevalence of gender-based violence contribute to a heightened sense of vulnerability. The simple act of walking home alone at night can be fraught with anxiety and a constant need for vigilance.

In this context, *Northern Lights, lights will guide you home* transcends its aesthetic appeal and technological innovation, becoming a subtle yet powerful intervention that addresses these very real concerns. The projected aurora, responding dynamically to the presence and movement of individuals, offers a unique form of symbolic companionship. For a woman or non-binary person navigating a dimly lit street, the ethereal dance of light that mirrors their steps can provide a sense of being seen, acknowledged, and less alone.

The interactive nature of the installation is crucial here. The fact that the aurora shifts and flows around the individual, particularly with unique effects designed for those with moving disabilities, creates a personalized and responsive presence. This digital mirroring can subtly counter the feeling of invisibility that can contribute to a sense of unease. It's not a physical protector, but rather a luminous echo, a visual affirmation of presence in a space that might otherwise feel indifferent or even threatening.

Drawing upon the rich tapestry of Old Norse myths and Scandinavian folklore, we can find echoes of protective and guiding female figures. While the Valkyries, as mentioned previously, were associated with battle and fate, other figures embody aspects of care and guidance. Consider Frigg, Odin's wife and the queen of Asgard, often depicted as wise and possessing foresight. Though not directly linked to the aurora, her role as a powerful and knowing presence resonates with the idea of a benevolent force watching over individuals. Similarly, the Dísir, female guardian spirits associated with fate and protection, were often invoked for aid and support.

The projected aurora can be seen as a contemporary, technologically mediated manifestation of these ancient archetypes—a luminous guardian that accompanies the solitary traveler. Unlike the often-male-centric narratives of old, this digital companion is inherently inclusive, responding to all individuals regardless of gender identity. Its presence is a silent affirmation, a visual reassurance in the darkness.

The "strange amalgamation of old and new" takes on an added layer of meaning when considering this gendered perspective. The ancient human desire for protection and companionship in the face of darkness finds a novel expression through cutting-edge artificial intelligence. The neural networks, trained on data that captures the fluid beauty of the aurora, become tools for generating a sense of safety and connection in a contemporary urban environment. It's a poetic irony that the most modern of technologies is employed to address a deeply rooted and often gendered experience of vulnerability.

The specific attention paid to individuals with moving disabilities further underscores the installation's commitment to inclusivity and care. The unique effects designed for this group ensure that the experience of digital companionship is accessible and responsive to a wider range of human experiences. This thoughtful design counters the often-homogenous and able-bodied assumptions that can inadvertently exclude marginalized groups from public art experiences.

The choice of Umeå, a northern city with long periods of darkness, as the site for this installation is particularly significant. The stark contrast between the potentially isolating winter nights and the vibrant, responsive aurora highlights the artwork's intention to counteract feelings of loneliness and vulnerability. The "lights will guide you home" not only in a literal sense but also in an emotional and psychological one, offering a sense of connection in a potentially isolating environment.

From a feminist art perspective, *Northern Lights, lights will guide you home* can be interpreted as a subtle act of reclaiming public space. By introducing a luminous and responsive presence, the installation subtly challenges the power dynamics that can make women and non-binary individuals feel like they are navigating a space designed primarily for and dominated by men. The interactive aurora offers a sense of agency and visibility, transforming the experience of walking alone at night from one of potential anxiety to one of unexpected connection and beauty.

The ephemeral nature of the projected aurora, while aesthetically pleasing, also carries a symbolic weight in this context. It is a temporary intervention, a fleeting moment of digital companionship. However, its impact can be lasting, leaving a trace of wonder and a subtle shift in the perception of the urban environment. It suggests that even temporary acts of beauty and connection can contribute to a greater sense of safety and belonging.

The integration of computer vision and AI, while seemingly detached, becomes a crucial element in this act of digital companionship. The technology acts as an unbiased observer, responding to human presence without prejudice. This neutrality can be particularly significant for individuals who may have experienced negative or threatening interactions in public spaces. The aurora's response is purely based on movement and presence, offering a non-judgmental and consistent form of interaction.

In conclusion, the updated interpretation of *Northern Lights, lights will guide you home* through a gender equality lens, reveals a profound depth and social relevance. The installation's innovative use of technology to create an interactive Aurora Borealis becomes a subtle yet powerful response to the pervasive issue of women's and non-binary individuals' safety and feelings of vulnerability in public spaces. By offering a luminous and responsive digital companion, the artwork subtly challenges the power dynamics of urban environments and provides a moment of connection and reassurance in the darkness. The "strange amalgamation of old and new" transcends mere aesthetics, becoming a poignant commentary on the enduring human need for safety and companionship, addressed through the innovative and inclusive potential of contemporary art and technology. The installation serves as a powerful reminder that art in public spaces can not only beautify but also subtly address pressing social issues, fostering a greater sense of security and belonging for all members of the community.

# Tyson Yunkaporta & Jack Manning Bancroft

## Kolab protocol for energetic and informatic relations

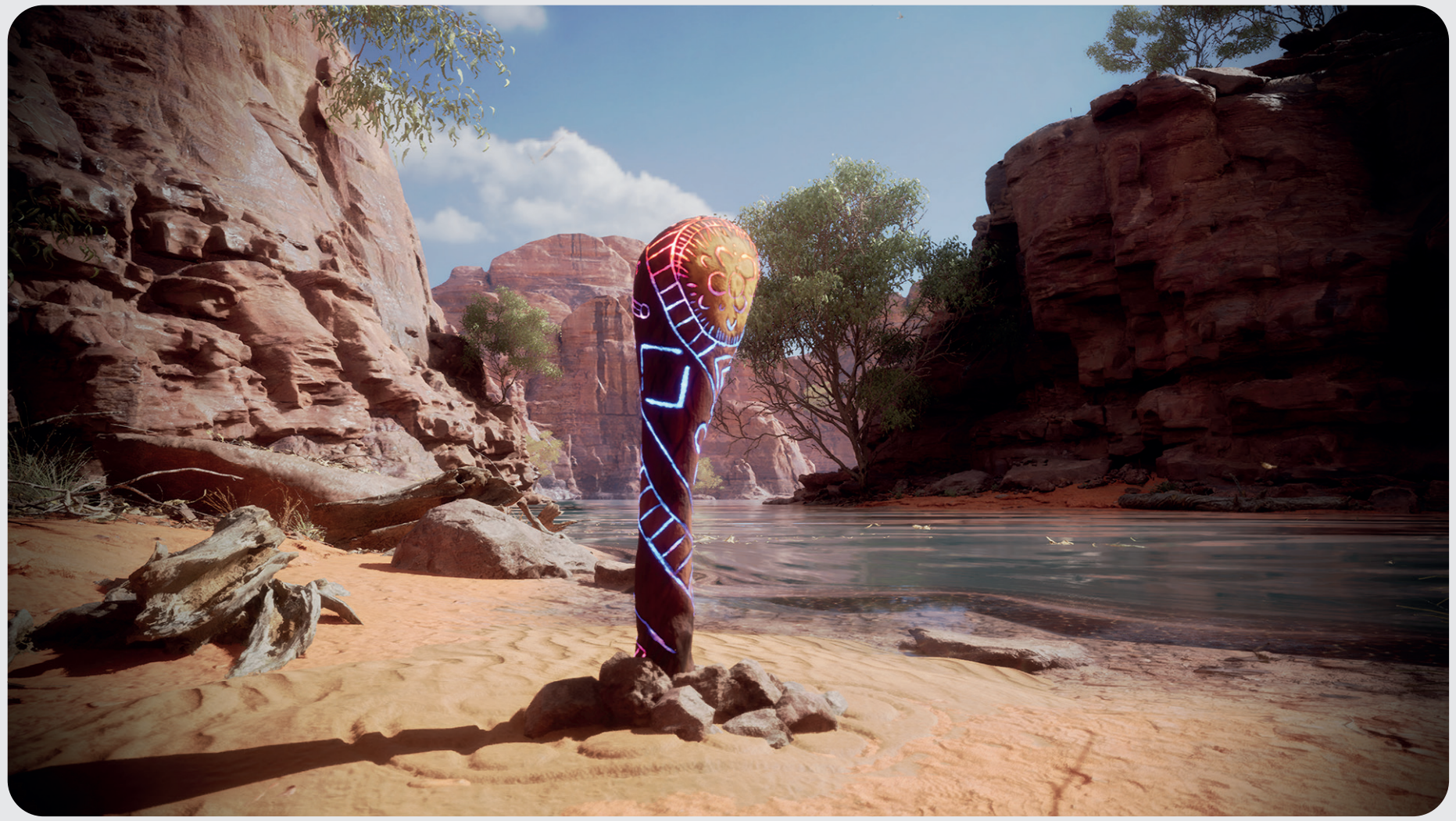

This work is a digital animation to communicate the thinking-feeling Indigenous Knowledge and Lore behind a system for research, analysis, design and management developed by the Indigenous Knowledge Systems Lab at Deakin University, Australia. It was created as an alternative to the skewed relations of neo-liberal 'partnership' models of co-management, co-design, and co-authorship that cause so many Indigenous programs to fail. It is Kriolised with a K replacing the C in these words to foster Kollaboration informed by bio-cultural landscapes of complexity and good relation, strong embassy.

| Ko-design Processes/ Relations | Part<br>*Autonomy* | Dyad<br>*Re(gu)lation* | Kind<br>*Governance* | Whole<br>*Embassy* |
|---|---|---|---|---|
| **Symbiotic Relation** (Non-zero sum exchanges) | ><<br>*-take-give-* | <><br>*-give-take-* | >><br>*-take-take-* | <<<br>*-give-give-* |
| **Kinship Relation** (Pronouns for group dynamics) | -I-<br>*Solitary tasks* | -us-two-<br>*Tasks in pairs* | -us-only-<br>*Tasks in groups* | -us-all-<br>*Whole team task* |
| **Informatic Relation** (Analytical lenses) | (=O=)<br>*Data points* | O=O<br>*Data connections* | (OXO)<br>*Data flows/loops* | =O=O=<br>*Data fields* |
| **Narrative Relation** (bio-cultural metaphor) | e.g. Ant signal<br>*Story to separate signal from noise* | Coming Rain<br>*Story to reveal phenomena* | Waterway Care<br>*Story to know processes* | Flood Behaviour<br>*Story to follow pattern* |

### Ko-design principles

A Ko-designer knows that they are just a node in a complex system, mirroring dynamics of land and community, moving in and out of diverse relations, from Part, to Dyad, to Kind, to Whole (see Table below). Group behaviour and governance patterns are communicated by non-human kin, through the physics and metaphysics of cooperative relations in local land-bases. Relations of Part and Kind nurture personal and group sovereignties and identities. In relations of Dyad and Whole, individual and group identities are de-emphasised in pairs and broad inclusive groups. Pairs are maybe unlikely matches that must negotiate boundaries and routines of exchange to regulate behaviour and remain in good relation. This is a smaller-scale process of the larger embassy dynamics that occur in the category of wholes: large class groups or groups of multiple classes, gathering with common purposes and stories, creating rituals, communities of practice, and collective meaning-making.

Symbiotic Relation describes the energetic exchanges that occur in natural systems (which include human social systems, in Indigenous ways of being and Knowing). Encounters between entities in these complex systems are categorised in western science as +-, -+, -- or ++ exchanges between predators, prey, parasites and hosts, framed as competitive and individualised relations in a series of zero-sum games. Ko-design relations include multiple entities in every singular encounter. Transactions are perceived as give and take in greater and lesser amounts across a pluriverse of relations, within a network of flows that are balanced in the aggre-gate. One species may be in a season of feasting or mating, but many other species are involved with signalling (colour, smell, movement, presence, display) to trigger multiple events that support and align with this activity.

Kinship Relation describes the social organisation and protocols of groups, using translations of Indigenous pronouns into English. These are: -I- to indicate actions of the self-in-relation; -us-two- to indicate events negotiated in pairs; -us-only- to indicate encounters in exclusive groups determined by relevant commonalities such as gender, identity, age, skill-set, complimentary knowledges and so forth; -us-all- to indicate whole group events, which are usually of ritual significance and not useful for completing tasks or decision-making. Note that each relation type has hyphens on either side —this is to indicate that even relations have relations! Each group or pair or individual is connected and networked far beyond the immediate field, and these relations must be respected as sources of knowledge, energy and resources.

Informatic Relation refers to various analytical lenses for understanding data, complex systems and design. At the first level (=0=) you focus on the data points or data sets in isolation, because reductionism is essential in many contexts and stages of knowledge processes, and is not some kind of dichotomous, oppositional enemy of holism. At the second level 0=0 you identify the connections between the different data points or nodes, being sure to distinguish which relations are correlative and which are causative. At the third level (0X0) you find the data flows, the energetic and relational dynamics of open loops, closed loops, positive feedback loops and regulatory loops. At the fourth level you examine the data fields by tracking the patterns formed by the previous three lenses combined.

Narrative Relation refers to the biocultural metaphors you can draw from your habitat or place of meaning, to frame your ko-design within living systems, no matter what the topic. These may be expressed as stories or images or exposition, created to separate signal from noise, reveal phenomena, understand processes and follow patterns.

# Art, Creativity & Design Vision

# Art and Design—New Leadership

## Juhani Risku

Juhani Risku was born into a bilingual Finnish-Swedish family in Vaasa. He was a child of the backyards and wastelands, a ragged boy who dug caves, built huts of sticks and wood, and even set a fire in a single acre of lawn while playing scout. School started on the Swedish side and continued the Finnish side, which, looking back, gave him cognitive dynamics. The good news about changing schools was that at first, he only understood sandbox Swedish, and later he did not really know Finnish. So, the school could not indoctrinate him. School did not suit him, so with natural kindness he always managed to get back to the backyards. A conscientious morning person, he was never late for school, and he was not absent a single day. Reading was also replaced by listening. The upper secondary school exam period proved the power of the wastelands: three board exams and direct admission to the University of Jyväskylä to study mathematics and physics.

The work at the power plant, which began during high school, led to the development of manual skills and machines, systems, and professionalism. There, every mechanic, and fitter acted as a master for Risku in the maintenance of the 300 MW turbine and generator. The power plant provided the skills for metalworking and maintenance of a large plant with both the British Parsons turbine and the German Siemens turbine.

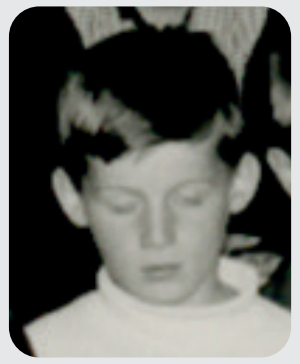

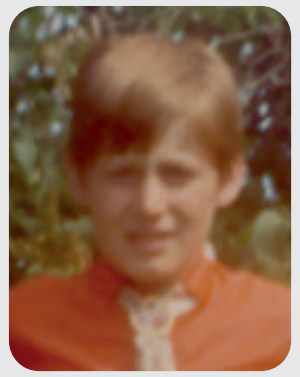

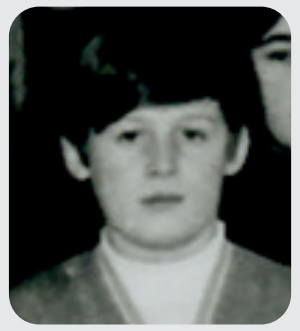

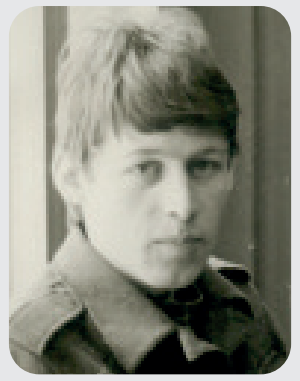

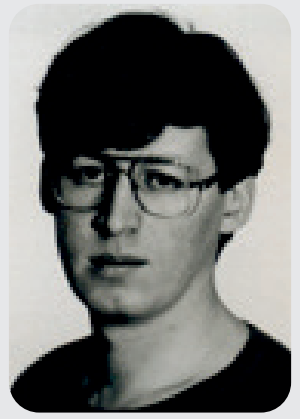

**School of Architecture**

However, mathematics and physics at the University of Jyväskylä did not seem to be his area of expertise, which Risku applied to the Department of Architecture at the University of Tampere, and he got in the first time, apparently completely by chance. The first year passed quickly, completing almost a third of the bachelor's and master's level courses. The first summer changed everything: he doubts that he realized that architectural abstraction is something different from what was taught at university—after all, they study quite ordinary bulk design of basic construction, which is faked as art. Therefore, he began to seek out masters, and to his surprise, they were happy to accept a first-year student. This began the development of architectural abstraction, in which arkhé and téktōn were the only guiding factors in studying and sketching. Later, arkhé and téktōn were added to the concept of poïesis, i.e., absolutely original, and previously non-existent development.

**Steam Turbine mechanic's work**
EPV Energy, Vaasa, Juhani Risku 1976–80
Siemens & Parsons

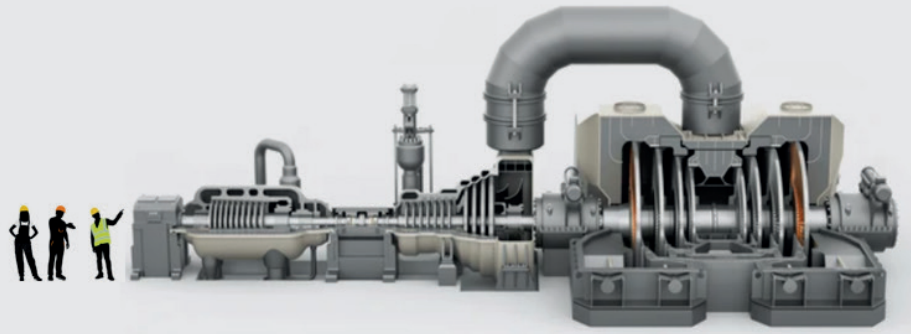

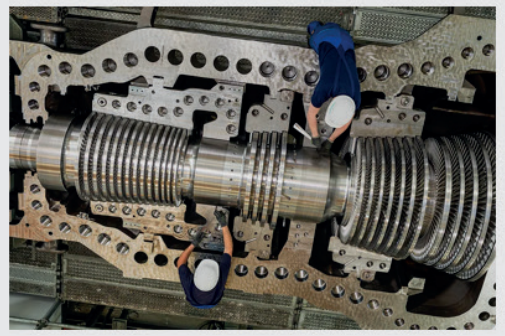

Risku worked as a mechanic apprentice in Vaskiluoto, Vaasa, on Siemens and Parsons steam turbines after high school. The work was manual and technically challenging. He became familiar with handcraft skills and the mechanics of structures.

**Army Chronicle, 200 pages**
Juhani Risku 1975 – 76

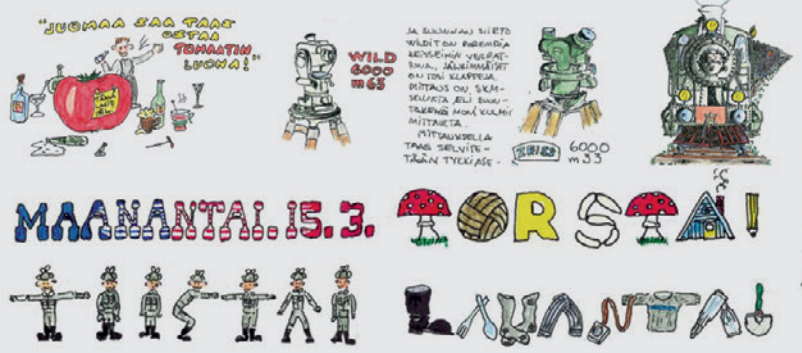

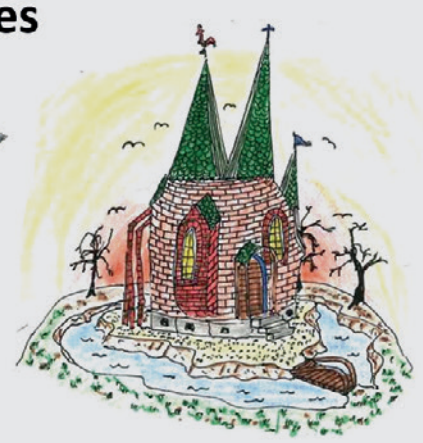

He served in the army in artillery, becoming a fireman and a regimental weapons officer. Risku wrote and illustrated a more than 200-page chronicle in the army, developing fonts and comics as a tool of power and entertainment. It is waiting to be printed as a book...

At the architecture school, Risku developed two methods that are still ongoing for him—the making of architectural sketches and the note-making method. An architecton, an architectural sketch, is a multidimensional physical sketch made simultaneously with a sketch, and a note-making is a conceptual entity based on a sketch. Drawing is not part of architecture, nor is making miniature models—sketches and architectons make architecture possible.

Two courses were overwhelming in this otherwise shoddy school: sculpture and painting, and photography! Risku took three years of each—nude model painting and sculpture, charcoal and oil paint, visual composition, all the colour theory, hand and eye sensitization, film-time light theory and photography, light painting, perspective correction and product photography. These courses taught both sensitivity and strength, self-confidence, and mastery of skills. Fortunately, he had the precision of a mechanic's hand and the responsibility of a power plant behind him.

The biggest disappointment of the architecture school was that it had gathered hundreds of mediocre people in addition to Risku himself. It would not have been possible to understand this without the hard work of the masters. Only one professor analysed and talked about architecture, the others pre-sented established working methods, talked about old buildings from the subjugated powers, falsely claiming them to be architecture, and presented their own poor, C-class buildings.

After graduating as an architect, Juhani Risku founded his own office and was allowed to design a few projects. The economic depression drove Risku and his family to Paris for five years.

**From Pispala to Paris**

The move to Paris took place from Pispala, the medieval Paris of Tampere. **Mika Kaurismäki's** *The Worthless* (1982) was a romantic ideal for Risku, which Paris turned out to be. His life in Paris, in turn, followed **Aki Kaurismäki's** *Bohemian Life* (1992) in all its daily and artistic activities: there was poverty, creativity, life in Montmartre, good food, cheese and even wine. Risku had a particularly romantic connection to Paris, having also lived a poor life in Pispala: no money, no job, an architect's degree, an ideal understanding of architecture, and his car was a Citroën 2CV4 from 1971. The car was good. It was the most beautiful graphite-grey vehicle and its own cast and enamelled SF plate, the country's symbol, was on the rear bumper. As a morning activity, he toured Paris on foot and from five o'clock on the metro with self-made tickets, collecting books, furniture, objects and firewood for the cold, damp apartment at 55, rue Lepic. Romance was distant, but ever present.

Paris turned out to be the perfect city for Juhani Risku: It was a place where he could be incognito or get in touch with Parisians. Paris is of course a historical place, but it also has the best modern architecture, which complements the pompous style. Living in Montmartre was perfect: the Arc de Triomphe and the Eiffel Tower could be seen from the kitchen window, and the Sacré Cœur from the living room. The landlady was a concert pianist who came to teach her students on the concert grand piano in the living room on Fridays.

In 1994, Risku wrote a preliminary theory book on architectural forms, which has not been published, but it has 260 pages already folded in InDesign. In Paris, he worked as a photographer, architect, and artist, conceptualizing numerous architectural plans around Paris. The photographs were stereo images on slide film, which accumulated about 5,000 pairs of images.

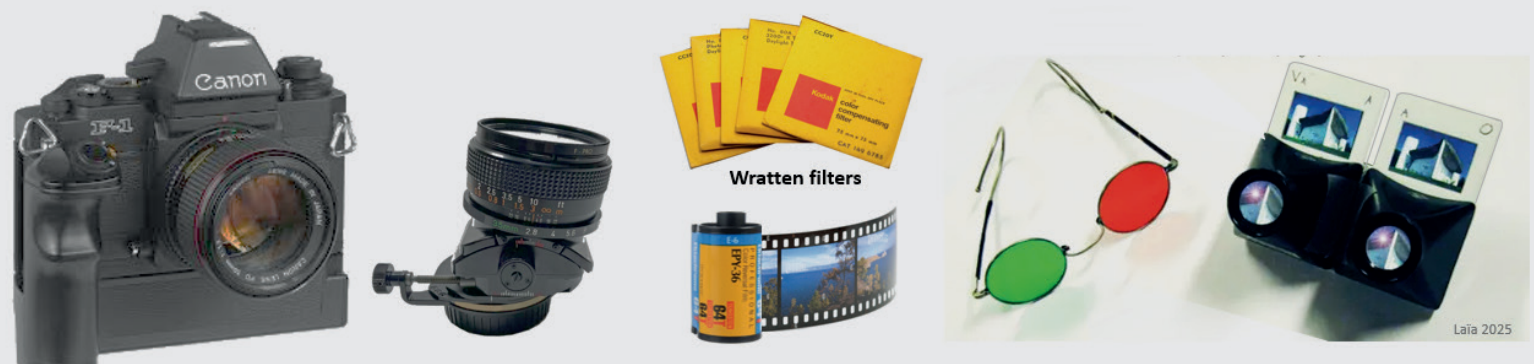

Professional camera | Tilt & Shift 35 mm lens | Colour slide film 35 mm | Black & White glasses | Stereo slide viewer 2 x Gucki

Juhani Risku began stereo photography on slide film in 1981. He has accumulated a total of about 25,000 pairs of images of architecture, nature, and various artifacts. Real film is still needed for stereo photography and the skills of the film era to expose correctly.

Atelier Risku Paris applied the Pizzi™ precast technology and manufacturing system in Paris and the casting was done with high-strength fibre concrete (K120). Risku was a member of Partek Dimensio™ team.

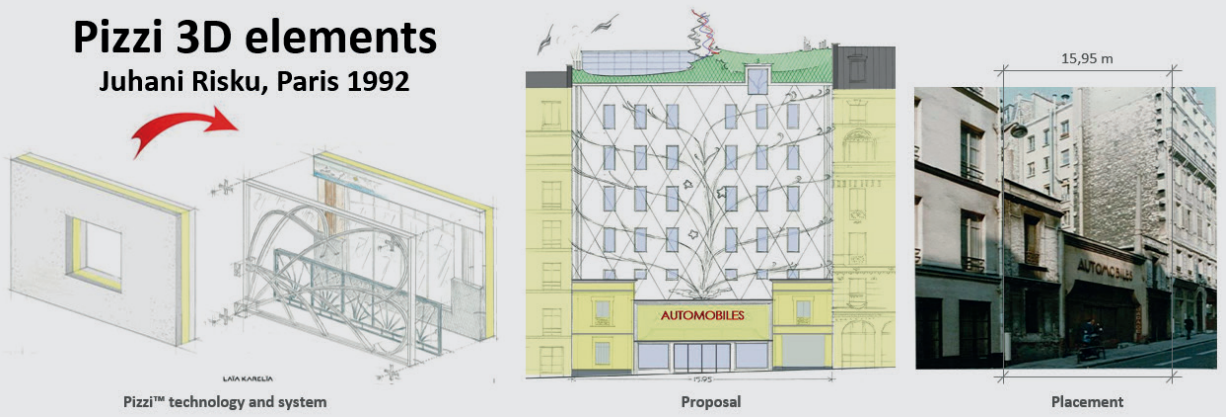

The Pizzi™ series was a collection of 3D elements to replace normal sandwich elements. The elements were "printed" on a powerful 3 x 6 x 2 m³ 3D printer. The draftsman carved the mould from artificial clay and the casting was done with high-strength fibre concrete (K120). Risku was a member of the Partek Dimensio™ team. Between 1987 and 1992, Risku created several architectural examples of 3D concrete elements and components around Paris. None of them were built.

Pizzi™ is a technology and manufacturing system for concrete elements. It was applied as a proposal by Atelier Risku Paris at 71 rue Blanche, Paris 9ème in 1994.

All essential concepts must be defined so that they can be handled and used appropriately in different contexts. Beauty is one of the most important, especially in the arts. Here is the definition of beauty in the form of a function, which Risku uses in his theory of architecture.

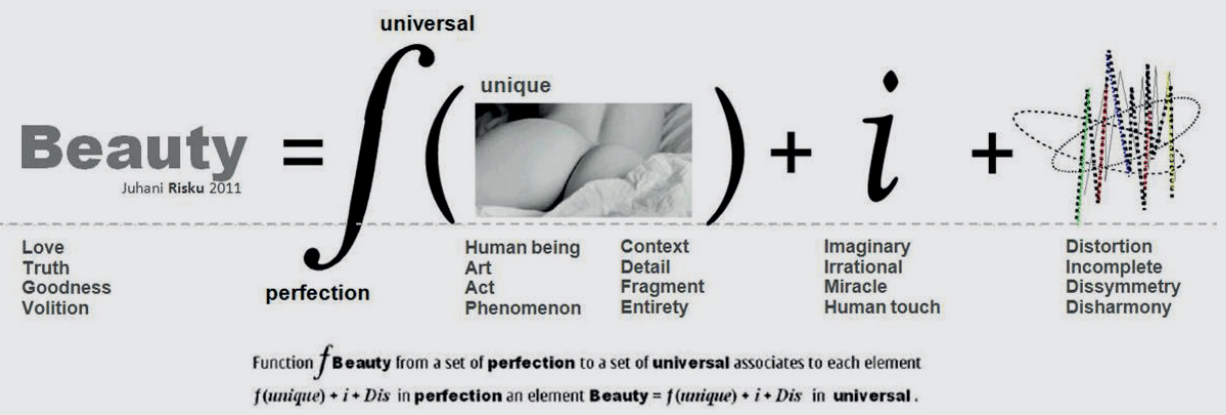

Four-dimensional phenomena have always interested Risku. That is why Cynefin, the sense making framework, by Dave Snowden, is well suited for the analysis and understanding of almost any phenomenon. Risku also developed his own Tetra theory, in which most phenomena are revealed to be four-dimensional. Here is Steve Jobs placed in Tetra, by his mindset.

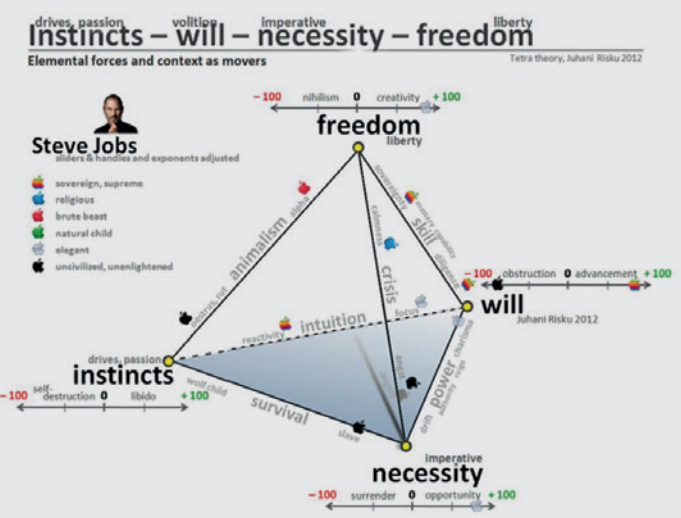

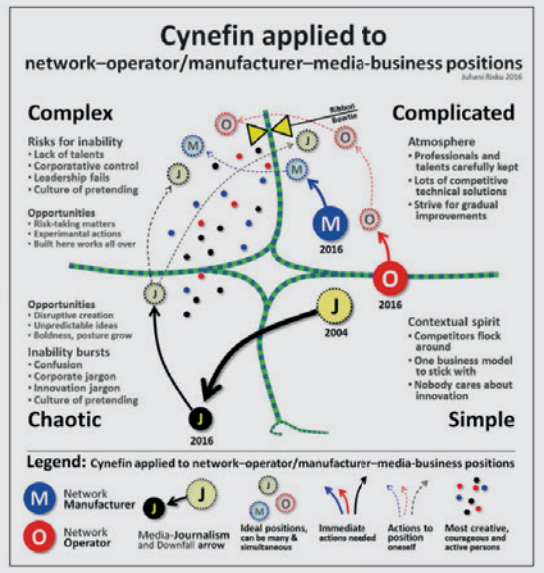

**Graphic Design**

Drawing has always been one of Risku's skills, but at a very young age, around ten years old, he began to design new objects, cars, and phenomena by visualizing them. This left drawing behind and conceptualization took over. He usually built devices from scraps he found, which, however, never worked. He applied graphic design to tables and diagrams related to design and architecture.

Risku's doctoral dissertation article “Software startuppers took the media’s paycheck. Media’s fightback happens through startup culture and abstraction shifts” presents the latest possible turn in media, where state-sized companies Google (Alphabet), Facebook (Meta), Amazon and Apple can be brought down with a new model of media and journalism. The basis is The Trio, which consists of a network operator, a network manufacturer, and a media company. They include all mobile device owners in their developer community and start producing new content, also receiving a salary for it. Below is a graphic representation of the media change.

**Architectural concepts**

Over the years, Juhani Risku has outlined and defined architectural abstraction theoretically and practically to meet the requirements of arkhḗ, téktōn and poïesis and skill building. This requirement is strict, as a designer should only design buildings that he or she can build. Only that brings credibility

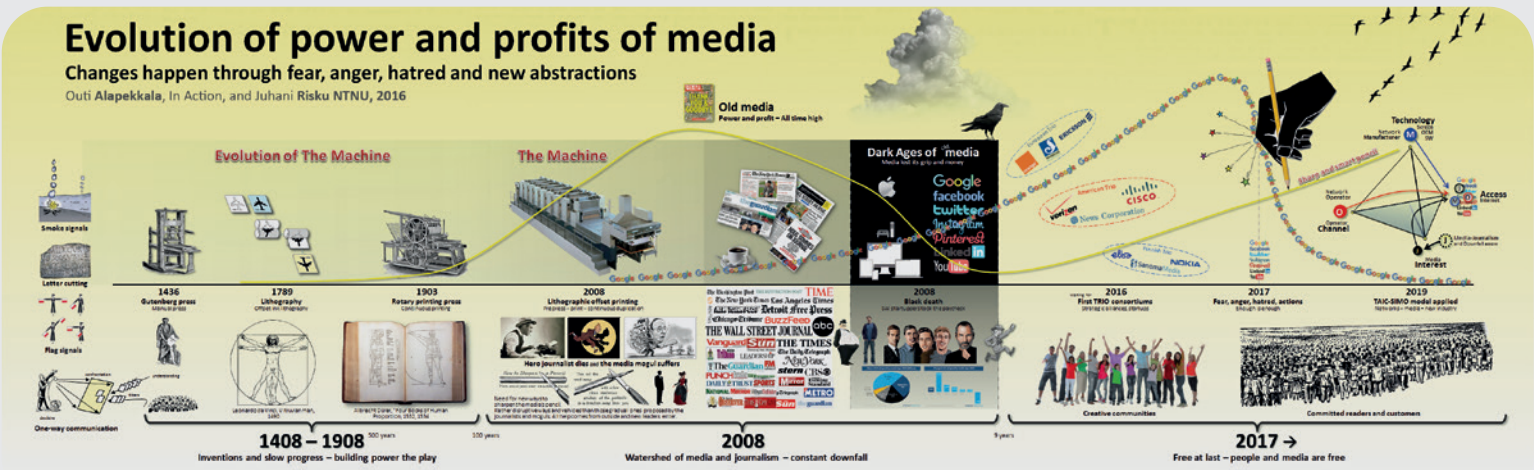

Illustration from Risku’s dissertation. It shows the evolution of the media industry in such a way that the largest companies must be displaced by new innovative ways of operating.

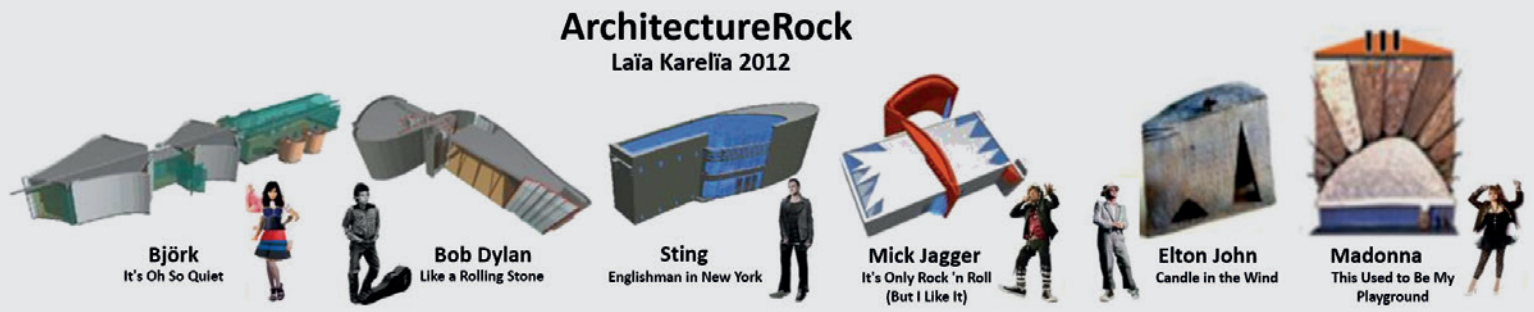

ArchitectureRock—better homes, better Rock ‘n’ Roll. Risku sketched out their own houses for six good rock artists. Usually, rockers have the most stupid style copies as their homes, which shows that they are ultimately cowards. ArchitectureRock makes it possible to be a civic hero for the first time.

to the designer. Otherwise, the plans may be just artificial compositions without architectural depth. Examples of this are **Frank Gehry, Zaha Hadid** and **Rem Koolhaas:** They only used flashy drawing, artificial intelligence, and their crazy privilege to produce beloved garbage. None of them could hardly do anything with their craftsmanship—a master builder, or architect, can.

**Nokia**

Juhani Risku applied for a job at Nokia Corporation and got a job as an industrial designer in the User Experience and Usability organization Nokia Networks. Working at Nokia was like a dream and a paradise compared to the drudgery, clay, mould, and poor design of the construction industry. He had already done his master's thesis as an architect for Nokia, Fazer and Valio, so getting into Nokia was an honourable thing. Nokia had good products, the best processes, high-quality engineers, and good leaders. Until… … everything started to collapse due to the combined effect of arrogance, too many incompetent leaders and finally a lazy and cowardly engineer. Risku changed organizations from Networks to the company's business development unit, from there to the Generation Nokia marketing organization and finally to smartphone design. His entire career at Nokia was a versatile work of an industrial designer.

Risku has made several thorough proposals since 2002. Below is the first one, Mobile Arena, which will renew the entire mobile infrastructure, services and media.

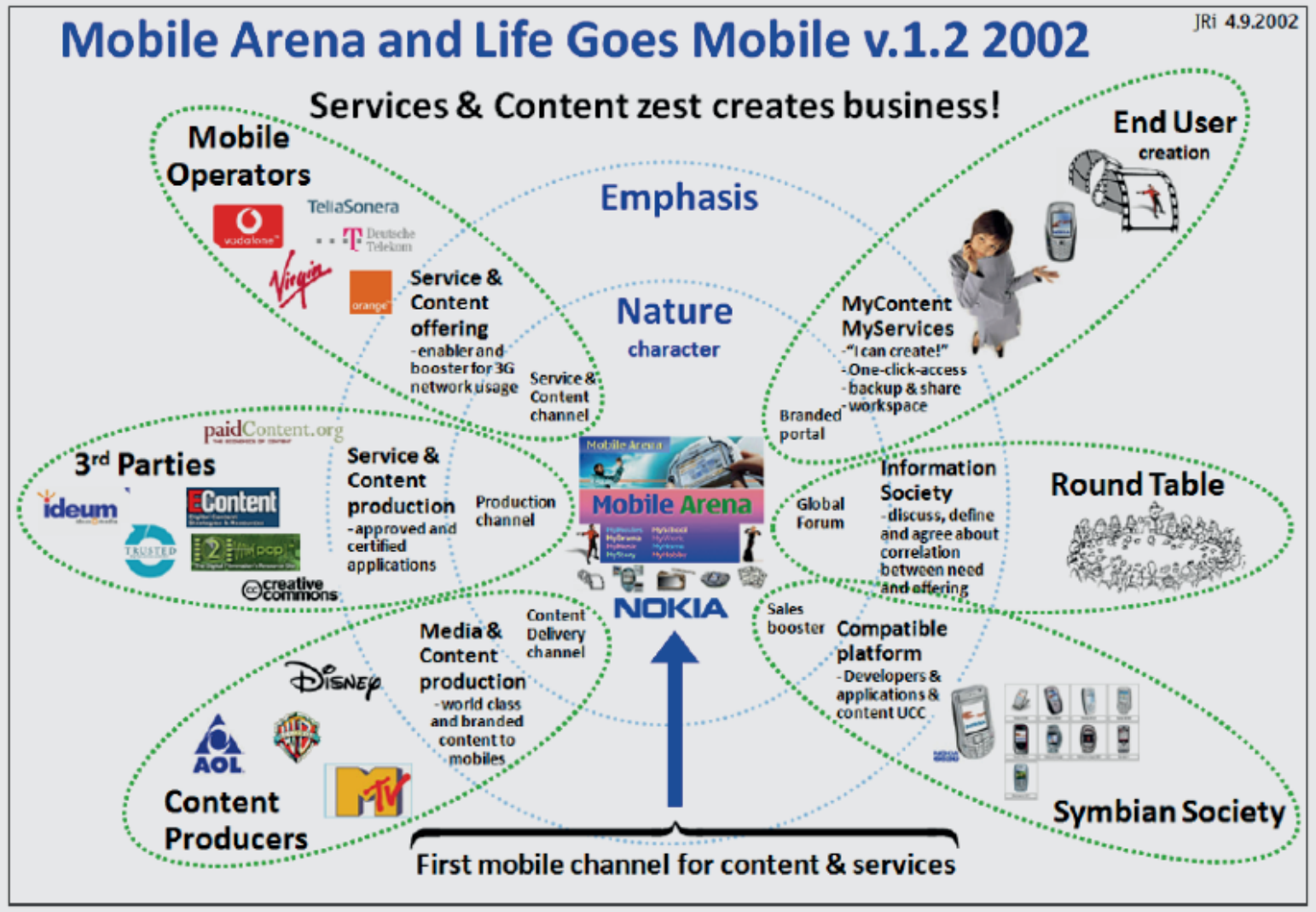

Mobile Arena version 1.2 in the fall of 2002. It would have been conceptualized in about six months by 20 people, the cost would have been 0.6–1.0 million €. With this investment, Nokia would have had a once-conceptualized mobile service package in its desk drawer and a virtual team that could have started implementing it quickly if the order had been given.

Mobile Arena 1.0 was a one-man, fifty-page PPT presentation of mobile services and solutions, which included location, music and video services, community, education, and creation services.

The presentation was rejected at Venture Achievement because it did not relate to the competition categories. In their assessment, the Venture directors NVO (Nokia Venture Organisation) said that mobile services and production are not in Nokia's interests.

Xseries is an overall concept published internally at Nokia on October 28, 2008, which is Risku's own concept in his role as Head of S60 User Experience Operations. Services and content have been re-merged in the Xseries design into segmented customer groups according to benefits, hobbies, interests, and human needs. Since the Xseries conceptualization worked with a new management model, it would have freed up Nokia to produce models faster to market and customize features for different companies, operators, and individuals. Nokia was notoriously slow in product development; the same device was available in different shells as amputated versions—only the plastic parts changed.

The Xseries would have been the new Nokia flagship model, launching the most significant innovations in the industry, technical innovations, and comprehensive services like the market leader, with which the user could finally do everything that had been expected of a mobile device for a couple of decades. According to the new product development model designed by Risku, the 18-month process from idea to platform would have been the same in three months. This would have made Nokia reach Apple's iPhone in its version 2.0. Nokia did nothing but slept through its luck and eventually died.

On his last day at work, Friday, July 31, 2009, Risku gave the then CEO a 95-page booklet that discussed what Nokia needed to do and who to lay off in order to improve its competitiveness. Apple had released the iPhone in the spring of 2007, which Nokia was three years late to release. The booklet was made in the same format that Nokia published a few times a

**Nokia Corp.**

**Juhani Risku, 2001–2009, Helsinki, Tampere, Espoo**

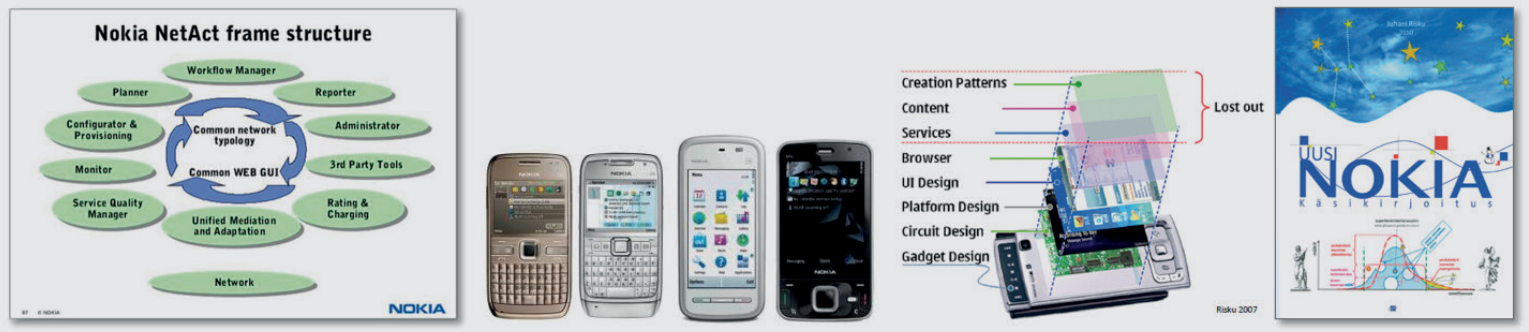

Nokia Networks' NetAct™ participation by Risku — Symbian S60 participation by Risku — Criticism by Risku — Nokia Book by Risku

The picture shows Juhani Risku's career at Nokia in four parts: 1. Nokia Networks and NetAct™, 2. Symbian S60 User Experience, 3. analysis and criticism of Nokia's contribution to device development only, and finally the book New Nokia—Manuscript in 2010.

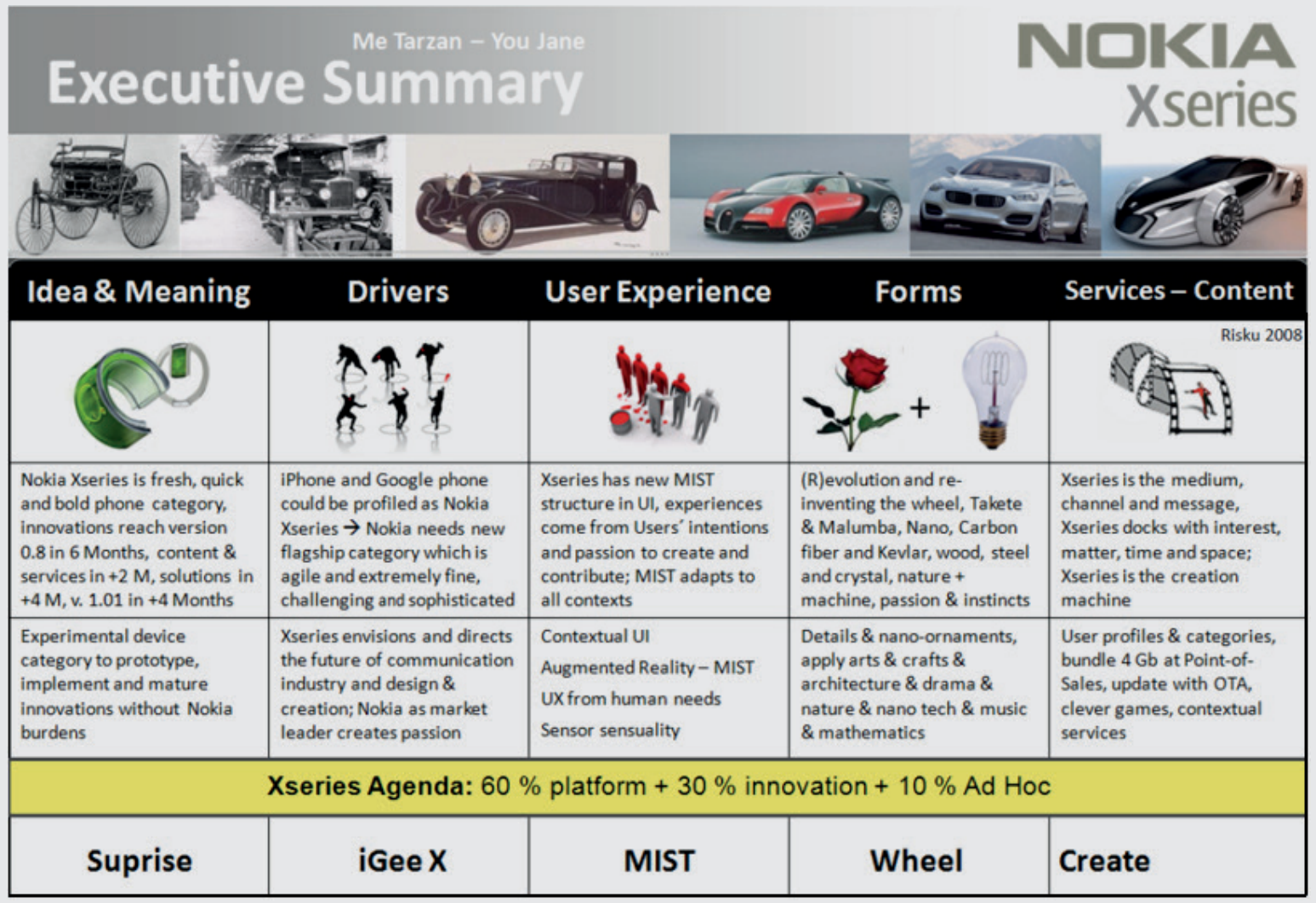

Xseries was both an experimental and revolutionary category, where innovations and concepts could be brought to market quickly without the burdens of Nokia's slowness. Xseries would have used new technologies and a user interface, which is a new generation of MIST UI. Technologies include SOM (Self Organizing Maps), sensors, Tactile Touch, projector, NFC (Near Field Communication).

year —in colour, with lots of pictures and analytical text. The booklet was also harsh and critical, criticizing weak leadership and arrogance. Nine years at a great company were behind him, after all, he had also done his master's thesis as an architect for Nokia, in addition to Valio and Fazer. The booklet was Risku's Exit Report, an analysis of why he was resigning and a proposal to fix Nokia into a good company again.

### French course turned into a getaway to Finnish Lapland

After Nokia's layoffs, Risku was left with nothing, so he had to find something to do. There were no architect jobs available, but he had saved up some money. His unemployed status offered him any course in Finland in any field. He chose an 80-day high school French course because five years in Paris was not enough. The Exit Report Risku wrote at Nokia in 2009

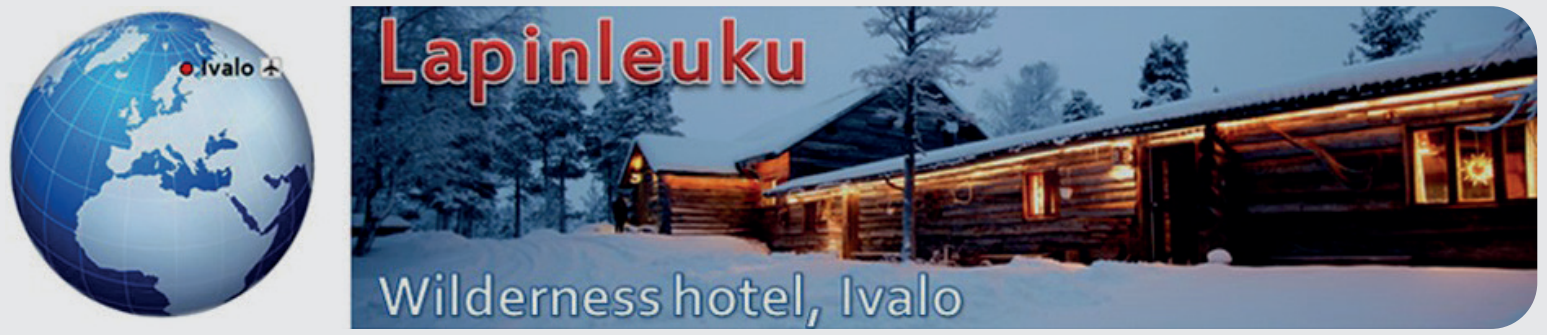

The wilderness hotel Lapinleuku was one of the finest and most mystical hotels in Finnish Lapland on the shores of Lake Inari. It offered Risku hotel studies, but also excellent conditions for creativity. The Nokia book, city models, Arctic formalism, several lectures, and participation in the sceonography of the local theater were born there.

was completed in a wilderness hotel in the spring of 2010 as the book New Nokia—Manuscript. It caused a stir in Finland and sparked a wide international debate. Because of the book, Risku's French course turned into a getaway to Lapland, Inari and Ivalo, as he was on the cover of the tabloid press with his picture more than 20 times, but also in quality media, such as the Wall Street Journal, New York Times, Daily Mail, Asahi Shimbun, Le Capital, BrandEins, Wired, Tech Crunch, FastCompany and The Register.

Instead of 80 days, he stayed for five years on the shores of Lake Inari, in an Arctic wilderness hotel. There he was able to make up a thousand beds, take hundreds of tourists on reindeer, husky, sled and hiking safaris, and speak German, French, Swedish and Norwegian. The best things about Lapland are the polar night and the -30°C frost: no mosquitoes! In addition, he was able to play ice hockey with the Tunturikiekko team and often visit the Siida Sámi Museum.

In the wilderness hotel in Inari, there was time to develop a new garden city concept, Arctic Garden City. Risku was assisted by Claude Nicolas Ledoux (1736–1806) and Ebeneser Howard (1850–1928). They had worked as urban planners in a special way, which made it worthwhile to team up with them. Risku's Arctic Garden City is a modern-day systemic city model that scales to all cities of approximately 200,000 with supplementary and conversion construction. New garden cities can also be built according to the model.

**Winterisation of Segway, trip to Murmansk 2010**

In Inari, Finnish Lapland, Risku designed a new northern form and architectural typology for the Arctic region. It is small-scale, local, and blends into nature. The Segway was made usable in severe frost and snow. Two Sámi women helped create its winter outfit.

**Industrial Design**

Juhani Risku's interest in industrial design began when he was working as a mechanic at a power plant. When servicing a turbine, various tools and mechanical devices had to be built separately, as there was no equipment or equipment for everything. After studying elsewhere outside of architecture school, he found product development

Arctic Garden City is a Ledoux–Howard–Risku urban model that is ready for immediate use in infill development, new city construction, and urban renovation. This urban model is a synthesis and development of the 21st century city. It scales from cold to hot conditions, from desert to mountain, from island to continent, from fields to backyards.

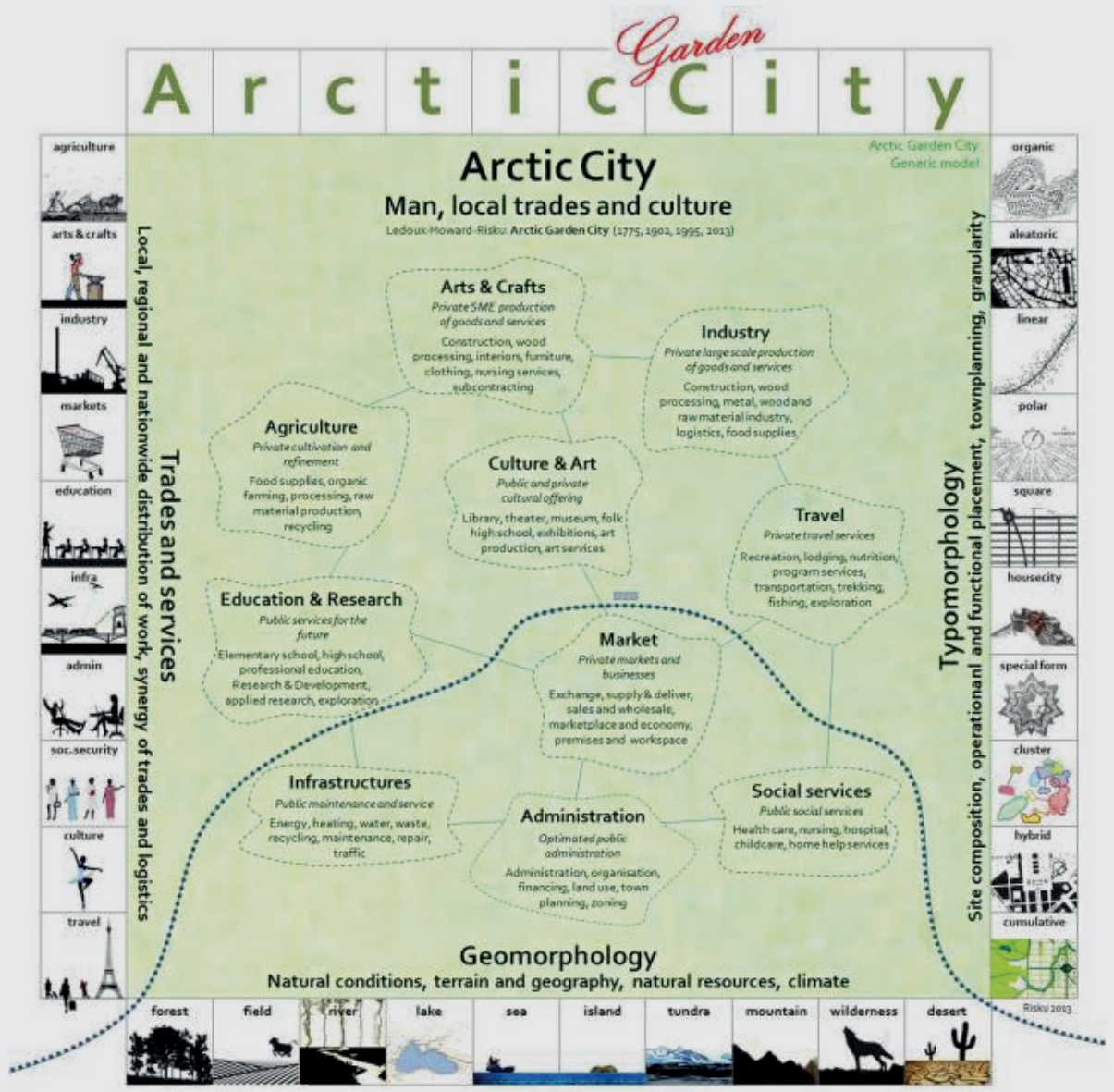

Arctic architecture is a concept developed in Lapland that is small-scale and built from local materials, making them suitable for human habitation in nature.

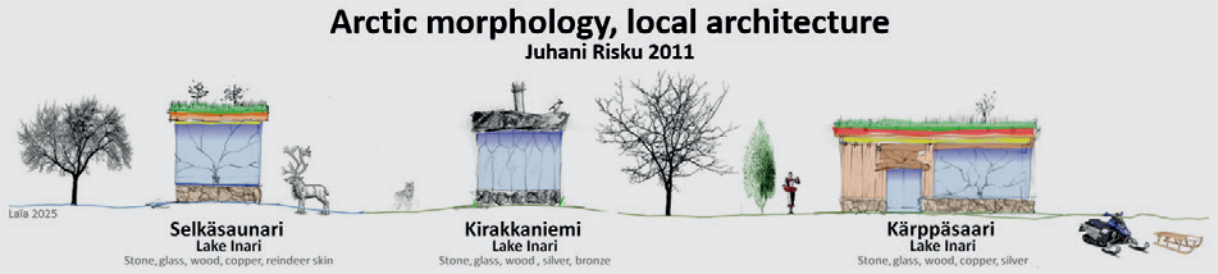

Risku's Segway comes with a winter gear package that includes Arctic Segway clothing to protect against the cold, studded tires, and riding gear to explore the northern wilderness. He went to Murmansk in the winter to ride this strange device.

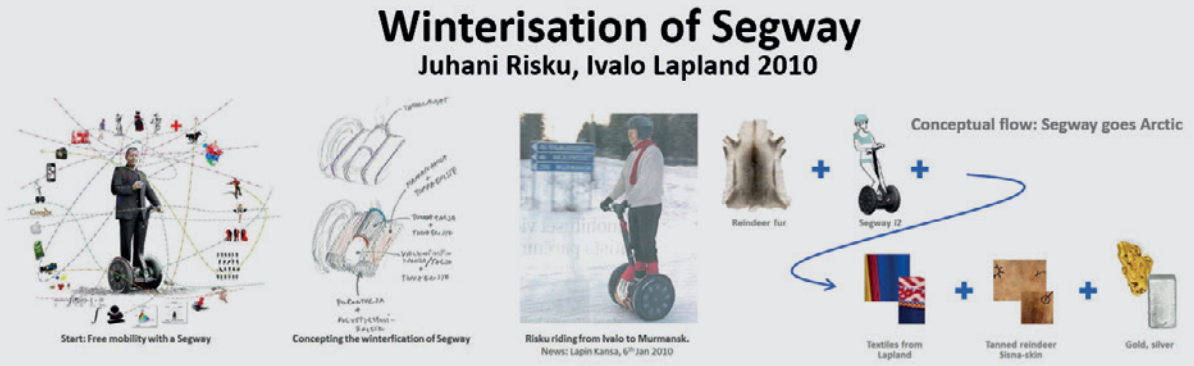

The Ledoux system is a system for large load-bearing structures that uses cast iron connectors with wooden columns and beams. The system allows for the connection of various functional elements as needed: benches, containers, lights and electrical appliances.

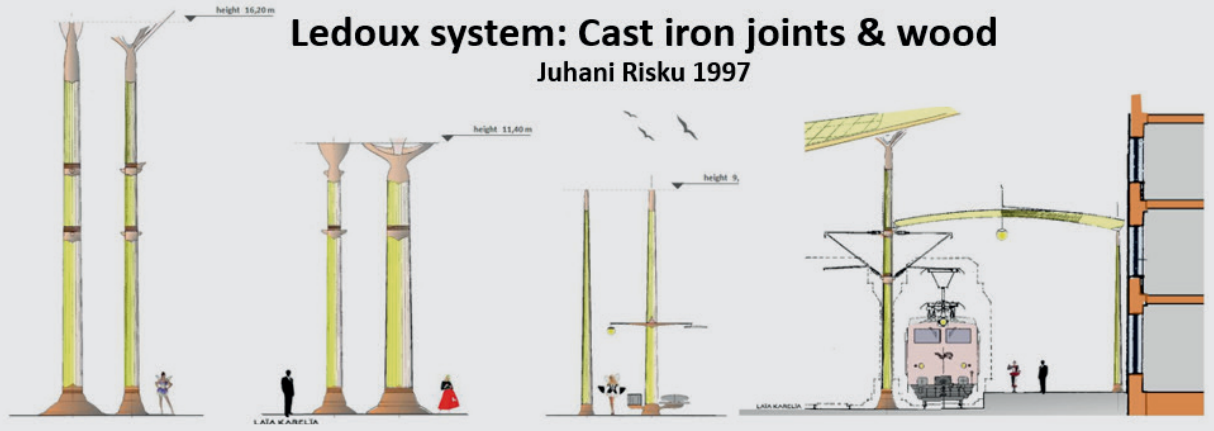

and various construction techniques to be normal activities—once he knew the manufacturing methods, he was able to implement more special solutions himself. After returning from Paris, he studied industrial design, 3D modelling and even fashion design at Aalto University, having worked as a teacher of scenography and costume at the Drama Studio of Tampere University and as a teacher at the Educational Theatre for five years.

The work at Nokia is clearly industrial design from systems to products, production development and brand design, precisely the topics that belong to this part of design.

Interest Machine™ is a future device solution consisting of mobile devices, computers, and televisions. All devices are interconnected and dynamically controlled. Interest Machine applies the next abstraction of computing and technologies such as virtual reality (VR), artificial intelligence (AI), augmented reality (AR) and new media abstractions such as dynamic visualization, knowledge and meaning-based search, sorting and cross-matrix computing. Usability, understandability, and relevance are the drivers of the next generation of devices and solutions, where data is transformed into information, information into knowledge, knowledge into truth and wisdom, truth and wisdom into drama and beauty, drama and beauty into meaning.

**Educational concepts**

Risku started designing educational sets and tools at the Sara Hildén Academy in Tampere in 1986. He had just graduated as an architect and worked at the academy for five years, alongside his own office. The students were children and young people aged 8 to 16, for whom architecture was a new abstraction. At that time, drawing and crafting had to be avoided, so conceptualizing, planning, and building architecture freed them from the indoctrination of visual arts.

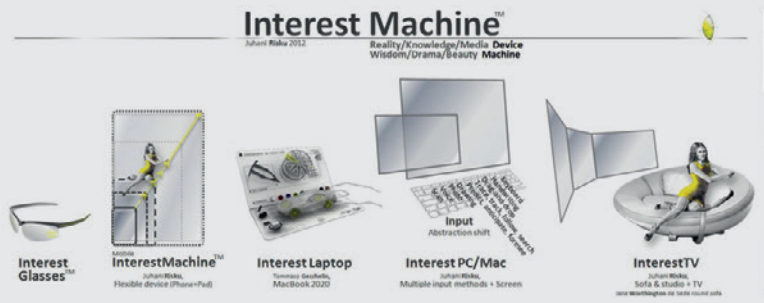

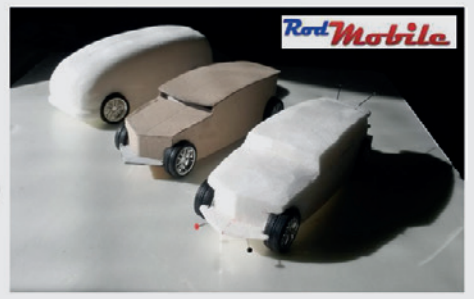

When Nokia did not take up Risku's Xseries concept, he later developed his own Interest Machine system. Similarly, RodMobile is an electric car system that can be built on a ready-made platform, either by buying parts from a store or by making them yourself.

The designer and architect series consist of basic geometric pieces, different human contexts, and objects collected by the designer and architect in their own box. It is also important for children and young people to touch objects and find new uses for them.

**Designer's & Architects Sets**
Juhani Risku, 1986– 2015

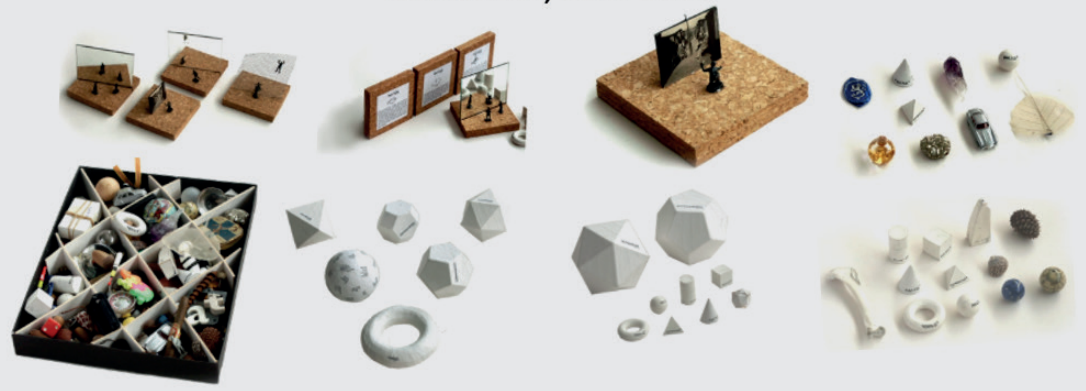

Risku has used the designer and architect series at the Universities of Tampere, Lapland and Jyväskylä, Aalto University, NTNU and Sara Hildén Academy. The exercise task was to make a box where found objects and various subjects from nature are collected. They can serve as sources of inspiration and provide ideas for both design and writing.

As a general observation, it can be stated that too many degrees at university can be completed simply by sitting and wondering. When the student workstation developed by Risku is based on active note-making, sketching and conceptualization, the student is already more active in completing his courses due to the new way of working. You do not draw during lectures; drawing is for children aged 2–12. Lectures are the most important centre of knowledge, skill and understanding in the academic world. In lectures, you must be able to trust what the lecturer says. However, Risku has already encountered too many incompetent train-

The lecture is the most important platform for academic study, which is human-friendly and reciprocal. The lecture is also dynamic, allowing interaction between the lecturer and students to promote learning and research. The lecture also aims to encourage students to work independently.

**Lecture is the student's workstation**
Juhani Risku, NTNU 2015

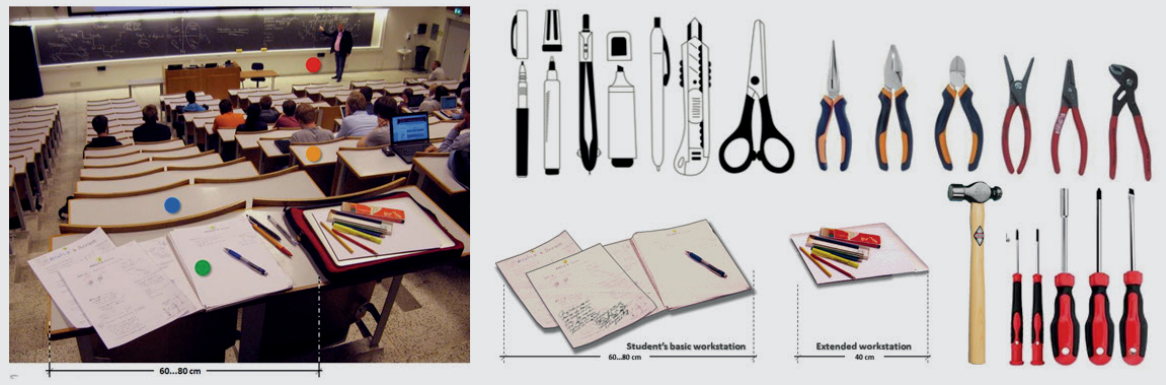

The first academic course in podcast format of podcast, Design Podcast5 credits, was arranged during 2021–2024 at the University of Jyväskylä. The themes were Design, Creativity, Form giving, Technologies, Service Design, Sustainability, Human being, Criticism, Education, Beauty, Design Thinking and Conclusion. All themes were connected to interests and practices in design contexts.

**Design Podcast, 12 parts à 1h 20min**
Juhani Risku, classroom 2018–2020 & Podcast 2021–2024

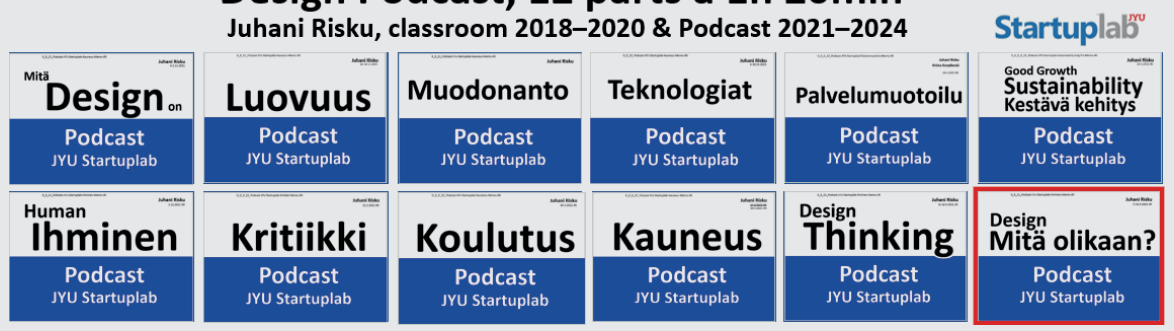

ers and pedagogues, so in the end, over 26,000 universities are sure to produce mediocre ones. The student's workstation changes everything: it makes the student skilled, goal-oriented, independent, and courageous. He or she also has thousands of pages of notes from his or her studies, which he or she can bind into books for himself or herself.

The Fuzzy Front End Design course Design Podcast series consisted of 12 episodes, each over an hour long. The discussions included design professionals from Finland, the Netherlands, and the UK. Two inspiring assistants from the Startup Laboratory acted as facilitators for the discussions. Almost a hundred students from all faculties of the University of Jyväskylä went through the podcast course. They were consistently so good at their tasks that the average grade was close to five on a scale of 1–5. The workload was heavy, and artificial intelligence was not allowed.

**Trondheim, Jyväskylä, Järvilinna**

Juhani Risku's academic career began in a special way: After writing the book New Nokia—Manuscript, he received a surprise invitation to come and meet Intel's IT Innovation Director for Europe. Risku packed the necessary supplies and his Segway i2 SE vehicle into his van in Lapland. The book had attracted a lot of attention around the world, so the keynote was perfectly timed. The inviter was a professor who also became Risku's dissertation supervisor and long-term collaborator. First, he went to Trondheim with the professor and then to Jyväskylä to study for a doctorate.

At NTNU in Trondheim, he began working on information systems science and also built a laboratory for the department for practical design and implementation needs. The student workstation now had an NTNU Makerspace alongside it. At the same time, two articles were completed for the upcoming dissertation. Soon, the opportunity opened to begin doctoral studies at the University of Jyväskylä, in information systems science.

When JYU StartupLab was established at the University of Jyväskylä, entrepreneurship, innovation, design, and creativity entered. StartupLab was an inspiring workplace for all its students, researchers, and staff. There, people did not just sit and read, but also came up with ideas and conceptualized services and systems, made plans and practically built devices and software-based products. Startup culture also created energy for students from other faculties of the university when they participated in the Fuzzy Front End Design course. The course encouraged the use of note-making, not note-taking. Note-making, like designers, includes sketching, conceptualizing, taking notes, critiquing, product design and free abstraction. The FFED course was made into a 12-part podcast course, which was the first design course for liberal arts universities to offer five credits. In these universities, art and design are usually absent, making FFED a solution to activate creativity and innovation in universities globally.

**Future**

Risku's doctoral dissertation was completed in 2021, dealing with design and creativity in software-based startups. In it, he was able to combine the experience he had gained during his professional career and become a freelance researcher. The dissertation was written in the most inspiring community, with the best resources and guidance. In his opinion, it is worth writing a dissertation in a new field for himself, as it allows him to combine already acquired knowledge and skills with new ideas and innovations. Now the dissertation had a perspective of creativity, design, startup culture and information systems science. On the other hand, creativity, design, and entrepreneurship were already fresh in his mind, but also information technology, such as 3D modelling, and Building Information Modelling (BIM) are commonplace in architecture. However, they are used for bulk construction and for decorating and strangeifying buildings (strangification), which has developed into wow construction following the classification of diseases and crimes.

After finishing his university career, Risku settled in the Järvilinna Art Centre, where he has his own workshop and adjacent workshops for carpentry, ironwork, and ceramics.

**The most important thing—to be an apprentice to many masters**

Juhani Risku considers two things to be the most important professional aspects: that he has been able to be an apprentice to masters who have welcomed the young man with pleasure and have guided him in everything selflessly, and that he was able to expand his understanding into new fields in his doctoral dissertation.

Juhani Risku's stage name is **Laïa Karelïa**, which are the French names of his father's and mother's birthplaces, Laihia and Karelia. The writing style has been found on a 16th-century map made by French cartographers. He now writes a theory of architecture based on the concepts of arkhḗ, téktōn and poïesis, full-stack Arts & Crafts skills, the management of skilled construction and an altruistic lifestyle. This whole thing Risku calls architecture an abstraction, which is required of a person who considers himself an architect.

Risku combines classical art with modern information technology in his visual art. His latest work is a double artwork, a sculpture, and a painting, CryptoDuo A², which is encrypted both physically and programmatically so that they cannot be forged. Both parts have both DNA protection and a documented blockchain, which are impossible to forge.

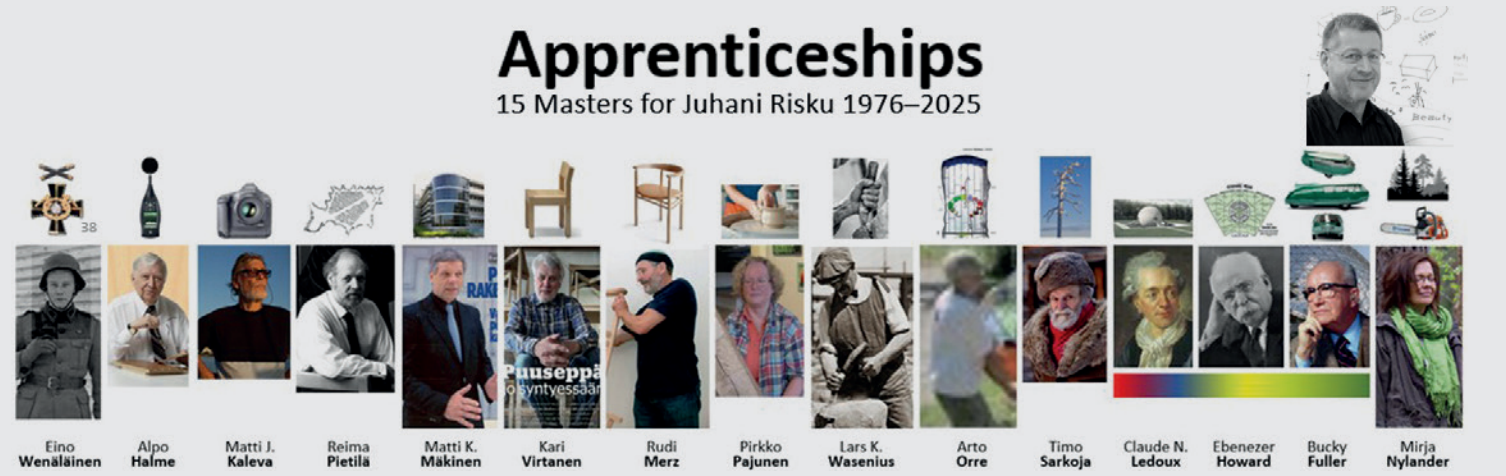

15 Juhani Risku's masters from 1976 to 2025, from left to right: turbine mechanic, acoustician, photographer, architect, carpenter, cabinetmaker, ceramist, stone carver, stained glass master, Arctic tourism master, garden city specialist, systemic systems planner, and forestry specialist. The collaboration with the masters has lasted from two to five years.

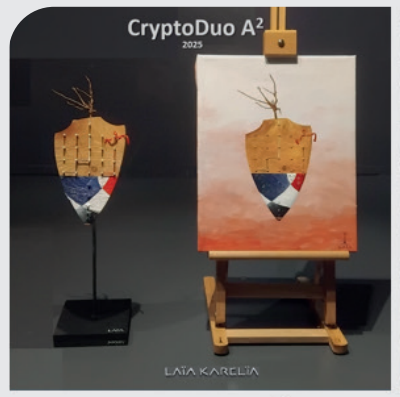

CryptoDuo A²

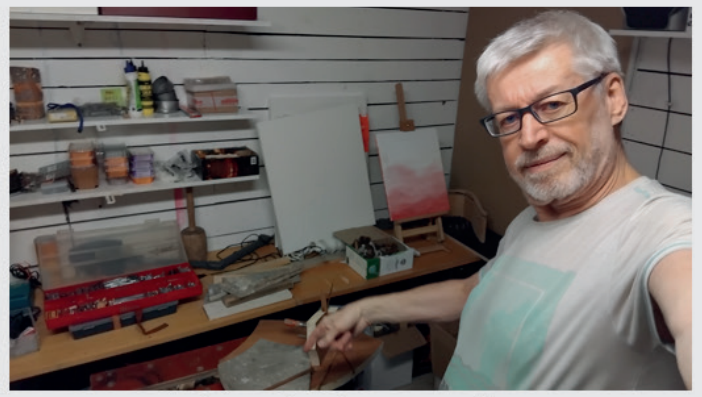

Juhani Risku a.k.a. Laïa Karelïa

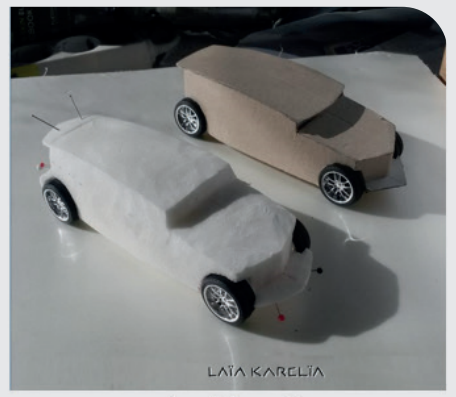

RodMobile A & B

A work that is impossible to fake CryptoDuo A, Risku in his workshop with CryptoDuo and RodMobile, an IKEA-style build-your-own electric car on a standard chassis. It is worth changing professions ten times, after which their knowledge, skills, understanding, criticism, and leadership are always in use.

# Creativity is.
## Thoughts on the essence of creativity, how it is experienced, and what it might become.

Anne Stenros

*The Earth* raised me
*The Water* supported me
*The Fire* hardened me
*The Wind* carried me
*The Void* set me free
and
*The Light* gave me
wisdom

**Introduction—Questions and answers**

My mother's creativity blossomed for 95 years—so mine must still be in its infancy. It was only after exploring her archive of thousands of sketches that I began to reflect on what creativity actually is. From her, I learned the foundations of creative work and life: be curious, strong, demanding, inspiring, and generous.

**The Aim** The Japanese concept of GODAI, meaning "Five Great Elements," defines five fundamental forces of life: Earth, Water, Fire, Wind, and Void. Sometimes a sixth is added: Awakening—or Light. I began to wonder: how do these forces manifest in creativity and creative work? The answer surprised me in its complexity and depth.

**The Method** Zeno of Elea is often considered the father of classical dialectics. Originally, dialectic referred to a form of philosophical dialogue used in Ancient Greece. Socrates further refined it in conversations about concepts like good and evil. A central idea of dialectics is *indirect proof*—showing that a view leads to contradictions, thus validating its opposite. Dialectics also became a developmental process, where truth gradually emerges. Aristotle later adapted Socratic dialectics into a method for uncovering truth or winning debates.

A didactic approach builds on constant questions and answers—thesis, antithesis, and synthesis. Through this exchange, truth unfolds. Yet truth itself is not a fixed destination, but an evolving, dynamic process. Opposites are vital forces that drive growth and clarity.

**Result** My approach to creativity is partly dialectical. This felt natural, as it helped me see and understand the deeper layers of creativity. By exploring opposites, I discovered entirely new dimensions in how we define it. My perspective deepened. In didactic thinking, wisdom doesn't come from having the right answers—it comes from the courage to ask the right questions. *The right question leads to the right answer.*

## GODAI—The Five Powers of Creativity

### I Earth: Creativity is not just inspiration —it is learned. *Curiosity*

My mother took me to the Milan Furniture Fair for the first time when I was just 16. We wandered the streets, eyes wide with curiosity. It became a tradition: she took me to fairs and together we scouted new trends, colors, and forms, all the while walking and talking. From her, I learned, quite literally through trial and error, what trend-scouting is, and how to stay interested in the new.

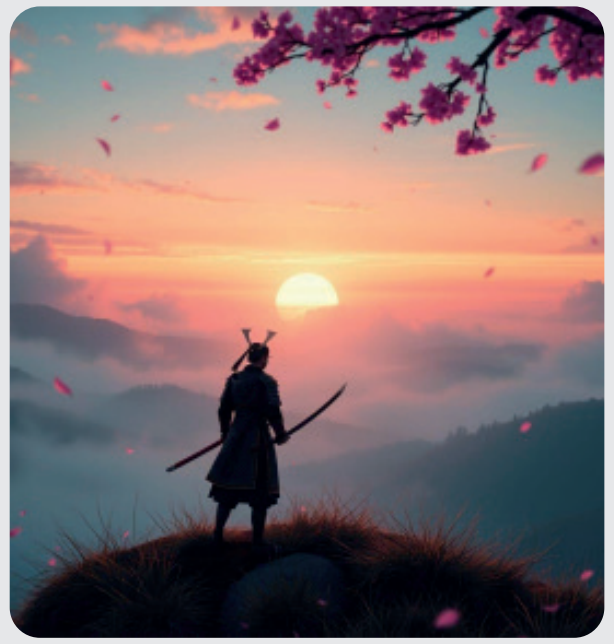

**EARTH: The path of creativity is a solitary one.**
A lone samurai on a hilltop. Made by NightCafe based on images I had made.

Eventually, I realized: you can't see deeply without learning. Curiosity demands time, effort, and a sharp eye. Without a broad base of knowledge and experience, creativity fades. It requires ongoing curiosity toward culture, people, society, and the environment. The engine of creativity is this very curiosity—a drive to explore phenomena, people, and ideas.

To truly grasp a rising trend, you must have sifted through enough information to see it. *Creativity's true fuel is a relentless curiosity.*

### II Water: Creativity is not just dreams —it is action. *Strength*

Practice makes mastery. Creativity is not just abstract thinking—it's grounded in skill that evolves over time. The more a creative mind sketches or explores, the more likely it is to spark profound ideas. Without skill, creativity remains a dream. Practice tests and refines ideas, bringing them to maturity. How often do we return to old themes, only to discover new depth with sharpened skill? The clearer the creative's routine, the faster they enter the flow of free thinking. In evolving their process, they refine both their expression and thought.

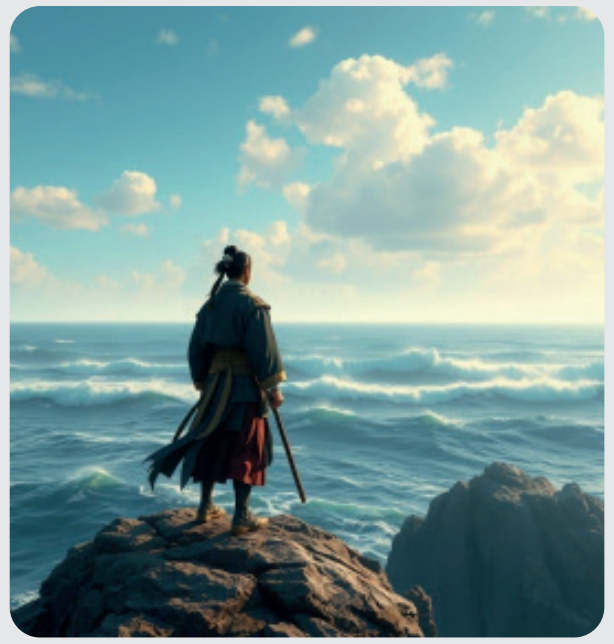

**WATER: The current of creativity is demanding and tests resilience.**
A samurai looks over the sea to the horizon. AS © 2025. Made with NightCafe.

When I was writing my doctoral dissertation, I never imagined at the outset where it would ultimately lead. It was the explorer's mindset: the general direction was known, but the details, opportunities, and realities emerged along the way. That first insight was not the destination—the journey went much further. An artist once said: By the time my work hangs in the gallery, I've already moved on. Creative thought always leads ahead. *Creativity becomes strong through dedicated practice.*

### III Fire: Creativity is not just intuition —it is passion. *Inspiration*

A visionary is a guide and a thought leader. At the heart of creativity lies renewal, transformation—and the ability to lead that change. Intuition inspires and guides us unconsciously, but when combined with creativity, it becomes transformational leadership.

Creative leadership requires a strong vision that sets the direction and defines the goals. Beyond expertise, a creative leader needs deep passion and the ability to inspire others to strive toward shared aims. They discover the means to the end, pursue goals, and trust the power of process on the path to something new. The creative leader naturally renews their thinking, remains curious about many things, and sees possibilities where others have ceased to believe. They inspire trust in colleagues and strangers alike, guided by intuition and faith in success.

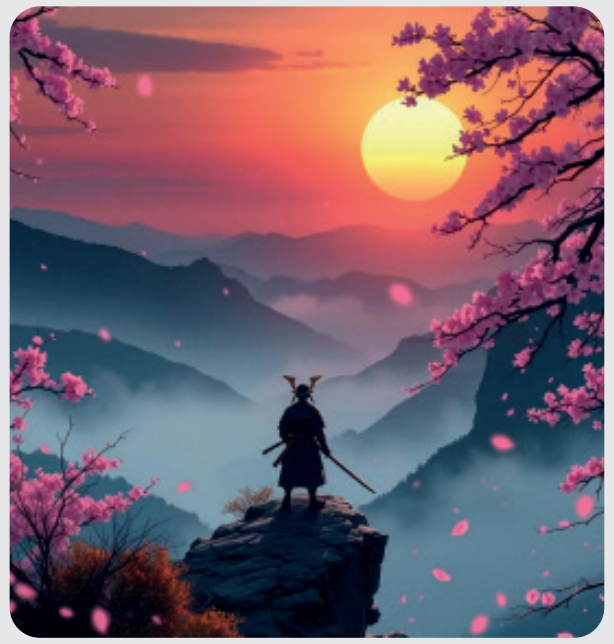

**FIRE: The flame of creativity never dies—it burns eternally.**
A lone samurai on a cliff. Made by NightCafe based on images I had made.

The charisma of a creative leader is often contagious and energizing. Creative energy is born from passion for the subject, the task, the mission. Creativity blooms through the power of inspiration. *The energy of creativity is its power to inspire.*

### IV Wind: Creativity is not just cleverness —it is diligence. *Demanding*

As a young woman, my mother worked two jobs: by day, she designed complex interiors for public spaces; in her spare time, she devoted herself to her true passion—furniture design. She drew and designed almost every waking hour. A sketchbook was always within reach, along with stacks of professional journals and books. She took notes daily, alongside her sketches, and our family holidays were centered around sites she wished to study and explore.

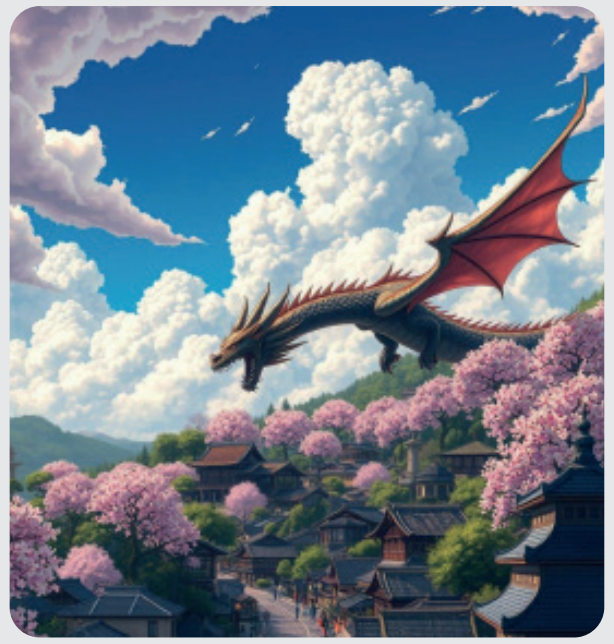

**WIND: Creativity challenges us to grow and bloom.**
A Majestic Dragon. Made by NightCafe based on images I had made.

My mother was a lifelong learner—curious and passionate. Lifelong learning came naturally to her. Cleverness may come easily, but mastery is hard. Creative growth demands constant practice and true diligence. Before the internet, my parents took me on study trips to Japan, the United States, and throughout Europe. Our destinations were architectural gems and the newest highlights of design.

A creative mind requires not only diligence but also effort, time, and total dedication. Creativity is not a job or a role—it is a lifelong calling, a spiritual and skillful pursuit. The goal is a complete inner vision. *The challenge of creativity lies in how much it demands of us.*

**V Void: Creativity is not random —it is purposeful. *Rewarding***

The fifth element of Godai, the void, represents insight and creative liberation from rules. In emptiness, creativity blossoms freely. A famous Japanese architect once said that as a child, he lived by a small lake. To meet his friends, he had to walk around it. Later, he intentionally designed a similar open space into his architecture—a symbolic "lake" at the center of his buildings.

This empty space becomes a home for the creative spirit we all pursue. Creativity does not unfold randomly; it has its own internal logic and direction. While its ideas may seem scattered at first glance, the purpose behind them is focused and clear.

This is where creativity's greatest reward lies: self-guidance and the sharpening of inner vision. A creative person knows when a piece is finished. They feel the moment when the work brings them home—to the idea or seed they set out to find. As my mother once said, "I stop the line before it begins to wander." *Creativity's reward is found in its destination.*

**VI Light: Creativity is not only for the gifted—but for the playful. *Enlightenment***

The sixth force is enlightenment—or light—which means seeing and understanding the true nature of things.

Research indicates that 98% of five-year-olds perform at the level of creative geniuses. However, by age ten, only 30% maintain that level of creativity. Additionally, studies demonstrate that children who play freely in nature are generally more creative in school—and later in life. Nature serves as an ideal playground for fostering creativity and imagination.

Climbing rocks, building tree forts, paddling rivers, collecting sticks, stones, berries, or mushrooms, wandering forest trails, identifying birds and plants, working with natural materials—all of these ground the child in nature, making them part of it. *Nature intelligence*—or "nature smart"—is an essential part of being human, and its importance continues to grow.

Creativity is deeply rooted in both nature and play—not in rare talent alone. It requires a playful mind that dares to test ideas, combine them, and apply them in new ways. The more experience one has with creativity, the more freedom they gain to break rules and invent anew. Only in play are we truly free—when we leave behind the need for certainty. *Creativity is born in nature and nurtured by play.*

*"I keep six honest serving-men
(They taught
me all I knew);
Their names are
What and Why and
When and How and
Where and Who."*
—Rudyard Kipling

**Questions—On Creative Future**

Creativity, as a human trait, a form of knowledge, and a skill, is one of the most essential elements as we build the future of humanity. Architect Cedric Price began his lecture in 1966 with the now famous line: "*Technology is the answer, but what was the question?*" This very same question still applies today and tomorrow.

Creativity is not the answer to everything—it requires the right questions. One of the most important future skills will be the ability to ask creative questions. Research shows that by 2030, creativity—along with curiosity, lifelong learning, and imaginative thinking—will be among the most vital skills in the workplace, alongside technology.

The true power of creativity lies in its diversity; the creative ecosystem is like a natural forest—it nourishes, networks, grows, renews, strengthens, and allows all plants to bloom. Nurturing nature's wisdom is part of both human creativity and human understanding. *Creativity is here and now—its goal lies in the future.*

### 1. Who?—The Creative Individual

Creative people are often highly disciplined and demand much of themselves and others. They envision the outcome or the work "complete" through their mind's eye and are willing to invest whatever it takes to realize that vision. But above all, *creativity requires courage* and the ability to think, act, and create independently and with purpose.

Meaning is what drives the creative individual—and gives the strength to face resistance and challenges. The creative individual naturally bears responsibility for *stewardship in creativity* and its visionary execution.

### 2. What?—The Creative Connection

At the heart of creative work is the *creative experience*—a state of mind filled with insatiable curiosity, determined perseverance, infectious passion, continual growth and learning, openness, tolerance, and an endless sense of play.

These qualities—*curiosity, determination, enthusiasm, growth, openness, and playfulness*—are the central ingredients of the creative experience. They empower the mind to imagine new ideas, projects, and solutions.

### 3. Where?—Creative Communities

Creative experiences multiply in *creative communities*, where collaboration and interdisciplinary exchange allow creativity to flourish. Togetherness and shared experience are growing trends—and will continue to gain importance. In today's digital and tech-saturated world, people crave physical presence and hands-on collaboration alongside virtual creative platforms.

In this kind of work, shared creation becomes tangible. The result is a *creative ecosystem* where digital and physical worlds and communities engage in real-time interaction.

### 4. How?—A Creative Society

The *democratization of creativity* is a powerful future trend. New AI-driven and digital tools lower the threshold for engaging in creative work and offer flexible new models and solutions. Interaction with AI quickly brings ideas to life and allows for rapid testing and iteration. Successful creative

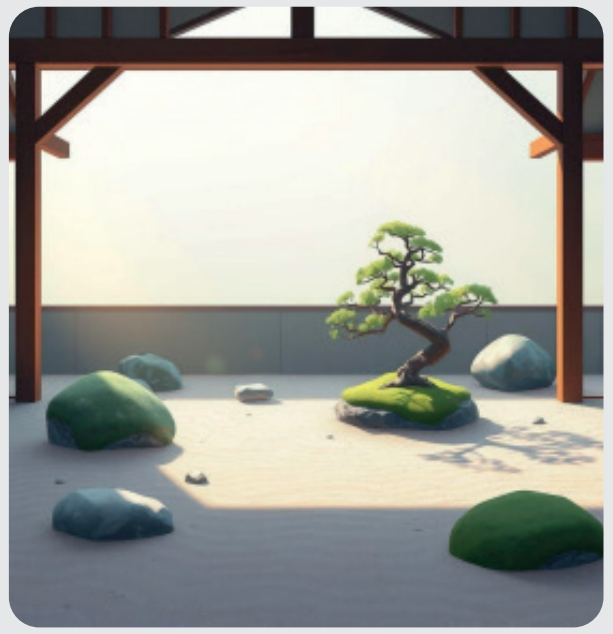

**VOID: The starting point of creativity is to begin from emptiness.**
A Serene Zen Garden. Made by NightCafe based on images I had made.

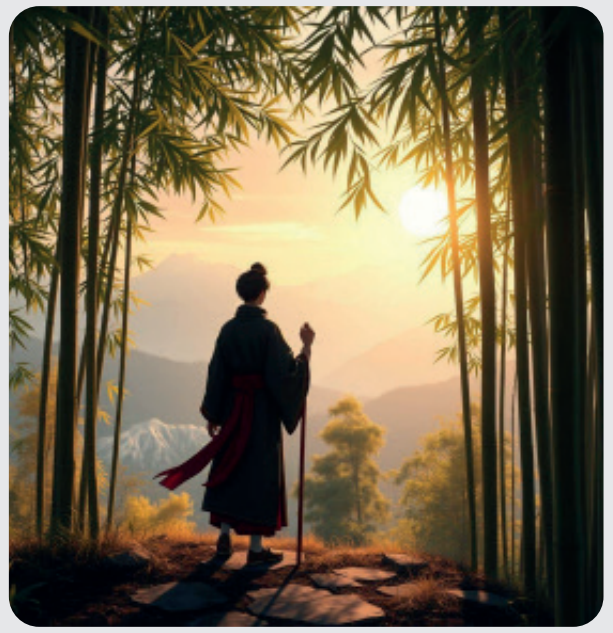

**LIGHT: The goal of creativity is to move toward the light of wisdom.**
A Lone Traveller. Made by NightCafe based on the images I had made.

work will increasingly be a co-creation between humans and machines.

The role of academia will be to expand creative research, address questions of digital ethics, copyright in AI-generated work, and promote creative entrepreneurship. New tools and opportunities require new knowledge, guidance, and ethical frameworks. A *creative society* must be consciously and collectively built to ensure it is inclusive and sustainable.

### 5. When?—Creative Entrepreneurs

Looking toward 2030, the influence is shifting from the *Creative Class* to independent creative individuals. We are entering the *Creator Economy*—a world where artists, influencers, and creators form direct relationships with audiences, bypassing traditional gatekeepers.

Creative individuals are building new virtual art markets and immersive digital art worlds. Augmented reality (AR) and virtual reality (VR) are enabling rich, interactive storytelling in art, fashion, and music—merging digital and physical into one seamless creative expression. By 2030, creative industries will stand at a crossroads: either walk the road of collaboration with technology—or part ways. Either way, *AI-powered creativity* will shape our future understanding of creativity.

### 6. Why?—Creative Understanding

Why is creativity so essential? Because we face unprecedented global challenges—the climate crisis, geopolitical tensions—without ready-made solutions. Creative thinking and doing are our tools for renewal.

*Ethical creativity* and *ethical design* are emerging frameworks with deep historical roots. Don Norman, the father of UX Design, now advocates for *Humanity-Centered Design*. Victor Papanek, in his book "Design for the Real World" (1971), already insisted that design must support inclusion, social justice, and sustainability. Creative thinkers, makers, and advocates have always stood at the frontier of the most meaningful questions. It takes creative courage to name these questions—and even more to search for brave, bold answers.

**Summary: Creative Superiority and Society**

In the end, I asked a large language model what it believed to be the most important qualities of creativity. I wanted to test myself—had I forgotten something essential? The answer came quickly: the qualitative dimension of creativity—*originality*. And that was number one on its list of five.

*"Originality—the ability to generate new, unique, or unconventional ideas that go beyond the usual patterns."* —ChatGPT

At the heart of creativity lies originality—the authenticity of the idea or insight. This is what defines *creative excellence*: when a solution or its application is completely new and one of a kind. The value and meaning of creativity stem from its uniquenesshow exceptional and novel the execution is.

Many creative innovations have transformed the course of human development through unprecedented solutions. A truly original creative mind is a deeply human trait and a distinctive aspect of personality. The idea of the "creative genius" is shaped by role models—acclaimed figures in art and science—who have left their mark on humanity through invaluable work. *Original thinking is the true strength of creativity.*

I also asked the AI what it sees as the benefits of creativity. Surprisingly, the list focused on individual gains: enhanced mental well-being, increased self-confidence and self-expression, and lifelong learning. It also noted communal benefits such as stronger collaboration, greater adaptability in the face of change and uncertainty, and the ability to identify opportunities in challenging circumstances. As a summary, the language model offered:

> *"In essence, creativity empowers you to shape your world, solve meaningful problems, and live more fully—both individually and collectively."* —ChatGPT

A creative society is made up of:

**1. Individual and collective creative intelligence,**
**2. Authentic, creative interaction,**
**3. An elevated level of societal innovation,** which helps solve key issues, and
**4. The community's capacity for creative transformation.** Together, these generate **creative abundance** and **impact** within society.

We could define the creative equation as one in which everything centers on the individual and collective **experience of creativity.** Its energy (level of innovation) is directly proportional to individual creative intelligence and the community's capacity for change. Together, they are the key to the future. *Creativity is humanity's power to renew itself.*

**Summa Summarum**

In closing, I want to highlight a few key insights about creativity. True creativity and creative action require *creative presence* and a lived *experience of creativity*. This is often described through the concept of *flow*—being carried by the current of creative energy. That sense of presence and creative immersion enhances our well-being and strengthens our capacity to face challenges. *Creative problem-solving* finds new opportunities even in difficulty—and awakens a sense of *creative hope*. It is through *creative acts* that we shape our future.

Yet above all, I want to emphasize *creative courage*. The greatest of all samurai, Miyamoto Musashi, was not only a warrior but also a skilled and revered *sumi-e* artist. The undefeated swordsman once said that *art, skill, and combat are the same*. Musashi believed that to master the art of battle and its strategy, one must also master many forms of art and craft. Both combat and creation, he said, require careful planning and anticipatory thinking. In both, courage plays a vital role—it is the spiritual resource of both the fighter and the artist. Without it, we cannot endure.

In memoriam Pirkko Stenros (1928-2024)

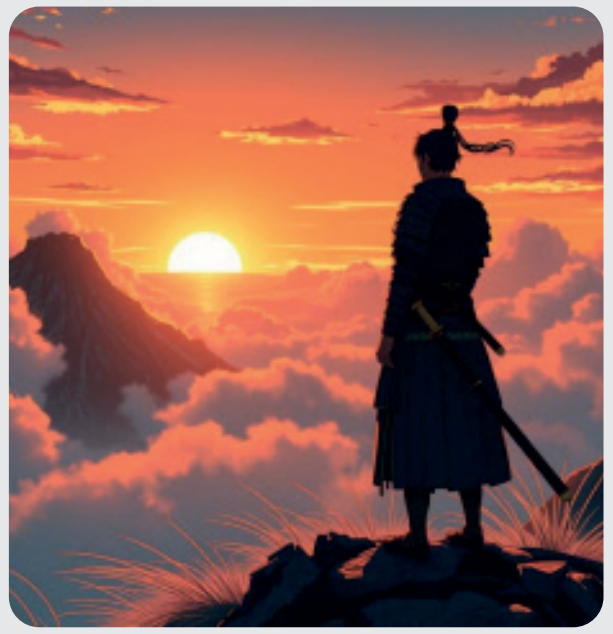

**Creativity always sees hope—its future is bright.**
A lone samurai on a mountain top. Made by NightCafe.

**Epilogue**

Every creative individual has a mission—a purpose greater than their work: to improve society in some meaningful way. Every creative person seeks to achieve something beyond themselves, to leave a legacy—not just a footprint, but a *handprint*, or even a *heartprint*.

Everyone's creativity is equally important, rich, and meaningful—if they dare to open up to it. Every truly creative act has its place and purpose, if one perceives it correctly—and every authentic creative act is about creativity itself and born of it. *Creativity is.*

## Post Scriptum

*"To every thesis there is an antithesis, and to this there is a synthesis. Truth is a never-dying process."* —Rollo May

After finishing the essay above, I stumbled upon a classic book on my bookshelf—*The Courage to Create*, written by the renowned psychoanalyst Rollo May in 1975. In his book, May outlines four distinct types of courage. The first is physical courage, often associated with strength, but May also interprets it as the ability to listen to and understand the body. The second is moral courage, rooted in the capacity to feel compassion toward the suffering of others. Third, he identifies social courage as the ability to form meaningful connections with another individual or a community. The fourth, and in May's view the most vital, is creative courage—the courage to build a new society based on new forms and models.

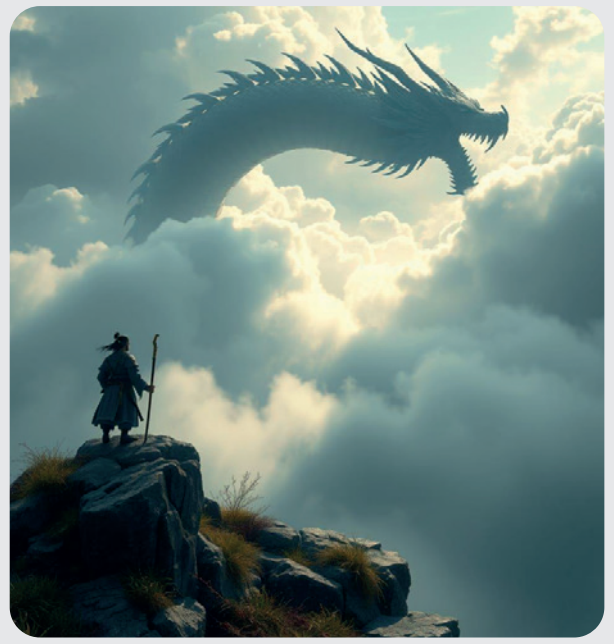

**Creative courage challenges old truths and changes the world.**
A lone samurai and a dragon. Made by NightCafe.

Creative courage is the act of challenging the prevailing status quo, and May believes artists are the first to lead the way. According to him, creative courage is our gateway to eternity—through it we gain insight into the fundamental themes of life and death. Reading May's words, it becomes clear that creative sensitivity and courage go hand in hand. Courage is required to see, imagine, and recognize; sensitivity is needed to feel, experience, and act. Both have their rightful place in creative thought and action. Creative courage is the transformative force of life—a current of truth that moves our understanding of humanity and existence forward. It is a natural force, the spirit and breath of creativity itself: always renewing, eternally alive.

# Butterfly and the Beastly —Perspectives on Intimacy, Privacy, Data and Human-Nature Ecologies

Rebekah Rousi

**Foreword**

This short piece reflects the phenomena resting at the heart of the "Emotional experience of privacy and ethics in everyday pervasive systems (BUGGED)" project, kindly funded by the Research Council of Finland. The project itself was inspired by personal experiences of privacy violation and the ways in which ethics themselves may be used as weapons of control, equally as they form the basis of our understanding between *good* and *evil*. The piece jumps between topics, all the time centring on human-to-human relationships either in consensual intimacy, or exploitation and violence. Yet, it is a reflection on how all these spiral concerns on the concrete consequences of relationship-driven data on social, cultural and natural environmental ecologies.

**Introduction**

The *Butterfly: Glo-cal Effects of Data, Energy and Industry* focuses on the butterfly effects of how we design, produce and engage with digital systems. This has a profound impact on not only environmental ecosystems, but also socio-cultural ecosystems. One perspective that has not been mentioned openly here, yet is pervasive within all the works present in this exhibition, is that of privacy and personal space. The works see and hear you. The works breathe and expand with your data—your movement, your emotions, and your voice. Perhaps the flow of the artworks indicates that they eb in response to your existence. Perhaps, they pleasantly ignore you. But, you have no idea as to whether or not they acknowledge you, part of you, or account for everything that you think.

Ironically, the more we talk about the outer-environment, the natural ecologies and how our activities—many driven by an irrational greed (not the sustainable business that we strive to embrace)—affect the forests, the water, the land, and now even space, the closer we get to our souls, or *spirits*. We are using these technologies to connect with others. A large portion of our data consumption and provision is generated by the use of social media and dating applications, to stay in touch with loved ones, to spearhead in front of the competition when it comes to work and financial success, and even to spark romantic relationships. In fact, it was thanks to the prospect of developing dating applications that the Internet really took off. The French Minitel, a social network specifically designed to facilitate romantic encounters in the 1980s (Castells, 2009), holds its claim to fame as making the Internet financially viable. This was simply a precursor to what would be some of the most lucrative industries online—mobile dating apps (casual sex apps, De Ridder, 2022) and *pornography* (Donevan, 2021).

### In-head and on-line perversion and ecological impact

Whether it is under your clothes, or under the bonnet (i.e., your deepest, darkest desires), the butterfly effect of data is firmly ingrained in our deepest fantasies. The beast within bursts out and flourishes when people continuously engage in anything and everything to do with sex. This relationship between human desire—the desire for intimacy, or the beastly (aka de-humanized and immoral fetishes that lead to series of abuse chains) and its concrete impact on the natural environment becomes ever more complex, the more entrenched we become in the details of the data. When once upon a time, not very long ago, concerns were raised over unsolicited photography and video-recording of sexual acts (themselves solicited or not, see Powell & Henry, 2017), or unsolicited use of media portraying individuals in these acts (Nicklin, Swain & Lloyd, 2020), these days all it takes is a photograph or two to create a deep fake porn video (Maddocks, 2020). You can be a movie star without even knowing it.

It then must make you wonder about the intentions and possibilities of *all* connected media, all sensors, and all videos that happen to catch you as you walk by. The giant data collecting black hole keeps absorbing astronomical amounts of data all the time, from what we actively upload, generate and even search. All the while, a lot of this data is driven in one way or another by our need for relationships—loving and affectionate, or cruel and controlling. Interestingly, when making a quick and dirty literature search on "internet porn + environmental impact" in Google Scholar (yes, sorry, I used Google with its steep rise of 50% in gas emissions over the past five years [actually more now, considering it is 2025], see Milmo, 2024; and 22.7 billion litres of fresh water in 2024, see Günyol, 2025) revealed that the word "environment" for the most part, refers to the conditions within which individuals are exposed to and consume pornography. Interestingly, there does not seem to be too much attention placed on either the environment in which porn is produced (i.e., the dire conditions of many 'actors' who suffer financially, socially, culturally and from drug-related problems, see Gabriel, 2025; Yaakobovitch, Bensimon & Idisis, 2024—or refer to Sustainable Development Goals 1 [no poverty], 3 [good health and wellbeing], 5 [gender equality], 8 [decent work and economic growth], 9 [industry, innovation and infrastructure], 10 [reduced inequalities], 12 [responsible consumption and production], and 16 [peace, justice and strong institutions], United Nations, n.d.) or the impact on the natural ecologies that multiply, ripple and wave from one corner of the globe to the other. Actually, this question in Google Scholar, "how much data does online porn generate?" does not produce precise results. It took performing the search in the *regular* Google to see the AI Overview that states that already in 2022 pornographic content accessible online equalled to over 10,000 terabytes.

### Relationships, data and privacy

So, this discussion has been jumping, from the human need for relationships and intimacy—trajectories for mass internet use that inevitably has profound effects on ecologies (social, cultural and natural environmental)—and then to an industry that on any level *is* tainted in terms of its environmental impact on human and natural ecologies. As I briefly mentioned above, our production and use of explicit content online has moved beyond the natural, and entered the realm of the *super* [natural] artificial, where for pleasure, for violence, control or revenge, or all of the above, the data we produce knowingly and unknowingly is used against us. We no longer need to be caught in the wrong act, or fall into the hands of the wrong partner, but that partner(s) may be anyone far and wide—one we have no idea we have chosen.

As the cameras trace our steps. As the scanners analyse our faces, and as we provide our portraits to social media, or even institutional websites, we never know who or how our data will be used. The footprints that our acts leave online, or on the earth (i.e., via Internet searches and media consumption) willingly or unwillingly, are an everyday fact of this data-driven reality. And, we never know how that data collection and use will transform us as social or spiritual beings. There is much concern about data privacy, laws and regulations are even being made, but what are we afraid of?

### Conclusion

Well, I will tell you what made me afraid. It is that hijacking of *self* and tarnishing of *spirit*. As someone who has suffered privacy infringement, who has been stalked, harassed, and threatened, I do not want to engage in the Internet for intimacy, let alone the *beastly*. I view my relationships offline as the ones to be cherished, even if, a world apart from my family, the best form of connection is WhatsApp. But, when you have someone or something enter your life, like a vampire without permission, willing to take everything you have and all you are, and *then some*, the feeling of data-driven technology—the type of data that is supplied by me—is, well, often the feeling of a giant molestation machine. Past violations remain like stains on your psyche, and taint all future encounters you have with any phenomenon or technology that remotely reminds you of the crime. Violations steal our minds. Google (and all the other large language models, and so-called artificial intelligence) steals our minds. Our minds are having concrete negative effects on natural ecologies—social and biological. Our minds, perverted desires, and media-driven acts to control others are literally killing our planet.

I want to feel comfortable with technology that learns from me, and can interact with me in natural and affective ways, but I feel infringed. Not only can my data be translated into the beastly, but I *see* the beast inside it all. Keep your self-driving cars, passport free border controls, social robots, and self-cleaning washing machines. I want to keep me, thanks. My fallen soul wants out and up.

# Educational Programme

## Educational programme
## —Five Victorian Frogs

**Designed and deployed by Toija Cinque and Peta White with QuiverVision**

Image © Toija Cinque and Peta White

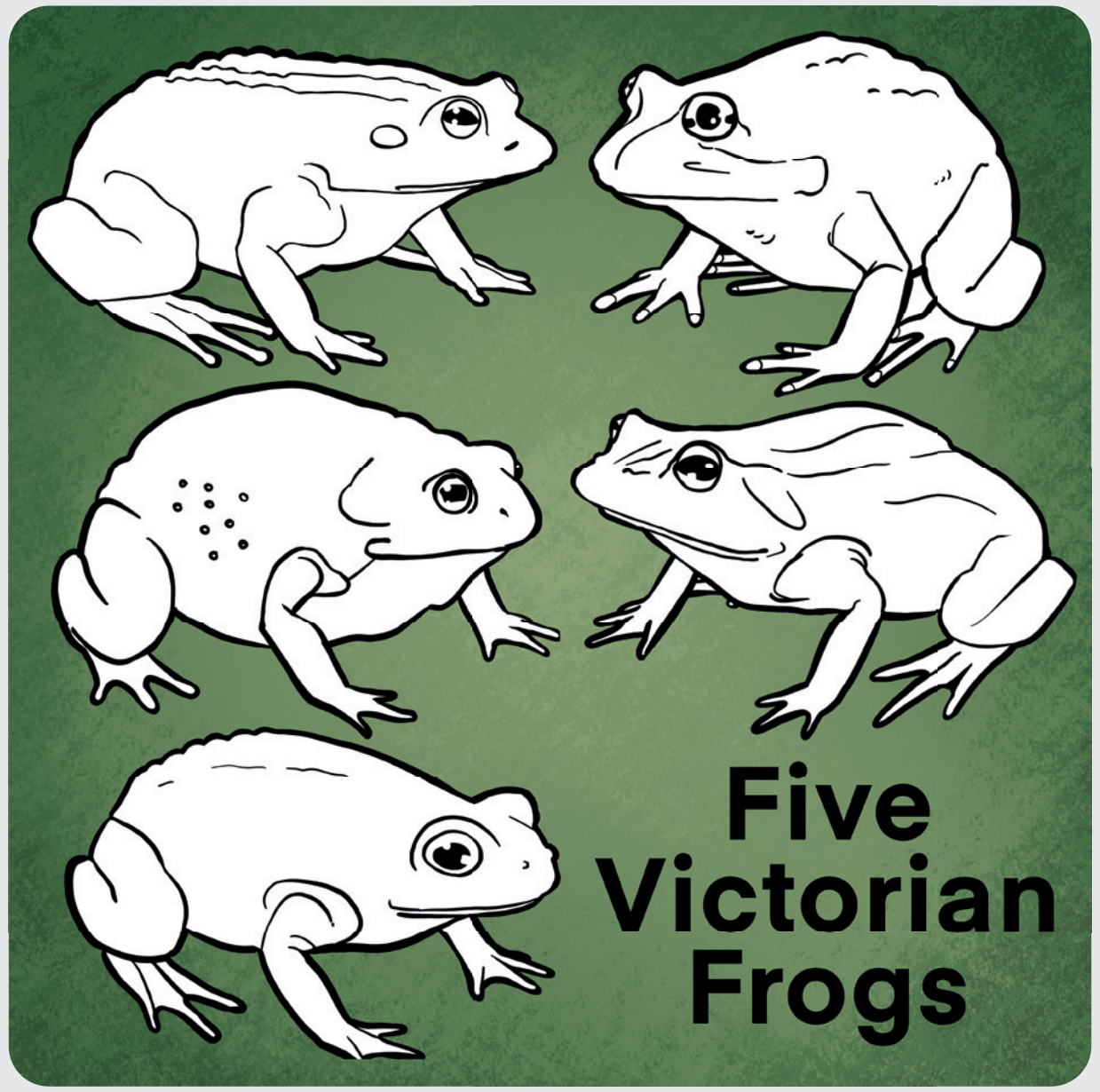

As part of the Butterfly educational programme, visitors aged five to ninety-five are invited to engage with the natural world through a guided exploration of the exhibition and the Five Victorian Frogs Immersive Learning Experience (ILE). This interactive AR installation brings five of Victoria's endemic frog species to life—each a vital bio-indicator of environmental change.

Following the experience, participants take part in a reflective discussion on ecological butterfly effects, climate change, and species interdependence. The session concludes with a short survey, encouraging critical thinking and personal connection to environmental futures. Designed to inspire curiosity, care, and climate consciousness, this learning experience blends digital media, scientific insight, and playful inquiry.

**To start with—what is a 'butterfly effect'?**

The Butterfly Effect shows us how small changes can lead to big impacts. Just like a butterfly flapping its wings might shift the weather far away, our everyday actions—good or bad—can ripple through the environment. The Butterfly Effect is, however, often misunderstood as a symbol of chaos—but meteorologist Edward Lorenz meant something deeper. He showed that within seeming disorder, there are patterns—and that small changes can lead to new, predictable futures if we pay close attention. In this spirit, the Five Victorian Frogs Immersive Learning Experience invites children and families to explore how tiny shifts in the environment—like cleaner water or a restored habitat—can ripple out into real, lasting change. Through augmented reality and playful roleplay, participants learn that their choices matter, and that creative action can shape a better world.

**Five Victorian Frogs: An Immersive Learning Experience (ILE)**

This work responds to the urgency of ecological awareness through a synthesis of digital art practice and scientific inquiry. Five Victorian Frogs: An Immersive Learning Experience (ILE) is a new media installation that leverages augmented reality (AR) to reanimate five endemic frog species of Victoria, Australia—each a bio-indicator of ecosystem health and environmental change. Through animated, interactive visualisations, audiences encounter the Baw Baw Frog, Bibron's Toadlet, Common Eastern Froglet, Growling Grass Frog, and the Pobblebonk Frog within layered ecosystems rich in scientific detail and sensory engagement. Drawing on research into immersive learning and science education, this installation is both a pedagogical and aesthetic intervention. It embodies a response to contemporary educational inequities intensified by global lockdowns, and to the broader disaffection from the natural world, often termed 'nature deficit disorder'. In doing so, it addresses gaps in place-based scientific education, aiming to reconnect young learners—and broader publics—with the biodiversity of their immediate environments. By integrating interactive technologies, ILE extends the potential of digital media art to foster care, understanding, and protection of fragile ecosystems.

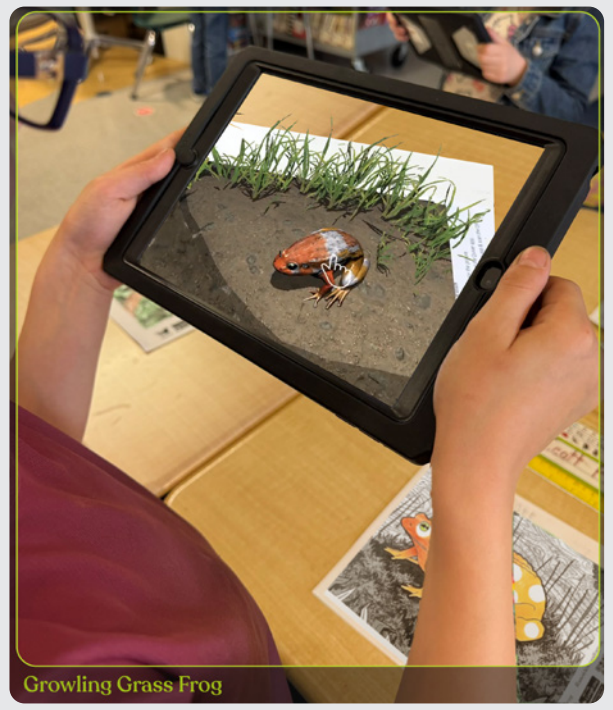

Image © Toija Cinque, Peta White (Deakin University) and QuiverVision

**Special note about the research behind the educational programme**

This research unites scientists and teachers to translate contemporary climate-related research practices into a novel curriculum approach that emphasises deep knowledge of science and scientific practices, values and decision making designed to prepare students for a 21st century marked by complex work futures and major socio-scientific challenges related to climate change. The research addresses the pressing national issues of developing students' STEM engagement and competencies for fast changing work futures, and for decision making and action regarding climate-related challenges.

Engaging scientists with school communities, researchers in Australia are exploring how Climate Change Education can be framed and developed within an expanded conception of science education to engage students in a) contemporary scientific R&D practices related to Climate Change; and b) decision-making and actions based on appraisal of scientific findings and solutions alongside broader knowledge and value systems. Learning sequences will be co-developed with the following features: video and media representation of scientists' research including their motivations and accounts of societal complexities framing the science; student engagement with scientific knowledge and practices utilising scientific papers and authentic data; the incorporation of innovative, guided inquiry pedagogies that promote student reasoning and learning; discussion of/reasoning about different values and stakeholder positions on sustainability issues framing the science and decision-making on personal and collective approaches to these issues, focused on developing student agency.

# Symposium & Workshop

# Extra Programme

## Symposium & Workshop

**Hybrid symposium & Workshop with keynote speakers Juhani Risku & Anne Stenros**

**Jointly organised by University of Vaasa, Deakin University, Umeå University & Tampere University**

As part of the academic program accompanying the Wasa Futures Festival's 2025 New Media and Performance Art Exhibition *Butterfly: Glo-cal Effects of Data, Energy and Industry* (11–31 August, Wasa Innovation Center), we invited submissions of extended abstracts for paper presentation at a one-day symposium. We especially encouraged scholars from a wide range of disciplines including engineering, business, communication, information systems, the arts and humanities, cognitive science and more. In particular, we sought papers that would adopt an artistic lens of inquiry, i.e., breaking the boundaries set within the respective disciplines, looking at the information available, and making sense of circumstances in new ways, to convey a holistic story of how our world is morphing through contemporary data, energy and corporate systems. The symposium was warmly opened with a keynote from Juhani Risku, followed by enriching presentations and discussions, and enhanced with another keynote by Anne Stenros on creativity and courage.

**Butterfly Symposium Context**

Not just a beautiful insect, the term 'butterfly' is used to describe how small changes in the starting conditions of a system can lead to large, unpredictable differences in outcomes (Palmer, Döring & Seregin, 2014). That is, even tiny modifications to one part of a system, may mean exponential, chaotic effects in other parts. 'Butterfly effect' was coined by Ed Lorenz, a meteorologist in the 1960s, but rather than referring to small changes creating unpredictable results, Lorenz actually meant that we could accurately predict the future in certain complex systems such as the atmosphere. Here, we adopt the term to both refer to sensitive dependencies on initial conditions as per Chaos Theory (Yorke & Li, 1975), to both describe how discrete differences in the starting state of a system can lead to radically different outcomes, as well as to describe how these radically different outcomes can somehow be predicted through artistic exploration. What Lorenz deemed as the "real butterfly effect" refers to the fact that there are limitations to the combinations of future unravelling that we may observe. In the case of the natural environment and globality, while we might find effects surprising, we can still predict alternative realities.

With emphasis on digital technology and data-driven systems (i.e., AI), we aimed to illuminate the omnipresence and physical implications of data and its potential for creative appropriation. The exhibition was positioned to reveal the untapped possibilities of digital realms, to reflect and enhance our ecological sensibilities, while converging the artists, their artworks, scholars and practitioners from across disciplines (including engineering, energy, computer science, business and communication). By exploring how human interaction with digital infrastructure can serve as a conduit for critical and creative practices that honour and advance ecological harmony, the exhibition asked audiences to contemplate the role of technology in a sustainable and equitable future. At its beating heart, it acted as a call to action—a reminder that in the quest for ecological balance, creativity and innovation can lead the way in transforming our collective consciousness and societal structures.

The symposium was a stage for open and interdisciplinary dialogue around how contemporary technologies and industries are reshaping the environment and human experience.

# Extra programme

13.8.2025

## Lights-Off Night

On Wednesday, August 13th, as part of the Butterfly Exhibition, the lights were turned off, but the evening was anything but dim. In that carefully curated moment of pause, guests stepped away from the constant glow of the digital world and into a space created for reflection, connection, and pause.

As electricity-powered brightness faded, performances and artefacts took center stage with each inviting participants to consider what life feels like beyond screens and circuits. It was a moment to reflect not just on silence and stillness, but on the real, physical cost of our digital lives. This is because, while our devices feel weightless, the infrastructures behind them are anything but. From the minerals mined for smartphones, to the vast energy demands of data centers powering cloud storage and AI, our digital convenience comes with an [in]visible footprint. Streaming a movie, sending an email, backing up photos—each action consumes electricity, water, and finite materials. These systems, though often hidden, leave a lasting mark on the Earth.

The event offered a chance to ask: What does it mean to give back? How do we balance innovation with responsibility?

Coinciding with the Wasa Future Festival Alumni evening and in the context of the Butterfly exhibition, the gathering became more than just an event—it was a shared reflection upon the costs of progress. For those who drifted over to Butterfly, the darkness was not empty—it was full of questions, insight, and a call to conscious living.

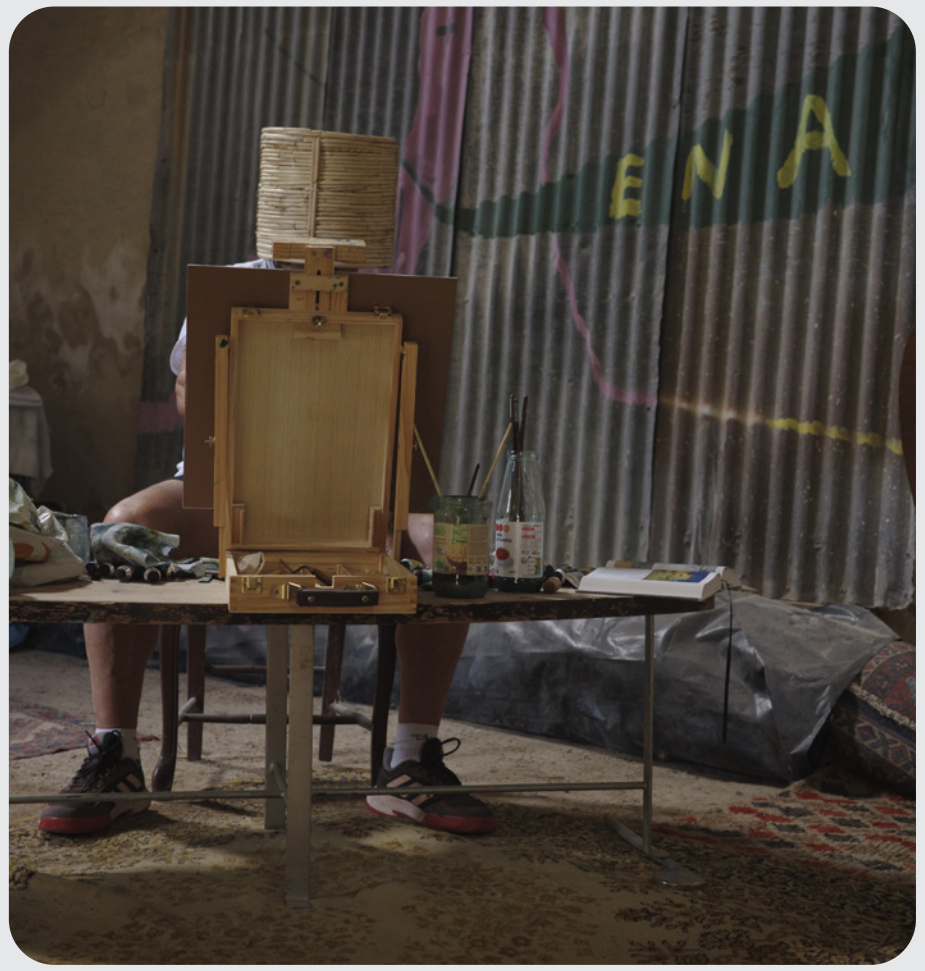

Domenico de Clario in performance with blind-folded piano playing sung paintings. © Domenico de Clario, photo taken by Jane Hirshfield.

14.8.2025 5:30pm—10:00pm

# Night of the Arts

In 2025, the City of Vaasa lit up with colour, sound, and creative possibility as Butterfly took centre stage during Night of the Arts—Finland's beloved celebration of artistic expression in all its wild, untamed brilliance. Part of the global constellation of White Night festivals (including Australia's iconic editions in Melbourne and Ballarat), this electrifying evening transformed the Wasa Innovation Centre into a living, breathing organism of sound, light, movement, and meaning. From twilight to midnight—and well beyond—audiences were swept into a dreamscape of immersive digital art, embodied performance, interactive storytelling, and unexpected encounters. The Butterfly exhibition unfurled its wings, inviting the public to explore the glo-cal reverberations of data, energy, and industry through art that shimmered, provoked, and inspired.

There were no velvet ropes or hushed galleries here. Only open doors, open minds, and open skies. Visitors danced with augmented frogs, listened to ecological lullabies, and lost themselves in the hypnotic loop of 360° visual worlds. Children giggled in glowing corners, artists performed among data streams, and scholars sipped wine while debating the poetics of critical infrastructures. It was more than an event—it was a happening, a rupture in routine, a shared dream of what art can do when given the city as a canvas and the night as its muse.

As part of the evening, the Butterfly exhibition remained open until 10:00pm, featuring durational performances by Domenico de Clario, whose work invited visitors to slow down and listen deeply. A special installation by the visiting artist collective Lava Cloud—featuring Pablo Kobayashi (Mexico) and Lucia Aumann (Argentina) further illuminated the evening. Their large-scale, glowing soft sculptures challenged perceptions of material and form, raising deeper questions about the systems and structures that underpin our built environments. The night culminated in a spontaneous, multi-site gathering that afforded cultural artistic exchange that extended Butterfly's reach across Vaasa. From one side of the city to the other, a mini-Butterfly emerged: spaces were activated, boundaries dissolved, and creativity flowed freely. And Butterfly? It didn't just take part. It soared.

## About VME at the University of Vaasa

The Virtual and Mixed Reality Environment (VME) at the University of Vaasa is a cutting-edge, open-access innovation hub where students, faculty, and industry partners explore the intersections of technology, creativity, and learning. Equipped with state-of-the-art VR/AR tools, eye-tracking systems, game design software, and robotics, the VME supports a wide range of educational programs and collaborative projects across disciplines—from game studies and human-computer interaction to brand management and energy systems simulation. As a dynamic space for experimentation, ideation, and hands-on development, VME bridges academia and real-world innovation, fostering a culture of digital exploration that is both inclusive and future-facing. It plays a key role in shaping the creative and technological landscape of the University and beyond.

## About Deakin's CDII

Deakin University's Critical Digital Infrastructures and Interfaces (CDII) Group is an interdisciplinary research collective at the forefront of investigating how digital systems shape—and are shaped by—social, cultural, ecological, and political forces. Fusing critical theory with technological inquiry, the group explores emerging technologies such as AI, data infrastructures, and human-machine interfaces, not just as tools, but as transformative agents in society. CDII's work challenges conventional narratives of innovation, offering expertise on digital ethics, emerging screens and digital interfaces, collaborative methodologies, all towards inclusively designing our socio-digital futures.

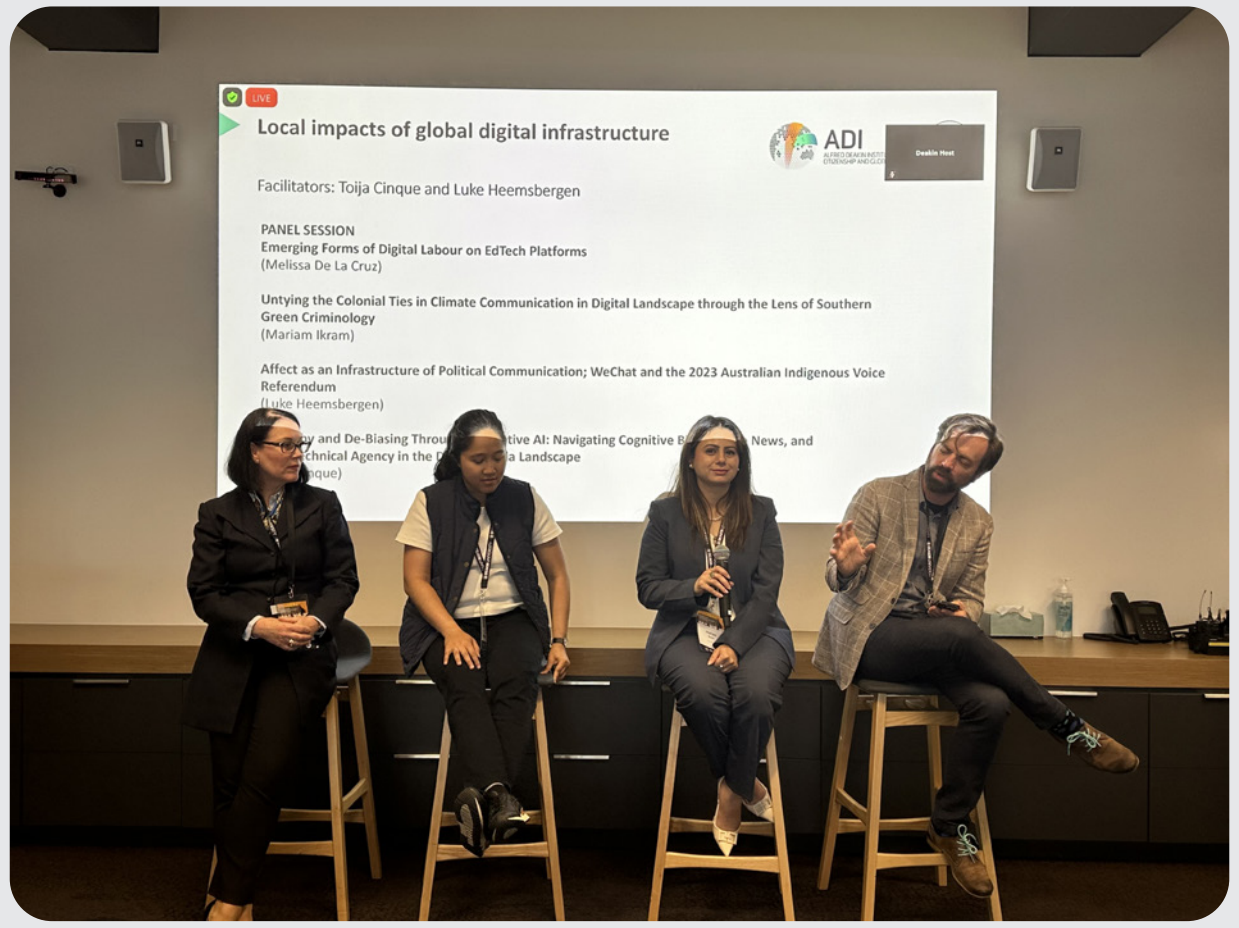

Co-Leads Toija Cinque and Luke Heemsbergen host a Global Digital Infrastructures symposium (2024). Image Credit: CDII

## Acknowledgements

*Special thanks goes to*

University of Vaasa

VME

Deakin University and Critical Digital Infrastructures and Interfaces Research Group

Game Research Lab, University of Tampere

Umeå University

Wasa Innovation Center

Pohjanmaan liitto

Aktia Foundation

Central Organization for Finnish Culture & Arts Associations—KULTA ry

City of Vaasa

Technobothnia

Platform

Elokuvakeskus Botnia

Malakta Center for Creativity, Culture, Arts and People

UPC Printing

348391 [Emotional experience of privacy and ethics in everyday pervasive systems (BUGGED)]

358714 [Multi-faceted ripple effects and limitations of human-AI interplay at work, business and society [SYNTHETICA])

## Partner focus: Platform

A Vaasan champion for contemporary arts, Platform is an artist-run initiative that has strengthened the local arts community since its foundation in 2000. In addition to hosting festivals, workshops and artist talks, Platform provides international artist residencies and an art space, assisting with crucial production, promotional and equipment support for exhibitions, events and projects.

Platform plays a significant role in creating impact through the arts by encouraging awareness and new interactions, by initiating dialogue between invited artists, community, activists, and theoreticians. A valued partner to Butterfly, Platform has generously introduced us to Lava Cloud—a visually spectacular feature of the Night of the Arts and kindly provided enormous exhibition support—we could not have had such an incredible event without this key local connection.

## Student contributions

We would also like to thank the Communication Studies students at the University of Vaasa for their hard work in establishing the social media outreach and foundations of this catalogue. So, from the media team, Maija Koskinen, Siiri Keltamäki, Olivia Salonen, Reetta Rajalin, and Laura Nissi—thank you very much! From the catalogue team, Alina Sydänlammi, Venla Kirvesmäki, Matilda Latvala, Mea Unkuri, Aliina Ahlfors, and Anni Kautto, thank you very much also! Both groups did a tremendous job in forming the foundations of Butterfly's communicational ethos. More appreciation for our students goes to our Wasa Future Festival interns Mette Rahiala, Ben Broere, and student photographer Alina Sydänlammi.

## Community contributions

We are truly grateful for all the support we have received from the University of Vaasa, and the University of Vaasa Foundation, as well as the Deakin University and the Critical Digital Infrastructures and Interfaces (CDII) research group (Australia), the Game Research Lab (University of Tampere), and the University of Umeå (Sweden). We are extremely thankful to the Regional Council of Ostrabothnia, the City of Vaasa, and the Aktia Foundation for their financial support. In terms of equipment we are extremely relieved and thankful to Technobothnia, Platform, and the Elokuvakeskus Botnia. On a personal level, we really want to extend our thanks to Juha Miettinen, Jenni Österlund, Kim Ketola, Pia Murto, Sami Pöntinen, Antti Kivikangas, Rutger Blom and Andre Vicentini for all of your work and professionalism in helping make Butterfly happen.

DEAKIN UNIVERSITY Critical Digital Infrastructures and Interfaces

UMEÅ UNIVERSITY

Vaasan yliopisto
UNIVERSITY OF VAASA

VAASA. VASA.

ÖSTERBOTTENS FÖRBUND
POHJANMAAN LIITTO

Tampere University GAME RESEARCH LAB.

WASA INNOVATION CENTER
Redefining modern life.

TECHNOBOTHNIA

KULTA

OPENINNOTRAIN
INDUSTRY 4.0 | CLEANTECH | FOODTECH | FINTECH

UPC print
COMMUNICATION CENTER

PLAT FORM

EFKC BOTNIA

Aktiastiftelsen i Vasa
Vaasan Aktiasäätiö

Suomen Akatemia

This project has received funding from the European Union's Horizon 2020 research and innovation programme under the Marie Skłodowska-Curie grant agreement No 823971

# WASA FUTURE FESTIVAL

11.-16.8.2025

Butterfly: Glo-cal Effects of Data, Energy, and Industry

An Interactive New Media and Performance Art Exhibition

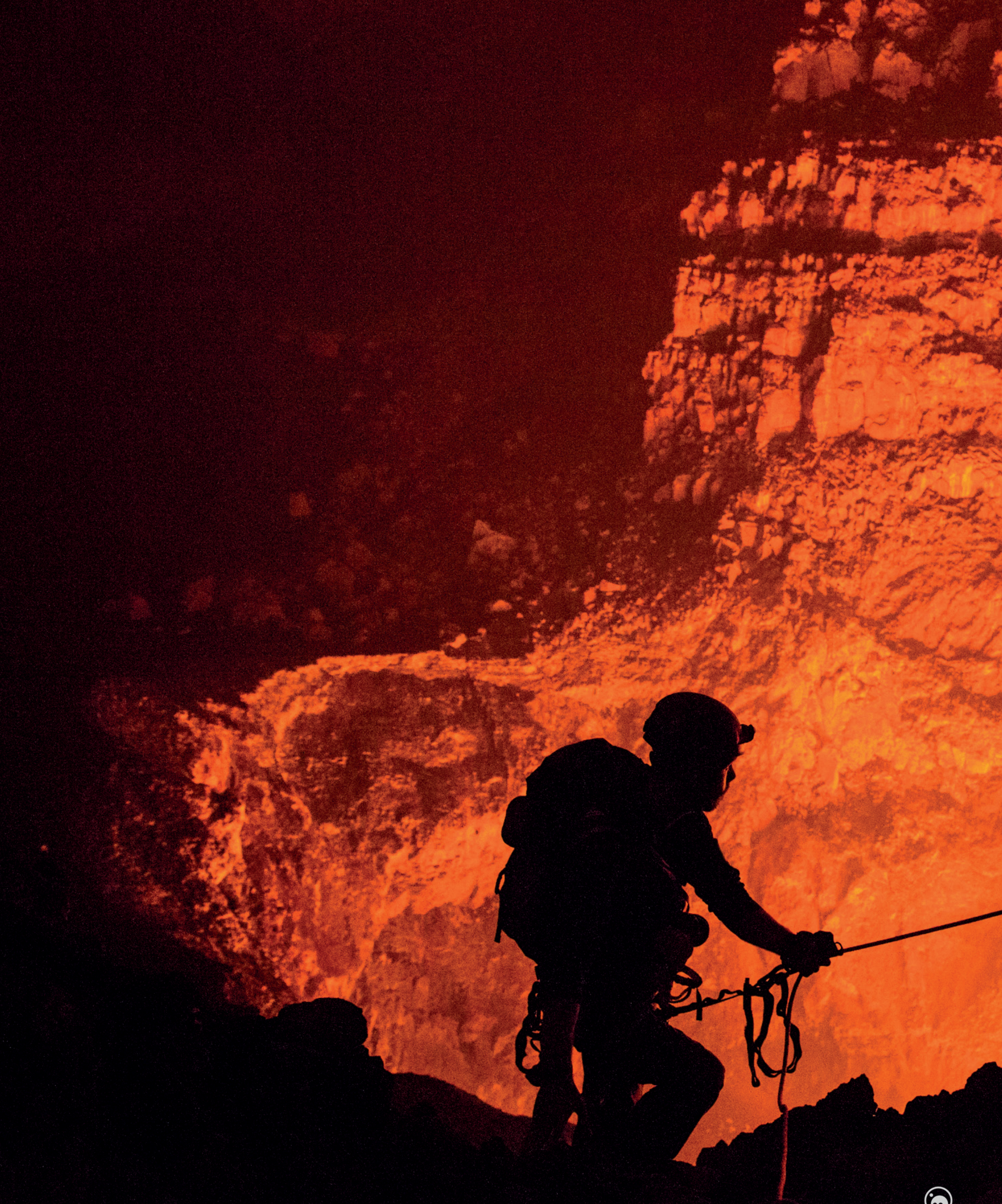

DEAKIN UNIVERSITY

UMEÅ UNIVERSITY

Vaasan yliopisto
UNIVERSITY OF VAASA

WIC
Redefining modern life.

Tampere University GAME RESEARCH LAB.